\documentclass[%
 article,
 reprint,
 superscriptaddress,
 amsmath,amssymb,
 aps,
 prd,
]{revtex4-2}
\usepackage{graphicx}% Include figure files
\usepackage{dcolumn}% Align table columns on decimal point
\usepackage{bm}% bold math
\usepackage{siunitx}
\usepackage{mathrsfs}
\usepackage{lineno}
\usepackage{xcolor}
\usepackage{comment}
\usepackage{adjustbox}
\usepackage[none]{hyphenat}
\usepackage[utf8x]{inputenc}
\usepackage{ucs}
\usepackage{lineno}
\usepackage{multirow}
\usepackage[normalem]{ulem}
\usepackage{subfigure}
\usepackage{rotating}
\usepackage{ragged2e} % Required for \RaggedRight command
\usepackage{xparse}
\usepackage{hyperref}

\newcolumntype{Y}{>{\centering\arraybackslash}X}

\NewDocumentCommand{\ncl}{ >{\SplitArgument{1}{-}}m }{\nclaux#1}
\NewDocumentCommand{\nclaux}{mm}{\ensuremath{\!{}^{#2}\mathrm{#1}}}
\NewDocumentCommand{\nclm}{ >{\SplitArgument{1}{-}}m }{\nclmaux#1}
\NewDocumentCommand{\nclmaux}{mm}{\ensuremath{\!{}^{#2\mathrm{m}}\mathrm{#1}}}

\begin{document}

\title{Signal Response Model in PandaX-4T}

\def\shKeyLab{School of Physics and Astronomy, Shanghai Jiao Tong University, Key Laboratory for Particle Astrophysics and Cosmology (MoE), Shanghai Key Laboratory for Particle Physics and Cosmology, Shanghai 200240, China}
\def\BUAA{School of Physics, Beihang University, Beijing 102206, China}
\def\BUAALab{Beijing Key Laboratory of Advanced Nuclear Materials and Physics, Beihang University, Beijing, 102206, China}
\def\zzu{School of Physics and Microelectronics, Zhengzhou University, Zhengzhou, Henan 450001, China}
\def\USTClab{State Key Laboratory of Particle Detection and Electronics, University of Science and Technology of China, Hefei 230026, China}
\def\USTCdep{Department of Modern Physics, University of Science and Technology of China, Hefei 230026, China}
\def\BUAALab{International Research Center for Nuclei and Particles in the Cosmos \& Beijing Key Laboratory of Advanced Nuclear Materials and Physics, Beihang University, Beijing 100191, China}
\def\pku{School of Physics, Peking University, Beijing 100871, China}
\def\YaLongSD{Yalong River Hydropower Development Company, Ltd., 288 Shuanglin Road, Chengdu 610051, China}
\def\IAP{Shanghai Institute of Applied Physics, Chinese Academy of Sciences, 201800 Shanghai, China}
\def\CHEPpku{Center for High Energy Physics, Peking University, Beijing 100871, China}
\def\SDUdep{Research Center for Particle Science and Technology, Institute of Frontier and Interdisciplinary Science, Shandong University, Qingdao 266237, Shandong, China}
\def\SDUlab{Key Laboratory of Particle Physics and Particle Irradiation of Ministry of Education, Shandong University, Qingdao 266237, Shandong, China}
\def\UMD{Department of Physics, University of Maryland, College Park, Maryland 20742, USA}
\def\TDLeeNew{New Cornerstone Science Laboratory, Tsung-Dao Lee Institute, Shanghai Jiao Tong University, Shanghai, 200240, China}
\def\TDLee{Tsung-Dao Lee Institute, Shanghai Jiao Tong University, Shanghai, 200240, China}
\def\MESJTU{School of Mechanical Engineering, Shanghai Jiao Tong University, Shanghai 200240, China}
\def\SYU{School of Physics, Sun Yat-Sen University, Guangzhou 510275, China}
\def\SYUSFI{Sino-French Institute of Nuclear Engineering and Technology, Sun Yat-Sen University, Zhuhai, 519082, China}
\def\NKU{School of Physics, Nankai University, Tianjin 300071, China}
\def\YTU{Department of Physics,Yantai University, Yantai 264005, China}
\def\FDU{Key Laboratory of Nuclear Physics and Ion-beam Application (MOE), Institute of Modern Physics, Fudan University, Shanghai 200433, China}
\def\USST{School of Medical Instrument and Food Engineering, University of Shanghai for Science and Technology, Shanghai 200093, China}
\def\SJTUSC{Shanghai Jiao Tong University Sichuan Research Institute, Chengdu 610213, China}
\def\SPEIT{SJTU Paris Elite Institute of Technology, Shanghai Jiao Tong University, Shanghai, 200240, China}
\def\NNU{School of Physics and Technology, Nanjing Normal University, Nanjing 210023, China}
\def\SYUzhuhai{School of Physics and Astronomy, Sun Yat-Sen University, Zhuhai, 519082, China}

%\affiliation{\shKeyLab}
\author{Yunyang Luo}\affiliation{\USTCdep}
\author{Zihao Bo}\affiliation{\shKeyLab}
\author{Shibo Zhang}\affiliation{\shKeyLab}
\author{Abdusalam Abdukerim}\affiliation{\shKeyLab}
\author{Wei Chen}\affiliation{\shKeyLab}
\author{Xun Chen}\affiliation{\shKeyLab}\affiliation{\SJTUSC}
\author{Chen Cheng}\affiliation{\SYU}
\author{Zhaokan Cheng}\affiliation{\SYUSFI}
\author{Xiangyi Cui}\affiliation{\TDLee}
\author{Yingjie Fan}\affiliation{\YTU}
\author{Deqing Fang}\affiliation{\FDU}
%\author{Changbo Fu}\affiliation{\FDU}
%\author{Mengting Fu}\affiliation{\pku}
\author{Lisheng Geng}\affiliation{\BUAA}\affiliation{\BUAALab}\affiliation{\zzu}
\author{Karl Giboni}\affiliation{\shKeyLab}
%\author{Linhui Gu}\affiliation{\shKeyLab}
\author{Xuyuan Guo}\affiliation{\YaLongSD}
\author{Chencheng Han}\affiliation{\TDLee} 
\author{Ke Han}\affiliation{\shKeyLab}
\author{Changda He}\affiliation{\shKeyLab}
\author{Jinrong He}\affiliation{\YaLongSD}
\author{Di Huang}\affiliation{\shKeyLab}
%\author{Yanlin Huang}\affiliation{\USST}
\author{Junting Huang}\affiliation{\shKeyLab}
\author{Zhou Huang}\affiliation{\shKeyLab}
\author{Ruquan Hou}\affiliation{\SJTUSC}
\author{Yu Hou}\affiliation{\MESJTU}
\author{Xiangdong Ji}\affiliation{\UMD}
\author{Yonglin Ju}\affiliation{\MESJTU}
\author{Chenxiang Li}\affiliation{\shKeyLab}
\author{Jiafu Li}\affiliation{\SYU}
\author{Mingchuan Li}\affiliation{\YaLongSD}
\author{Shuaijie Li}\affiliation{\YaLongSD}\affiliation{\shKeyLab}
\author{Tao Li}\affiliation{\SYUSFI}
\author{Qing Lin}\email[Corresponding author: ]{qinglin@ustc.edu.cn}\affiliation{\USTClab}\affiliation{\USTCdep}
\author{Jianglai Liu}\email[Spokesperson: ]{jianglai.liu@sjtu.edu.cn}\affiliation{\TDLeeNew}\affiliation{\shKeyLab}\affiliation{\SJTUSC}
\author{Congcong Lu}\affiliation{\MESJTU}
\author{Xiaoying Lu}\affiliation{\SDUdep}\affiliation{\SDUlab}
\author{Lingyin Luo}\affiliation{\pku}
\author{Wenbo Ma}\affiliation{\shKeyLab}
\author{Yugang Ma}\affiliation{\FDU}
\author{Yajun Mao}\affiliation{\pku}
\author{Yue Meng}\affiliation{\shKeyLab}\affiliation{\SJTUSC}
\author{Xuyang Ning}\affiliation{\shKeyLab}
\author{Binyu Pang}\affiliation{\SDUdep}\affiliation{\SDUlab}
\author{Ningchun Qi}\affiliation{\YaLongSD}
\author{Zhicheng Qian}\affiliation{\shKeyLab}
\author{Xiangxiang Ren}\affiliation{\SDUdep}\affiliation{\SDUlab}
\author{Nasir Shaheed}\affiliation{\SDUdep}\affiliation{\SDUlab}
\author{Xiaofeng Shang}\affiliation{\shKeyLab}
\author{Xiyuan Shao}\affiliation{\NKU}
\author{Guofang Shen}\affiliation{\BUAA}
\author{Lin Si}\affiliation{\shKeyLab}
\author{Wenliang Sun}\affiliation{\YaLongSD}
\author{Andi Tan}\affiliation{\UMD}
\author{Yi Tao}\email[Corresponding author: ]{taoyi92@sjtu.edu.cn}\affiliation{\shKeyLab}\affiliation{\SJTUSC}
\author{Anqing Wang}\affiliation{\SDUdep}\affiliation{\SDUlab}
\author{Meng Wang}\affiliation{\SDUdep}\affiliation{\SDUlab}
\author{Qiuhong Wang}\affiliation{\FDU}
\author{Shaobo Wang}\affiliation{\shKeyLab}\affiliation{\SPEIT}
\author{Siguang Wang}\affiliation{\pku}
\author{Wei Wang}\affiliation{\SYUSFI}\affiliation{\SYU}
\author{Xiuli Wang}\affiliation{\MESJTU}
\author{Xu Wang}\affiliation{\shKeyLab}
\author{Zhou Wang}\affiliation{\shKeyLab}\affiliation{\SJTUSC}\affiliation{\TDLee}
\author{Yuehuan Wei}\affiliation{\SYUSFI}
\author{Mengmeng Wu}\affiliation{\SYU}
\author{Weihao Wu}\affiliation{\shKeyLab}
\author{Yuan Wu}\affiliation{\shKeyLab}
%\author{Jingkai Xia}\affiliation{\shKeyLab}
\author{Mengjiao Xiao}\affiliation{\shKeyLab}
\author{Xiang Xiao}\affiliation{\SYU}
%\author{Pengwei Xie}\affiliation{\TDLee}
\author{Binbin Yan}\affiliation{\shKeyLab}
\author{Xiyu Yan}\affiliation{\SYUzhuhai}
%\author{Jijun Yang}\affiliation{\shKeyLab}
\author{Yong Yang}\affiliation{\shKeyLab}
%\author{Yukun Yao}\affiliation{\shKeyLab}
\author{Chunxu Yu}\affiliation{\NKU}
\author{Ying Yuan}\affiliation{\shKeyLab}
\author{Zhe Yuan}\affiliation{\FDU} %
\author{Youhui Yun}\affiliation{\shKeyLab}
\author{Xinning Zeng}\affiliation{\shKeyLab}
%\author{Dan Zhang}\affiliation{\UMD}
\author{Minzhen Zhang}\affiliation{\shKeyLab}
\author{Peng Zhang}\affiliation{\YaLongSD}
\author{Shu Zhang}\affiliation{\SYU}
\author{Tao Zhang}\affiliation{\shKeyLab}
\author{Wei Zhang}\affiliation{\TDLee}
\author{Yang Zhang}\affiliation{\SDUdep}\affiliation{\SDUlab}
\author{Yingxin Zhang}\affiliation{\SDUdep}\affiliation{\SDUlab} %
\author{Yuanyuan Zhang}\affiliation{\TDLee}
\author{Li Zhao}\affiliation{\shKeyLab}
%\author{Qibin Zheng}\affiliation{\USST}
\author{Jifang Zhou}\affiliation{\YaLongSD}
\author{Ning Zhou}\affiliation{\shKeyLab}\affiliation{\SJTUSC}
\author{Xiaopeng Zhou}\affiliation{\BUAA}
\author{Yong Zhou}\affiliation{\YaLongSD}
\author{Yubo Zhou}\affiliation{\shKeyLab}
\author{Zhizhen Zhou}\affiliation{\shKeyLab}
\collaboration{PandaX Collaboration}
\noaffiliation

%\linenumbers

% \author[a,b]{Yunyang~Luo,}
% \author[a, b]{Qing~Lin%
% \note{Corresponding author.}}
% % \author[c]{K. Ni}

% \affiliation[a]{State Key Laboratory of Particle Detection and Electronics, University of Science and Technology of China, Hefei 230026, China}
% \affiliation[b]{Department of Modern Physics, University of Science and Technology of China, Hefei 230026，China}
% \affiliation[c]{Department of Physics, University of California San Diego, La Jolla, CA 92093, USA}


% \emailAdd{qinglin@ustc.edu.cn}
% \emailAdd{taoyi92@sjtu.edu.cn}

\begin{abstract}
PandaX-4T experiment is a deep-underground dark matter direct search experiment that employs a dual-phase time projection chamber with a sensitive volume containing 3.7\,tonne of liquid xenon. 
The detector of PandaX-4T is capable of simultaneously collecting the primary scintillation and ionization signals, utilizing their ratio to discriminate dark matter signals from background sources such as gamma rays and beta particles.
The signal response model plays a crucial role in interpreting the data obtained by PandaX-4T. It describes the conversion from the deposited energy by dark matter interactions to the detectable signals within the detector. 
The signal response model is utilized in various PandaX-4T results.
This work provides a comprehensive description of the procedures involved in constructing and parameter-fitting the signal response model for the energy range of approximately 1\,keV to 25\,keV for electronic recoils and 6\,keV to 90\,keV for nuclear recoils. 
It also covers the signal reconstruction, selection, and correction methods, which are crucial components integrated into the signal response model.    
\end{abstract}

%Use showkeys class option if keyword
\maketitle

\section{Introduction}

%%%%%%%%%%%%%%%%%%%%
%% DM & PandaX Introduction
%%%%%%%%%%%%%%%%%%%%
Extensive astronomical evidence, as documented in the literature~\cite{rubin1980rotational, clowe2006direct, ade2016planck, zwicky1933rotverschiebung}, strongly indicates the ubiquitous presence of dark matter (DM) in the Universe.
The nature of DM remains elusive, with various theoretical frameworks proposing that it consists either entirely or partially of unknown particles~\cite{bertone2005particle, jungman1996supersymmetric, feng2010dark}. 
Among these hypotheses, the Weakly Interacting Massive Particle (WIMP)~\cite{jungman1996supersymmetric} stands out as one of the most promising candidates.
In recent years, significant advancements in sensitivity have been achieved by DM direct search experiments conducted in deep underground laboratories~\cite{xiao2015low, tan2016dark, collaboration2018dark, akerib2017results, agnese2018results, petricca2020first, meng2021dark, aprile2023first, lux2023first}. 
PandaX-4T, established in May 2020, has emerged as one of the world-leading experiments of this kind. With accumulated data spanning over 92 days dedicated to WIMP search, the PandaX-4T detector has played a pivotal role in advancing our understanding in this field.
Utilizing a 0.63-tonne-year exposure (Run0) conducted from November 2020 to April 2021, PandaX-4T has attained the most stringent constraint on the WIMP-nucleon spin-independent cross section at that time~\cite{meng2021dark}. 
Subsequently, following a campaign for impurity removal in the summer of 2021, the PandaX-4T experiment resumed stable operations and has since acquired more than 164 days of additional data (Run1).

%%%%%%%%%%%%%%%%%%%%
%% PandaX TPC brief intro
%% mention the signal model
%%%%%%%%%%%%%%%%%%%%
The PandaX-4T experiment utilizes a dual-phase liquid xenon (LXe) Time Projection Chamber (TPC) technique, as outlined in Ref.~\cite{meng2021dark}. 
In this configuration, the TPC enables the detection of both prompt scintillation ($S1$) and ionized electrons generated by energy depositions.
The ionized electrons undergo drift towards the top of the TPC under the influence of an applied electric field. They are subsequently extracted from the liquid phase into a thin gaseous xenon layer, where they experience a stronger amplification field. Through the process of electron luminescence, the ionized electrons are converted into secondary scintillation signals ($S2$).
The $S1$ and $S2$ signals are collected by photomultiplier tubes (PMTs) positioned at the top and bottom of the TPC.
Leveraging the time difference between the $S1$ and $S2$ signals, as well as the signal patterns observed on the PMTs, the longitudinal and horizontal positions (referred to as $z$ and $x$-$y$ positions) of an interaction vertex can be reconstructed. 
This positional information is crucial for various purposes, including distinguishing interactions originating from material or external radioactivity (e.g., those occurring near the TPC edges) and identifying interactions caused by neutrons that may exhibit multiple distinct interaction vertices within the TPC.
In addition, TPCs employing LXe as the target material possess excellent discrimination capabilities between DM-induced nuclear recoils (NRs) and background-induced electronic recoils (ERs) based on the ratio of $S2$ to $S1$. 
Consequently, the signal response model, which relates the deposited energy to the observable signals, plays a crucial role in interpreting DM signals within TPC-based experiments.

%%%%%%%%%%%%%%%%%%%%
%% Paper layout
%%%%%%%%%%%%%%%%%%%%
This paper provides a comprehensive description of the signal response model employed in the analysis of PandaX-4T results~\cite{meng2021dark, ma2023search, li2023search, gu2022first, zhang2022search}.
The signal response model utilized in PandaX-4T encompasses the conversion from deposited energy to the observable signals ($S1$ and $S2$). 
This includes the signal production within the LXe medium, the subsequent signal collection within the TPC, and the signal reconstruction, correction, and selection during the data analysis stage.
Given the inherently stochastic nature of these processes, a fast Monte Carlo (MC) simulation-based approach with the help of GPU boosting is adopted for the signal response model. 
This framework draws inspiration primarily from the NEST (Noble Element Simulation Technique) framework~\cite{szydagis2011nest, NESTv2}, as well as other similar methodologies found in the literature~\cite{collaboration2019xenon1t}.
Sec.~\ref{sec:signal_production} describes our modeling of the intrinsic signal production in LXe.
The signal collection, reconstruction, correction, and selection procedures are described in Sec.~\ref{sec:signal_collection}, ~\ref{sec:signal_reconstruction}, ~\ref{sec:signal_correction}, and ~\ref{sec:signal_selection}, respectively, which are the essential detector effects in the signal response model.
Furthermore, Sec.~\ref{sec:fit_to_data} presents the outcomes of parameter fitting to ensure the alignment of the model with the calibration data from PandaX-4T. The obtained results from this fitting procedure are reported.
Finally, a concise summary and a discussion of the findings are provided in Sec.~\ref{sec:summary}.

\section{Signal production in liquid xenon}
\label{sec:signal_production}

When particles interact with LXe, they transfer momentum to recoiling particles, which can be a shell electron in the case of ERs or a xenon atom in the case of NRs. 
These recoiling particles subsequently lose kinetic energy through elastic scattering (thermalization) and inelastic scattering (excitation and ionization) with the surrounding atoms.
The total number of detectable quanta (denoted as $N_q$), which includes excited xenon atoms and ion-electron pairs, is directly related to the deposited energy $\xi$. 
This relationship is governed by the work function, denoted as $W$. 
The work function represents the average energy required to produce a single detectable quantum in the LXe:
\begin{equation}
    N_q = N_i + N_{\textrm{ex}} = \textrm{B}(\xi / W, L),
\end{equation}
where $N_i$ and $N_{\textrm{ex}}$ are the number of ion-electron pairs and excited atoms, respectively.
We take a constant $W$=13.7\,eV~\cite{szydagis2011nest} in the signal response model.
$L$ is the Lindhard factor~\cite{Lindhard1963} characterizing the degree of heat quenching in the detection process.
For ER, the Lindhard factor has a value of 1.
To account for the probabilistic nature of the detection process, the expression $B(\xi/W, L)$ is utilized. 
Here, $B$ represents a randomly sampled number generated from a binomial distribution with the number of trials being $\xi/W$ and the success probability being $L$.
Upon interaction with the surrounding medium, the excited xenon atom combines with a neighboring atom to form a dimer, which subsequently undergoes a decay process with a lifetime of about 4\,ns or 22\,ns~\cite{hitachi1983effect}. This decay process results in the emission of a photon with a wavelength of $\sim$175\,nm.
A fraction of the ion-electron pairs formed during the interaction can recombine and form a dimer with a surrounding atom, leading to the emission of the 175-nm ultraviolet light as well.
The rest of the ionized electrons do not participate in the recombination process.
The numbers of the ion-electron pairs $N_i$, the emitted photons $N_{\textrm{ph}}$, and the escaped electrons $N_e$ can be written as:
\begin{equation}
\begin{aligned}
    N_i & = \textrm{B}\left(N_q, \frac{\alpha}{1+\alpha}\right), \\
    N_e & = \textrm{B}\left(N_i, 1-r\right), \\
    N_{\textrm{ph}} & = N_q - N_e, 
\end{aligned}
\end{equation}
where $\alpha$ is the mean ratio of the numbers of the excited atoms to ion-electron pairs.
The recombination fraction, denoted as $r$, exhibits intrinsic fluctuations based on previous discussions~\cite{akerib2017signal}. 
In the fast MC simulations, the recombination fraction is sampled from a Gaussian distribution, denoted as $G(\langle r \rangle, \Delta r)$, where $\langle r \rangle$ represents the mean fraction and $\Delta r$ represents the fluctuation of recombination.
The energy dependence of the mean recombination fraction, $\langle r \rangle$, is traditionally described by Birk's law~\cite{birks1951scintillations} in the high-energy region, typically around the order of 10 keV.
In the low-energy region, the energy dependence of $\langle r \rangle$ is considered to follow the Thomas-Imel model~\cite{Thomas1987}.
However, the existing measurements show a deviation of the $\langle r \rangle$ from the Thomas-Imel model in this low-energy region, and a global model (NEST model)~\cite{szydagis2011nest} fitting existing data is usually used in the community.
The availability of measurements for the mean recombination fraction ($\langle r \rangle$) and its fluctuation ($\Delta r$) is limited in this low-energy region due to the difficulty of getting keV and sub-keV energy depositions in a dense detector.
Consequently, the nominal values provided by NESTv2~\cite{NESTv2} have uncertainties associated with them.
In the PandaX-4T experiment, we have performed further tuning of the model parameters using our own calibration data to refine the values of $\langle r \rangle$ and $\Delta r$. 
The details of this tuning will be presented in Sec.~\ref{sec:fit_to_data}.
% To compare the performance of the NESTv2 nominal model with the PandaX-tuned model (referred to as P4-NEST), we present the mean photon yields $N_{\textrm{ph}}/\xi$ and charge yields $N_e/\xi$, along with the corresponding $\Delta r$, as functions of energy for both ER and NR. 
% These results are depicted in Figure~\ref{fig:ly_cy}.
% The detailed description of the empirical parameterization of $\langle r \rangle$ and $\Delta r$ as a function of electric field and deposit energy is given in Sec.~\ref{sec:fit_to_data}.

% \begin{figure}
%     \centering
%     \includegraphics[width=0.95\textwidth]{plots/PandaX-4T_LYCY.pdf}
%     \caption{
%     The light yield, charge yield, and recombination fluctuation $\Delta r$ as a function of the deposit energy are shown in the top, middle, and bottom panels, respectively.
%     The left and right panels correspond to the results for ER and NR, respectively.
%     The blue and red lines give the corresponding results for Run0 and Run1 electric field configurations.
%     The dashed and solid lines represent the nominal predictions from NESTv2~\cite{NESTv2} and the tuned results in this analysis.
%     Small panels beneath each major panels give the uncertainty bands of the residual difference between the tuned results and nominal NEST prediction.
%     The black dashed lines indicate the $\pm1\sigma$ uncertainties of the nominal NEST predictions.
%     }
%     \label{fig:ly_cy}
% \end{figure}

\section{Signal collection}
\label{sec:signal_collection}

TPC detects the primary scintillation $S1$ signals and the secondary scintillation $S2$ signals with different features.

The prompt scintillation signals, $S1$, are collected shortly after the particle interaction, typically within a time scale of 10 to 100 ns. 
However, the collection of $S1$ signals is associated with a success probability, typically around 0.1 to 0.2 in dual phase TPCs.
This success probability, also known as the photon detection efficiency (PDE), is spatially dependent and influenced by various factors. 
These factors include the coverage of PMTs within TPC, as well as the collection efficiency and quantum efficiency of the PMTs themselves.
The PDE is also affected by the purity level of the LXe, which can influence the absorption length of scintillation photons, and the reflection properties of the Polytetrafluoroethylene (PTFE) reflectors used in the TPC.
The number of collected photons $N_{\textrm{det}}$ is related to the number of photons generated $N_{\textrm{ph}}$ and  can be written as:
\begin{equation}
    \centering
    N_{\textrm{det}} = B\left( N_{\textrm{ph}}, \varepsilon_{\textrm{PDE}}(x,y,z) \right),
    \label{eq:pde}
\end{equation}
where $\varepsilon_{\textrm{PDE}}$ represents the spatially dependent PDE.

When a photon is detected by a PMT and converted into a photoelectron (PE) inside the PMT, there is a phenomenon known as double PE emission (DPE). 
This phenomenon refers to the emission of multiple photoelectrons from the PMT photocathode as a result of the initial detection of a single photon.
In the case of 175-nm ultraviolet (UV) light, it has been observed that there is an approximately $20\%$ probability for a single detected photon to generate two photoelectrons within the PMT~\cite{faham2015measurements}.
The number of the PEs $N_{\textrm{pe}}$ that are generated inside PMTs can be expressed as:
\begin{equation}
    \centering
    N_{\textrm{pe}} = N^{\prime}_{\textrm{det}} + B(N^{\prime}_{\textrm{det}}, p_{\textrm{dpe}}),
    \label{eq:dpe}
\end{equation}
where $p_{\textrm{dpe}}$ represents the probability of DPE.
We usually take $g_1 = \langle \varepsilon_{\textrm{PDE}} \rangle \cdot (1+p_{\textrm{dpe}})$ as a characteristic parameter for $S1$ detection.
$ \langle \varepsilon_{\textrm{PDE}} \rangle$ is the average PDE inside the fiducial volume (FV).
The spatial dependence of $\varepsilon_{\textrm{PDE}}(x, y, z)$ is obtained using calibration data, which will be illustrated in Section~\ref{sec:signal_correction}.
Note that the $N_{\textrm{det}}^{\prime}$ in Eq.~\ref{eq:dpe} represents the number of photons that are successfully clustered during signal reconstruction, which will be detailed in Section~\ref{sec:signal_reconstruction}.

\begin{figure*}[htp]
    \centering
    \includegraphics[width=\textwidth]{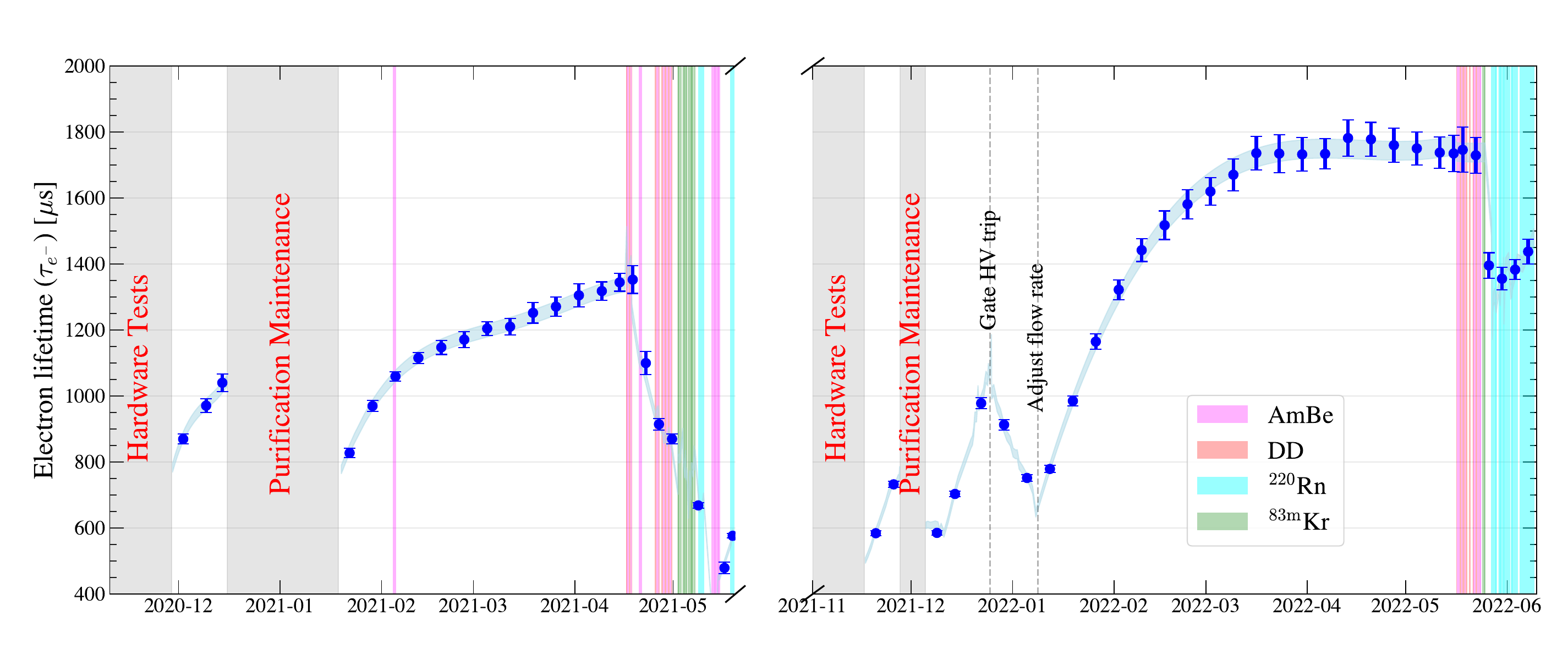}
    \caption{
    The electron lifetime evolutions in Run0 and Run1 scientific data taking periods.
    The gray shaded regions indicate the time periods for detector operations.
    % The green shaded regions indicate the time for detector calibrations.
    The period of calibration runs are present as well, including \ncl{Am-241}Be (magenta), DD (red), \ncl{Rn-220} (cyan), and \nclm{Kr-83} (green). 
    }
    \label{fig:tau_e}
\end{figure*}

The $S2$ signal in the PandaX-4T detector is obtained by the detection of secondary scintillation light emitted when the ionized electrons undergo drift in the sensitive volume and reach the gaseous xenon region. 
During the drifting process, electron attachment can occur, when some of the drifting electrons become bound to electro-negative impurity molecules present in the LXe. 
The probability of electron attachment is dependent on factors such as the concentration and type of impurity.
In the PandaX-4T detector, the most prevalent and dominant electro-negative impurity is oxygen.
The number of electrons that survive the drifting process and reach the gaseous xenon layer, denoted as $N_{\textrm{drift}}$, can be written in the form:
\begin{equation}
    \centering
    N_{\textrm{drift}} = B(N_e, e^{-z/(\tau_e v_{\textrm{drift}})}),
    \label{eq:tau_e}
\end{equation}
where $v_{\textrm{drift}}$ is the constant drift velocity of the electrons in LXe, and $\tau_e$ is the electron lifetime which is an indicator of the impurity level.
Throughout Run0 and Run1 of the PandaX-4T, the operation of the TPC was subject to various procedures and incidents, such as power outages.
These events led to the introduction of impurities into the TPC, despite the continuous circulation and purification of the LXe.
The evolution of the electron lifetime was monitored during this period using residual $\alpha$ events originating from \ncl{Rn-222} decays (as shown in Fig.~\ref{fig:tau_e}), as well as X-ray events resulting from the decays of neutron-activated \nclm{Xe-129} and \nclm{Xe-131}.

\begin{figure*}[htp]
    \centering
    \includegraphics[width=\textwidth]{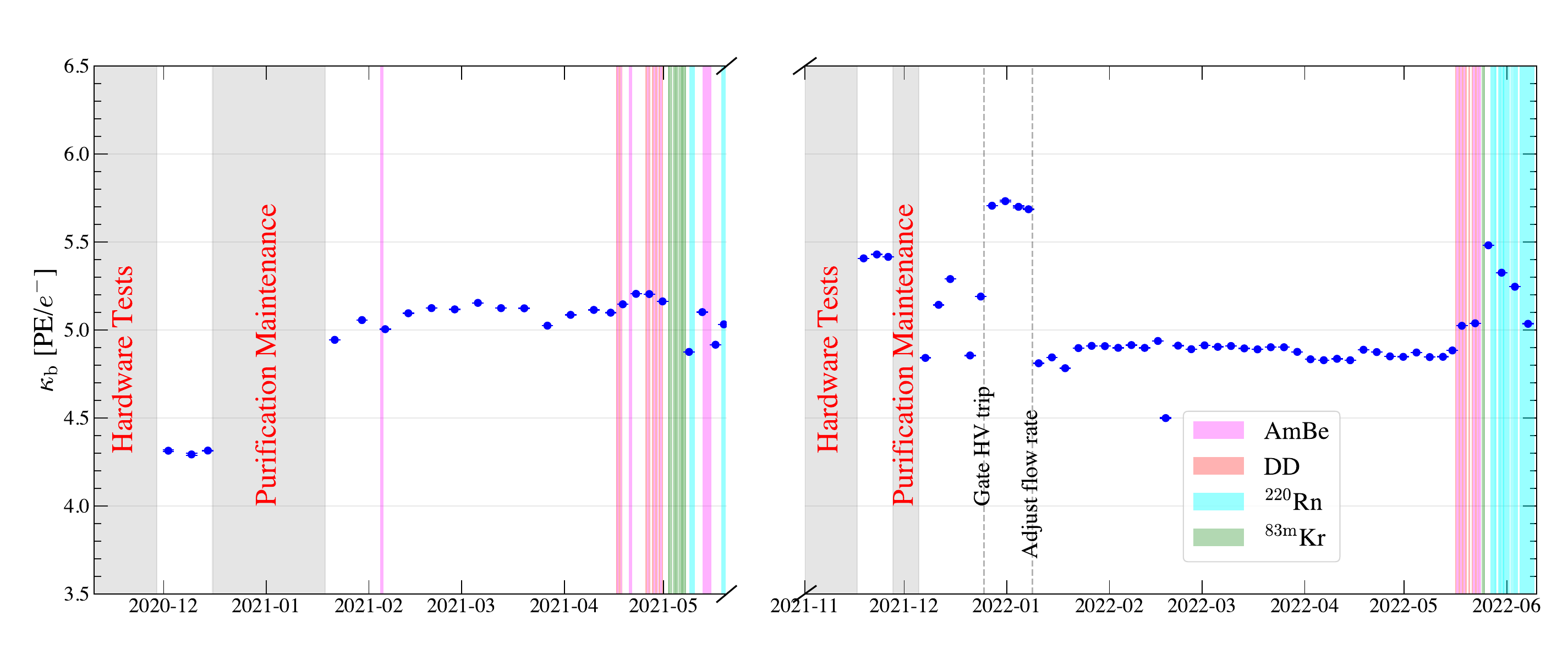}
    \caption{
    The evolution of $\kappa$ in Run0 and Run1 data taking periods.
    The gray shaded regions indicate the time periods for detector operations.
    % The green shaded regions indicate the time for detector calibrations.
    The period of calibration runs are present as well, including \ncl{Am-241}Be (magenta), DD (red), \ncl{Rn-220} (cyan), and \nclm{Kr-83} (green).
    }
    \label{fig:seg_evolution}
\end{figure*}

The drifted electrons are subsequently extracted into the gaseous xenon layer of the TPC.
However, it is worth noting that if the extraction electric field strength is insufficient, a fraction of the electrons may fail to be extracted, leading to a signal loss.
The number of extracted electrons $N_{\textrm{ext}}$ can be written as:
\begin{equation}
    \centering
    N_{\textrm{ext}} = B\left(N_{\textrm{drift}}, \varepsilon_{\textrm{ext}}\right),
    \label{eq:extraction_efficiency}
\end{equation}
where $\varepsilon_{\textrm{ext}}$ represents the extraction efficiency.
Once extracted, the electrons pass through the gaseous xenon medium, inducing excitation in the surrounding xenon atoms, which subsequently emit 175-nm ultraviolet (UV) light. 
These light signals are collected and converted into PEs by the PMTs.
The number of PEs of these light signals $N_{\textrm{prop}}$ then is modelled as
\begin{equation}
    \centering
    N_{\textrm{prop}} = G\left( \kappa (x,y) N_{\textrm{drift}}, \Delta \kappa \sqrt{N_{\textrm{drift}}} \right).
    \label{eq:electron_amplification}
\end{equation}
The overall amplification gain, denoted as $\kappa$, is determined by the combined factors of light collection efficiency, the gas gap thickness, and the strength of the amplification field.
However, the amplification process is subject to non-uniformities in the amplification field, leading to fluctuations in the total gain.
These fluctuations are quantified by the parameter $\Delta \kappa$, which typically exhibits magnitudes on the order of 20\% to 40\% of $\kappa$.
To describe the amplification of the charge signal, a commonly employed parameter is defined as $g_2=\langle \kappa \rangle \varepsilon_{\textrm{ext}}$ where $\langle \kappa \rangle$ is the average of $\kappa$ within the FV.
The spatial dependence of $\kappa(x,y)$ is illustrated and given in Sec.~\ref{sec:signal_correction}.
In order to have time-independent charge amplification in the data, the temporal feature of $\kappa$ is corrected (discussed in later Sec.~\ref{sec:signal_correction}) and shown in Fig.~\ref{fig:seg_evolution}.
Note that in the PandaX-4T analysis, the $S2$ signals from bottom PMTs are used.
The $S2$ distribution on the bottom PMTs is more spread than the one on the top PMTs, reducing the chance of PMT saturation for $S2$s.
Therefore, we use the corresponding parameter $g_{2\mathrm{b}}=\langle \kappa_\mathrm{b} \rangle \varepsilon_{\textrm{ext}}$ as the parameter for describing the $S2$ gain.

\begin{figure}[htp]
    \centering
    \includegraphics[width=0.95\columnwidth]{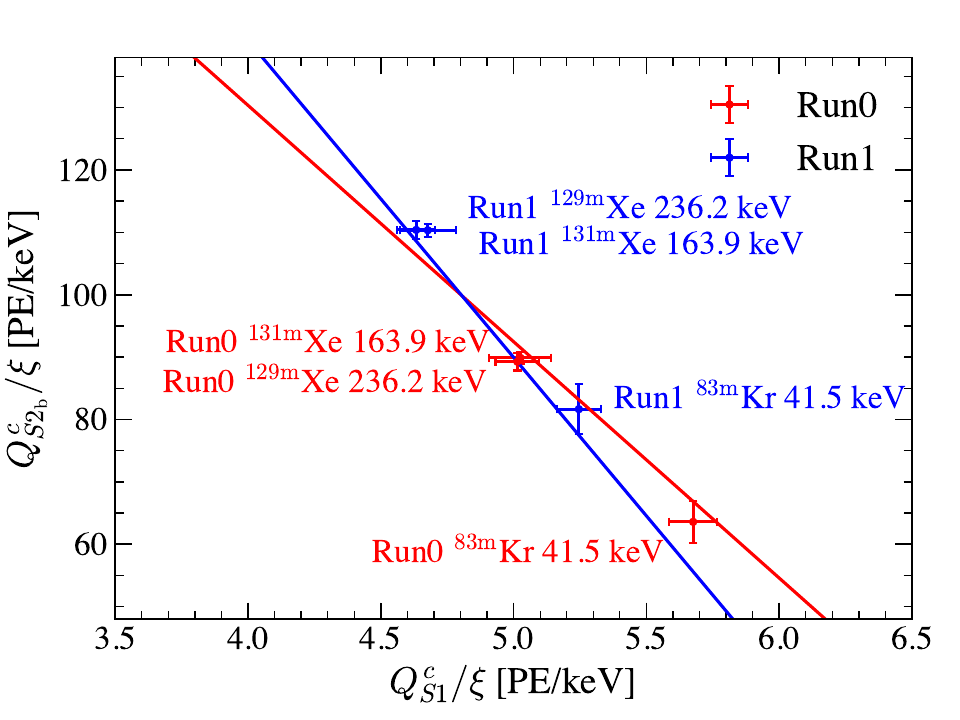}
    \caption{
    The corrected $S1$ and $S2_\mathrm{b}$ yields obtained from \nclm{Kr-83}, \nclm{Xe-131}, and \nclm{Xe-129} in Run0 (red) and Run1 (blue) data.
    A Run0-Run1 combined fit is performed using these corrected $S1$ and $S2_\mathrm{b}$ yields to obtain the $g_1$ and $g_{2\mathrm{b}}$, with a projection line being plotted for both runs. The best fit ($g_1$, $g_{2\mathrm{b}}$) values for Run0 and Run1 are
    ($0.103\pm0.005$, $3.9\pm0.4$) and ($0.093\pm0.004$, $4.7\pm0.5$), respectively.
    % ($0.09975\pm0.00085$, $3.953\pm0.110$) and ($0.09548\pm0.00067$, $4.524\pm0.077$), respectively.
    }
    \label{fig:doke_plot}
\end{figure}

Both $g_1$ and $g_{2\mathrm{b}}$ can be determined by analyzing calibration data. 
Following the charge correction steps discussed later in Sec.~\ref{sec:signal_correction}, $g_1$ and $g_{2\mathrm{b}}$ are fitted by ER peaks of ~\nclm{Kr-83} (41.5~keV), \nclm{Xe-131} (163.9~keV) and \nclm{Xe-129} (236.2~keV), according to the energy reconstruction formula:
\begin{equation}
\label{eq:energy_recon}
    \xi = W \left(\frac{Q^c_{S1}}{g_1}+\frac{Q^c_{S2_\mathrm{b}}}{g_{2\mathrm{b}}}\right)
\end{equation}
where $Q^c_{S1}$ and $Q^c_{S2_\mathrm{b}}$ are the corrected $S1$ and $S2$ charges, respectively.
The lower $\mathrm{b}$ indicates the $S2$ charge is obtained from only the bottom PMTs.
The correction here refers to the correction for signal's spatial non-uniformity, which will be detailed in Subsection~\ref{subsec:signal_spatial_correction}.
In order to establish a unified signal response model for the entire data-taking period of PandaX-4T, the parameters ($g_1$, $g_{2\mathrm{b}}$) for Run0 and Run1 data are considered to differ by a factor due to variations in operating conditions, specifically electric fields and liquid levels.
Using the $\alpha$ events from $^{222}$Rn decay, we obtain that the $g_1$ in Run1 is 9\% smaller than that in Run0, and $g_{2\mathrm{b}}$ 22\% larger.
% \textcolor{red}{ (QL to remove:
% The scaling ratio is determined on a run-by-run base by analyzing the $S1$ and $S2$ signals from consistent \ncl{Rn-222} $\alpha$ events present in the detector throughout the data-taking period. 
% These events serve as a reference to calibrate and normalize the signal response, as well as to correct the $S1$ and $S2$ for the detector's time dependence. This will be detailed in Sec.~\ref{sec:signal_correction}.
% }
Next, the corrected $S1$ and $S2_\mathrm{b}$ yields, defined as the number of detected PEs (corrected) in the $S1$ and $S2_\mathrm{b}$ signals per unit of energy, are measured for six ER peaks originating from \nclm{Kr-83} (41.5 keV), \nclm{Xe-131} (163.9 keV), and \nclm{Xe-129} (236.2 keV) in both Run0 and Run1 data sets.
The distributions of corrected $S1$ and $S2_\mathrm{b}$ yields, often referred to as Doke plots, in Run0 and Run1 are simultaneously fitted to obtain the $g_1$ and $g_{2\mathrm{b}}$.
% By fitting the data points obtained from the Doke plots, $g_1$ and $g_{2\mathrm{b}}$ are obtained.
Fig.~\ref{fig:doke_plot} illustrates the distributions of corrected $S1$ and $S2_\mathrm{b}$ yields from the six ER peaks, and the best fit of $g_1$ and $g_{2\mathrm{b}}$.

\section{Signal reconstruction}
\label{sec:signal_reconstruction}

The impact of signal reconstruction on the signal response model is substantial due to several factors. 
The region of interest (ROI) that we focus on is characterized by low deposit energy, making it susceptible to various sources of fluctuations introduced during the trigger, reconstruction, and correction processes. 
Additionally, the presence of noise, PMT afterpulsing, photoionization effects, and delayed electrons complicates the data selection process. 
These factors collectively contribute to the considerable influence of signal reconstruction on our model's accuracy.
In this section, we provide a detailed description of the sequential steps involved in signal reconstruction and present an effective model for signal reconstruction within the signal response model.

\subsection{Hit finding and clustering}

Following the collection and amplification of the 175-nm light signal by the PMTs, the resulting signal is passed to the V1725 fast analog-to-digital converter (FADC).
To reduce data size, we employ the Zero Length Encoding (ZLE) mode within the FADC. 
This mode involves setting a ZLE threshold, which is determined as approximately one-third of the single PE amplitude (20\,ADC units).
Only waveform segments that exceed this threshold are recorded.
For each PMT channel, waveform segments are first subtracted by their respective baselines, which are calculated segment-by-segment.
% The single hit is then constructed once the amplitude exceeds the 2.44~mV (20\,ADC) threshold with a segment length of over 400\,ns.
The single hit is then constructed once the amplitude exceeds the $2.44$~mV ($20$\,ADC) threshold and continues until there are $80$~ns of continuous sampling points whose amplitudes are below this threshold.
To keep enough duration for baseline calculation, the shortest duration of one segment is set to be $400$~ns.
% Subsequently, a single hit (with a window size of \textcolor{red}{xxx}\,ns) is constructed when the amplitude exceeds a threshold of \textcolor{red}{xxx}\,mV.
These single hits serve as the fundamental units within the entire data structure throughout the PandaX-4T data analysis process.
Based on these single hits, we further define the concepts of signal and physical event.
The single hits across the PMT channels are further clustered into a single pulse, once the time difference of any two adjacent hits in the clustered pulse is less than 60\,ns.
% A single hit is the fundamental unit of the entire data structure throughout PandaX-4T data analysis, based on which we further define the concepts of signal and physical event.
% Waveform segments from each PMT channel are aligned properly and each single hit is  found and divided sequentially.
% Timestamp, duration and channel id of each single hit are recorded, together with the baseline information.

\subsection{Pulse classification}

% \textcolor{red}{(@TY to fill)}

The clustered pulses are further categorized into $S1$ and $S2$ pulses based on their different characteristics.
$S1$ lights are generated almost instantly after the incident particle collides with the target nucleus, resulting in a narrow pulse.
Additionally, the $S1$ lights are generated in the LXe region below the liquid-gas surface. 
Due to the total internal reflection effect at the liquid-gas interface, the $S1$ signal received by the bottom PMTs is greater than that received by the top.
On the contrary, the generation of $S2$ signals involves the electron drifting which is affected by diffusion, and the continuous acceleration and extraction of electrons in the gaseous layer under a stronger electric field, resulting in a larger amplitude and a more spread time profile of the $S2$ lights.
Since the $S2$ lights are generated in gaseous xenon and closer to the top PMTs, the top PMTs receive more lights.

With these considerations, the classification of $S1$ and $S2$ pulses is primarily based on the pulse shape and its distribution across the PMTs.
To characterize the charge partition between the top and bottom PMT arrays, we define the Top-Bottom Asymmetry (TBA) as $\mathrm{TBA} = (q_\mathrm{t}-q_\mathrm{b})/(q_\mathrm{t}+q_\mathrm{b})$, where $q_\mathrm{t/b}$ represents the accumulated charge of the pulse from the top/bottom PMTs.
Additionally, the full width of the pulse is defined as the difference of the reconstructed left and right boundaries (over-threshold times with some buffer) of the pulse, which is affected by the afterglow effect.
To better characterize the pulse duration while mitigating the afterglow effect following the major pulses, we also introduce a pulse width metric called CDF width.
This width is determined by the time interval that encompasses the cumulative charge from 10\% to 90\% of the pulse and is denoted as $w_{\textrm{CDF}}^{90-10}$.
In the data processing of PandaX-4T, the analysis program iterates over all clustered pulses and assigns different types to each pulse.
The classification discussed in this work primarily focuses on the low-energy region ($<25$\,keV for ER or $<90$\,keV for NR). 
Before the classification, noise and discharging pulses are pre-identified based on their abnormal waveform shapes and concentrated distributions.
% Specifically, a pulse with negative height is identified as noise.
Signals larger than $10^6$\,PE with narrow widths and negative TBA values, or signals larger than $10^4$\,PE with one PMT channel accounts for more than $40\%$ of the total signal charge, are determined to be due to discharging.
Subsequently, the $S1$ and $S2$ pulses are identified with relatively loose filtering conditions compared to the final data selection.
For $S2$ pulses, a minimum total charge of 15\,PE is required, along with triggering of at least five PMTs and a TBA above a charge-dependent lower limit.
Furthermore, the full width of $S2$ pulses must be larger than 320 ns, and $w_{\textrm{CDF}}^{90-10}$ must exceed 240 ns.
Regarding $S1$ pulses, the total charge is accepted down to 0.3\,PE (corresponding to the ADC threshold of a single hit), with a charge-dependent upper limit for TBA.
Low-energy $S1$ pulses are also required to have $w_{\textrm{CDF}}^{90-10}$ less than 320 ns.

\subsection{$S1$-$S2$ Pairing}
\label{subsec:pairing}

In a TPC detector, a typical physical event is characterized by the presence of one $S1$ pulse accompanied by at least one associated $S2$ pulse. 
In the case of multiple scatters (MSs), multiple $S2$ pulses are present.
To define the time window for a physical event, we primarily consider a window extending 1~ms before and after the start time of the first arrived $S2$ pulse, which serves as the anchor $S2$. 
The anchor $S2$ is subject to updating if a subsequent $S2$ pulse with three times larger total charge is found within the event window.
In this case, the window's right boundary is extended to 1~ms after the updated anchor $S2$, and the left boundary is synchronously redefined as 1~ms before it.
This updating procedure is iterated until no more update on the anchor $S2$ is needed.
The 1-ms window is chosen so that it is sufficiently larger than the maximum drift time of the TPC, which are approximately 840 and 850\,$\mu$s in Run0 and Run1, respectively.

\begin{figure*}[htp]
    \centering
    \includegraphics[width=0.99\textwidth]{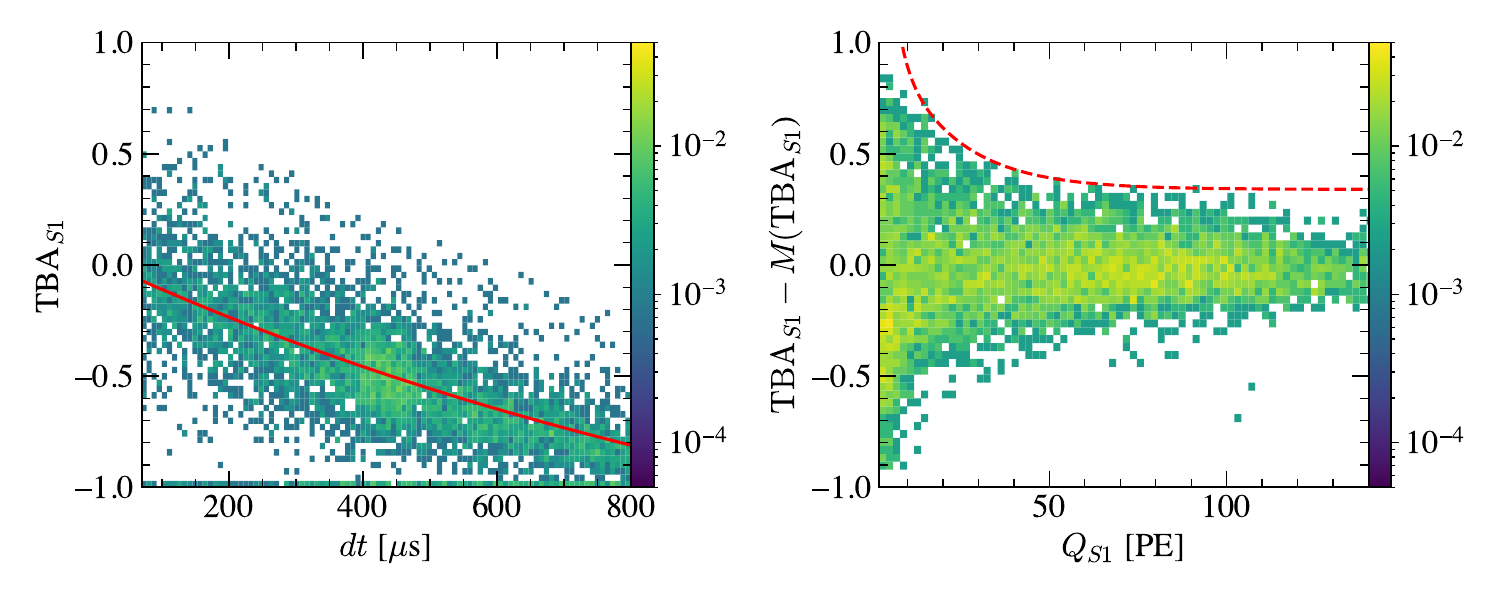}
    \caption{Left: the distribution of low energy ER and NR calibration events presenting the relation between TBA$_{S1}$ and drift time. 
    The red solid curve illustrates the median of TBA$_{S1}$ (ie. M(TBA$_{S1}$)). Right: Median-centering for TBA$_{S1}$ as a function of $Q_{S1}$. Any $S1$ above the red dashed curve is not treated as a major $S1$.}
    \label{fig:s1_tba}
\end{figure*}

Once the event window is determined, the major $S2$ pulse is identified as the pulse with the largest total charge. 
This pulse is considered to be the primary $S2$ signal associated with the event.
To enhance the pairing efficiency between $S1$ and $S2$ signals in physical events, several selection criteria are applied to the major $S1$ signal. 
These criteria aim to distinguish it from single electron (SE) fragments that may be incorrectly tagged as $S1$ pulses.
The major $S1$ pulse is required to be ``clean'', meaning that no other signals should appear in its temporal vicinity (within 400~ns before and after). 
Additionally, the major $S1$ pulse should not have more than 5 peaks~\footnote{The number of peaks in a signal waveform refers to the count of times the signal exceeds 1/3 of its maximum amplitude and then falls below 1/10 of its maximum amplitude.}, and its $w_{\textrm{CDF}}^{90-10}$ is constrained to a maximum of 240\,ns.
Furthermore, to distinguish the major $S1$ pulse from pulses due to discharging, it is required that the light distribution of the major $S1$ pulse is not excessively concentrated on the top and bottom PMT arrays.
We use the area of a single hit ($A_\mathrm{hit}$) in the waveform to quantify the concentration of the light signal, and set upper limits to the largest hit area detected from top and bottom PMT arrays, respectively.
% \textcolor{red}{The TBA range of good $S1$s is determined through calibration to be xxx to xxx, and we require a $S1$ pulse to have TBA in range of xxx to xxx.}
Since the TBA distribution of S1 (TBA$_{S1}$) has been tested to be consistent between waveform simulation and calibration data~\cite{pandax4t_wf}.
The $99.5\%$ quantile derived from waveform simulation in the parameter space of TBA$_{S1} - M$(TBA$_{S1}$) as a function of $Q_{S1}$ is applied as an upper bound, where $M$(TBA$_{S1}$) denotes the median values with respect to the drift time $dt$, as shown in Fig.~\ref{fig:s1_tba}.
The largest of the $S1$ pulses that satisfy the criteria within the physical event window before the first arrived $S2$ is determined as the major $S1$.

\subsection{$S2$ reclustering}
\label{subsec:reclustering}

In the analysis of PandaX-4T data, an additional reclustering procedure is applied to the largest and second-largest $S2$s in a physical event.
This step is necessary to address the problem of the initial clustering method, which combines two adjacent hits into one cluster if they are less than 60\,ns apart in time.
It has been observed that this approach can lead to incorrect fragmentation of hits that actually belong to an $S2$ signal.
To overcome this issue, the reclustering algorithm utilizes the $S2$ width relation with the $S2$ vertical position due to electron diffusion. 
For each instance of the largest and second-largest $S2$ waveform, the clustering width is redefined and adjusted based on factors such as the vertical position and the $S2$ size.

In the reclustering process of PandaX-4T, the determination of whether to merge each nearest neighbor pulse into a major $S2$ signal is performed based on whether the inclusion would result in a $S2$ $w_{\textrm{CDF}}^{90-10}$ that complies with the expected $S2$ width due to diffusion.
% The reclustering is performed iteratively until a convergence condition is met. 
Based on the distribution of the $S2$ $w_{\textrm{CDF}}^{90-10}$ as a function of drift time $t$, the standard deviation $\sigma_{\rm ori}(t)$ of the $S2$ $w_{\textrm{CDF}}^{90-10}$ as a function of drift time is derived.
The nearby small signal is merged if its resulted $w_{\textrm{CDF}}^{90-10}$ increment is less than $1.5\sigma_{\rm ori}(t)$.
For $S2$ signals smaller or larger than 1000 PE, the reclustering is forced to stop when the merged $S2$ signal's $w_{\textrm{CDF}}^{90-10}$ deviates by $5\sigma$ or $3\sigma$, respectively, from the expected $w_{\textrm{CDF}}^{90-10}$ of the initial $S2$ before the reclustering.
Additionally, the horizontal distance between signals ($< 200$~mm for $Q_\mathrm{signal} > 10$ PE) is taken into account to ensure spatial proximity, suggesting a common origin for the merging candidates.

% For the largest and second-largest $S2$ waveforms in PandaX-4T data, an extra reclustering is needed.
% This is because it bas been found that the initial clustering method with a fixed width of 60~ns would incorrectly split some hits that originally belong to one $S2$ signal, causing the $S2$ waveform to be fragmented erroneously.
% The reclustering algorithm is based on the diffusion relations between the width of $S2$ and drift time obtained from Monte Carlo simulations.
% For each instance of largest or second largest $S2$, the clustering width is redefined and adjusted based on the vertical position and the magnitude of the photocharge in the instance.

\subsection{Position reconstruction}
\label{subsec:pos_recon}

\begin{figure}[htp]
    \centering
    \includegraphics[width=0.9\columnwidth]{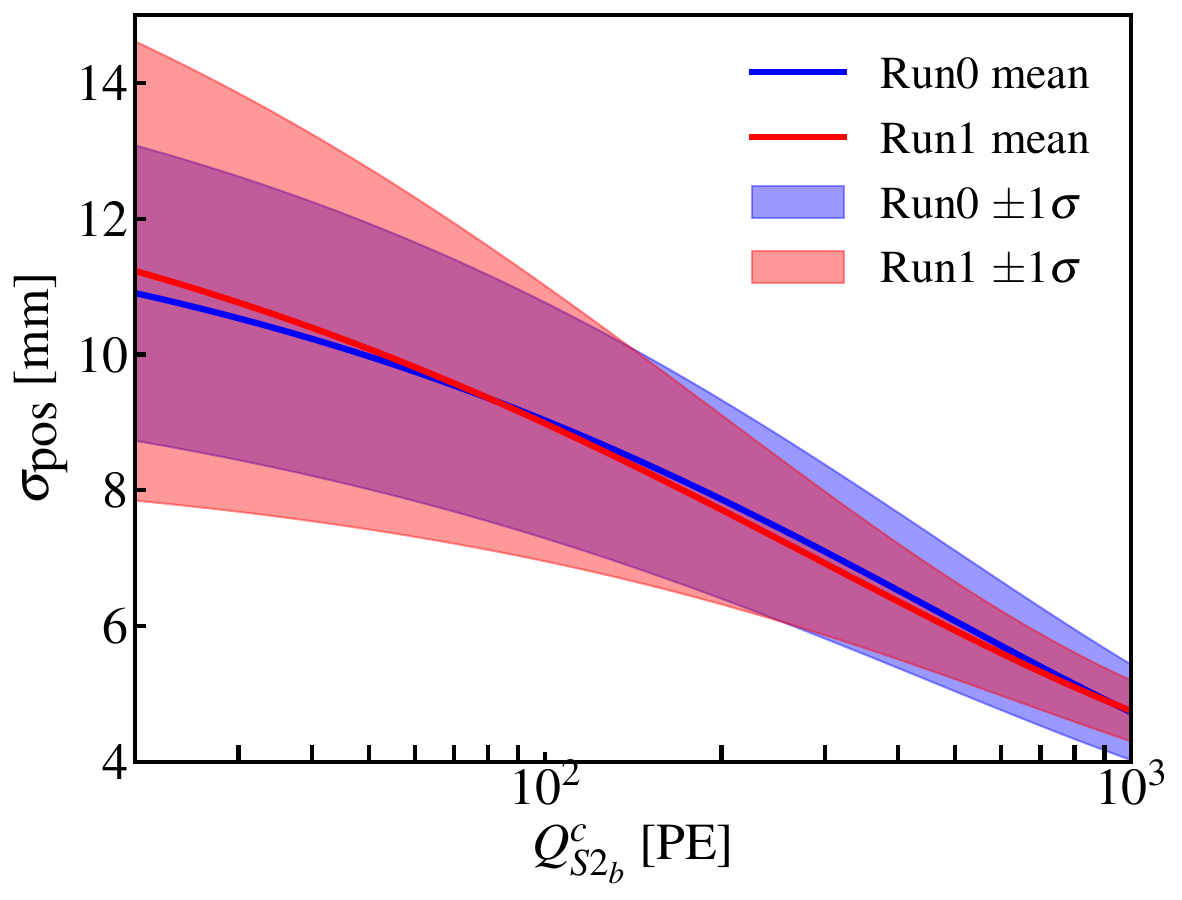}
    \caption{Position reconstruction resolution as a function of the corrected $S2$ charge from bottom PMTs.}
    \label{fig:position_resolution}
\end{figure}

The vertical position in the PandaX-4T experiment is determined by multiplying the drift velocity by the time difference between the $S1$ and $S2$ signals.
For horizontal position reconstruction in the PandaX-4T experiment, two algorithms have been developed: the template matching (TM) and the photon acceptance function (PAF) methods. These algorithms are designed to determine the scattering position of each event based on the signals collected by the top PMT array during the $S2$ signal. 
More details of the position reconstruction can be found in Ref.~\cite{zhang2021horizontal}.
The reconstruction quality is influenced by the statistical fluctuation of $S2$ hit pattern, and depends on the $S2$ charge.
% Fig.~\ref{fig:position_resolution} shows the standard deviation of the reconstructed position, denoted as $\sigma_{\textrm{pos}}$, as a function of the corrected $S2$ charge, which is estimated using optical simulation. 
The position resolution (Fig.~\ref{fig:position_resolution}) is estimated conservatively by taking the data-driven edge events ($S2$ from \ncl{Po-210} alpha), by a series Gaussian fit on the radius distribution of different $S2_\mathrm{b}$ ranges.

% The vertical position is simply the product of drift velocity and the time difference between $S1$ and $S2$.
% For horizontal position reconstruction, two algorithms are developed for the PandaX-4T experiment, namely the template matching (TM) and the photon acceptance function (PAF)~\ref{Gu202x in prep}.
% Both methods aim to determine the scattering position of each event based on the photocharges of $S2$ collected by the top PMT array. 

Non-functioning PMTs (``off-PMT'') lead to topological defects in the charge pattern on PMTs, leading to offsets in horizontal position reconstruction.
Especially when several adjacent PMTs are malfunctioning, this offset effect becomes more pronounced, which results in significant charge loss.
To reduce the reconstruction uncertainty at and close to ``off-PMT'' regions, the brightest PMT channel center is used as the prior position of the reconstructed algorithm. 
Relaxed $S2$ TBA selection criteria to the scatter events located at this region is applied to reduce the acceptance loss. 
% Fig.~\ref{fig:position_kr83m}

\subsection{Model of signal reconstruction in signal response model}

\begin{figure*}[htp]
    \centering
    \includegraphics[width=0.9\textwidth]{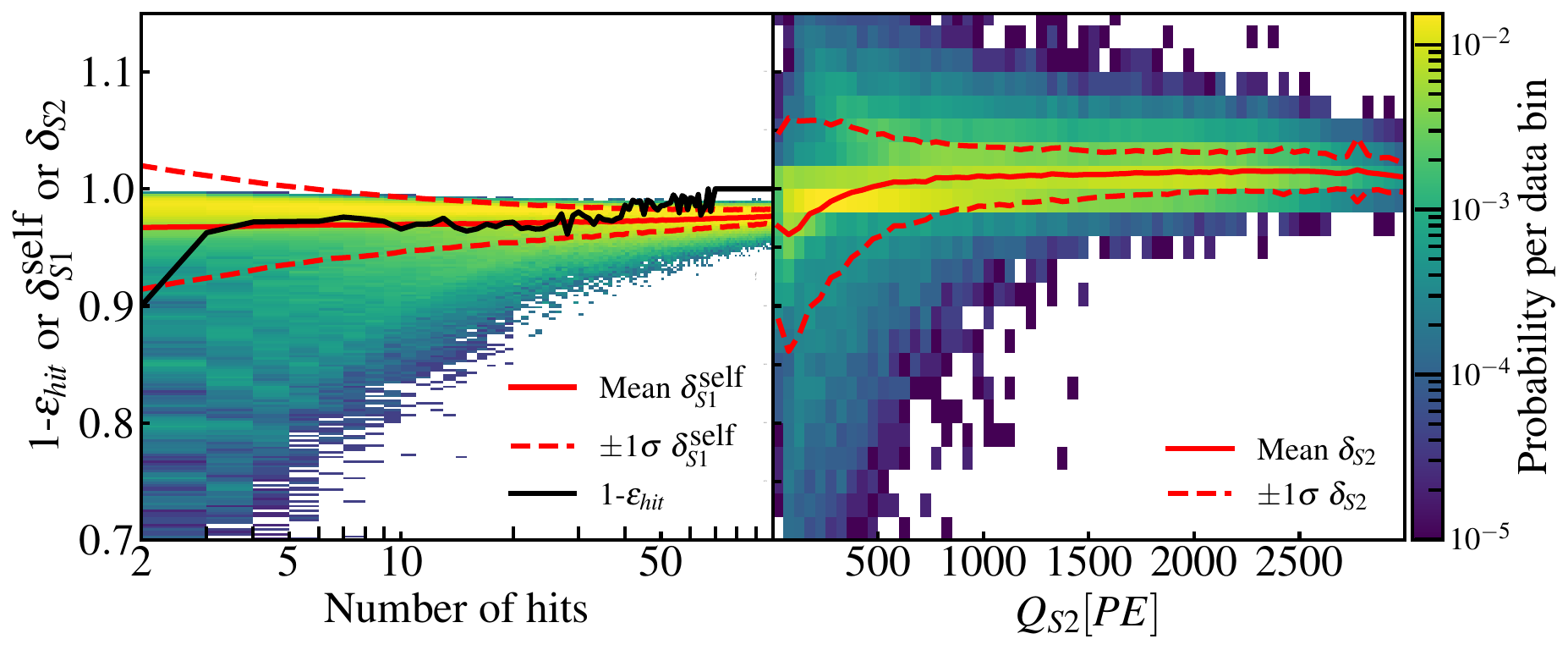}
    \caption{
    The hit surviving probability 1-$\varepsilon_{\textrm{hit}}$ is shown in left panel in black solid line, along with the probability distribution of self-trigger bias as a function of number of hit for the $S1$.
    In the right panel, it shows the probability distribution of $S2$ mean bias as a function of $S2$ charge $Q_{S2}^c$.
    The red solid and dashed lines display the mean and $\pm$1$\sigma$ of the contour, respectively, in both panels.
    }
    \label{fig:s1_s2_bias}
\end{figure*}

The PE waveforms, shaped by PMTs, undergo processing procedures described in the previous subsections. 
These detected hits are organized into clusters, which are further classified as either $S1$ or $S2$ signals based on their respective pulse widths.
The identified $S1$ and $S2$ signals are subsequently paired together to form physical events. 
However, it is important to note that both $S1$ and $S2$ signals can be subject to biases during the clustering and classification processes. 
For example during the clustering process, the hits are assigned to clusters if the time difference between any two hits is less than 60 ns.
Considering the photon propagation in LXe and reflection on PTFE surface, the efficiency loss caused by the clustering process could be non-trivial.
This effect mainly influences the low-energy region, and the number of photons that survive the hit clustering is modeled as
\begin{equation}
    N_{\textrm{det}}^\prime = B\left(N_{\textrm{det}}, 1-\varepsilon_{\textrm{hit}}\right), 
    \label{eq:hit_clustering_loss}
\end{equation} 
where the $\varepsilon_{\textrm{hit}}$ is the loss probability of 1\,hit during the clustering and is dependant on the number of hit $N_{\textrm{det}}$, shown in Fig.~\ref{fig:s1_s2_bias}.
The hit loss probability is estimated using the PandaX-4T waveform simulation framework~\cite{pandax4t_wf}.
% Using the waveform simulation~\cite{pandax4t_wf}, the $\varepsilon_{\textrm{hit}}$ is estimated to be about \textcolor{red}{xx} and \textcolor{red}{xx} with  $N_{\textrm{det}}$ being 2 and 3, respectively.
% With $N_{\textrm{det}}$ larger than 3, the hit loss can be negligible.
% Additionally, the scintillation lights generated within the TPC require some time to travel before they can be collected by the PMTs.

Furthermore, it is important to acknowledge the potential bias introduced to the $S1$ charge measurement due to the self-trigger threshold of the digitizers.
The self-trigger threshold is set at 20\,ADC, and as a result, single hits below this threshold will be discarded. 
Consequently, the $S1$ obtained after hit clustering may be underestimated compared to the true $S1$ value.
% To estimate the impact of this bias, two scenarios are compared: one without setting a threshold and one with a 20-ADC threshold. 
Externally triggered data without such 20-ADC self-trigger threshold from the LED light calibration is utilized to determine the efficiency of such 20-ADC self-trigger threshold to $S1$s with various signal sizes.
Applying the self-trigger threshold allows for the successful recording of approximately 90\% of the single PEs, contributing to the overall bias in the $S1$ charge measurement. 
The $S1$ charge is modeled in the signal response model as
\begin{equation}
    Q_{S1}  = N_{\textrm{PE}} G\left(1+\delta_{S1}^{\textrm{self}}, \Delta \delta_{S1}^{\textrm{self}}\right), 
    \label{eq:self_trigger_bias}
\end{equation}
where $\delta_{S1}^{\textrm{self}}$ and $\Delta \delta_{S1}^{\textrm{self}}$ represent the $S1$ mean bias caused by the self-trigger and its associated fluctuation, respectively.
The bias caused by the self-trigger threshold is overlaid in the left panel of Fig.~\ref{fig:s1_s2_bias}.

The aforementioned factors also collectively contribute to a slight bias in the $S2$ charge after the signal reconstruction.
The combined biases arising from the clustering, classification, and pairing procedures for the $S2$ signal ($\delta_{S2}$), along with their corresponding fluctuation ($\Delta \delta_{S2}$), are assessed through a dedicated waveform simulation, as detailed in Ref.~\cite{pandax4t_wf}.
Fig.~\ref{fig:s1_s2_bias} illustrates the mean and fluctuation of these biases for $S2$ signals as a function of the $S2$ charges.
The $S2$ charge ($Q_{S2}$) is then modelled as:
\begin{equation}
    \centering
    \begin{aligned}
    Q_{S2} & = N_{\textrm{prop}} G(1+\delta_{S2}, \Delta \delta_{S2}),
    \end{aligned}
    \label{eq:s2_bias}
\end{equation}

\begin{figure*}
    \centering
    \includegraphics[scale=0.45]{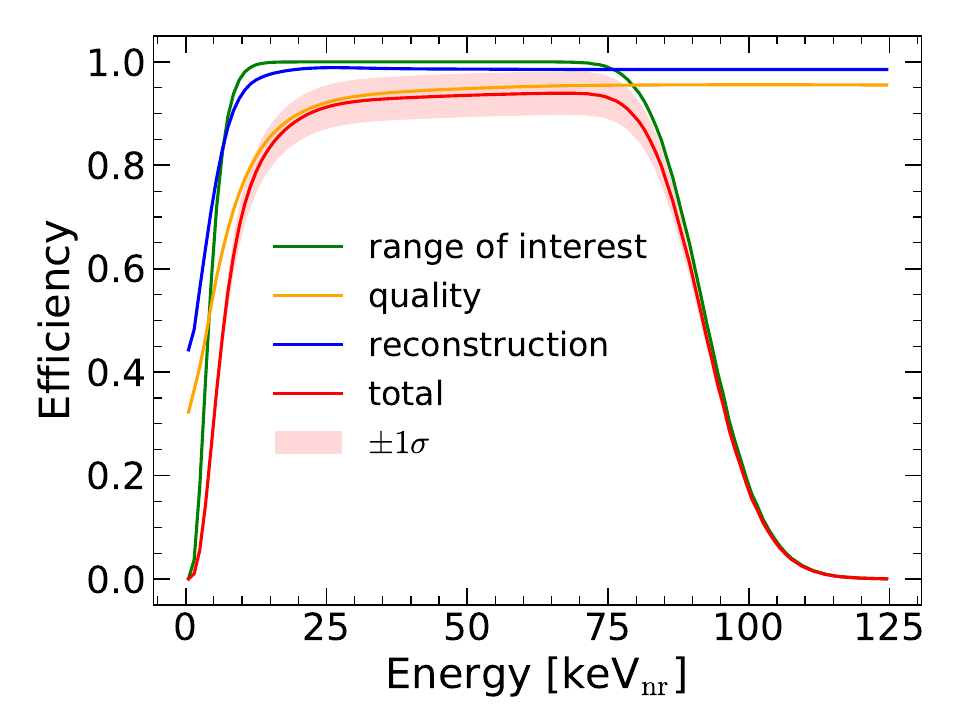}
    \includegraphics[scale=0.45]{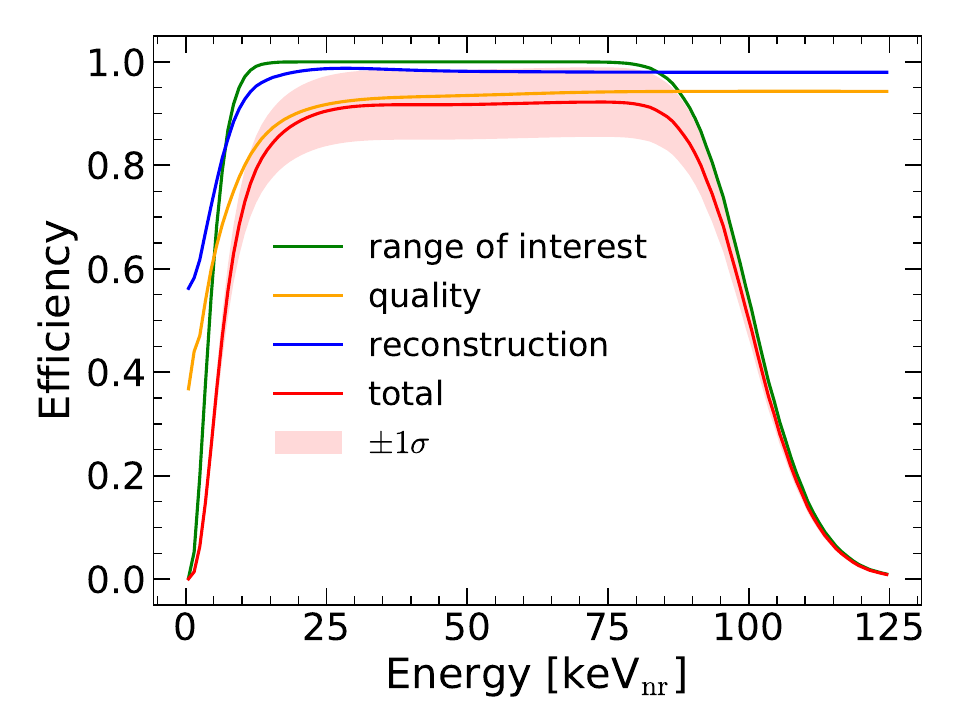}
    \includegraphics[scale=0.45]{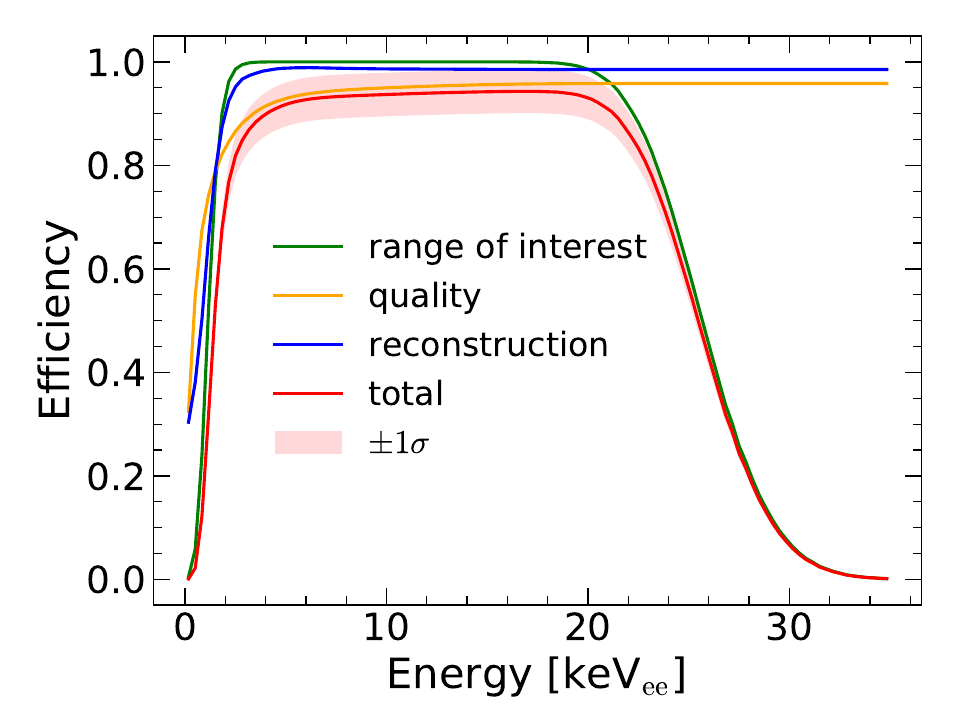}
    \includegraphics[scale=0.45]{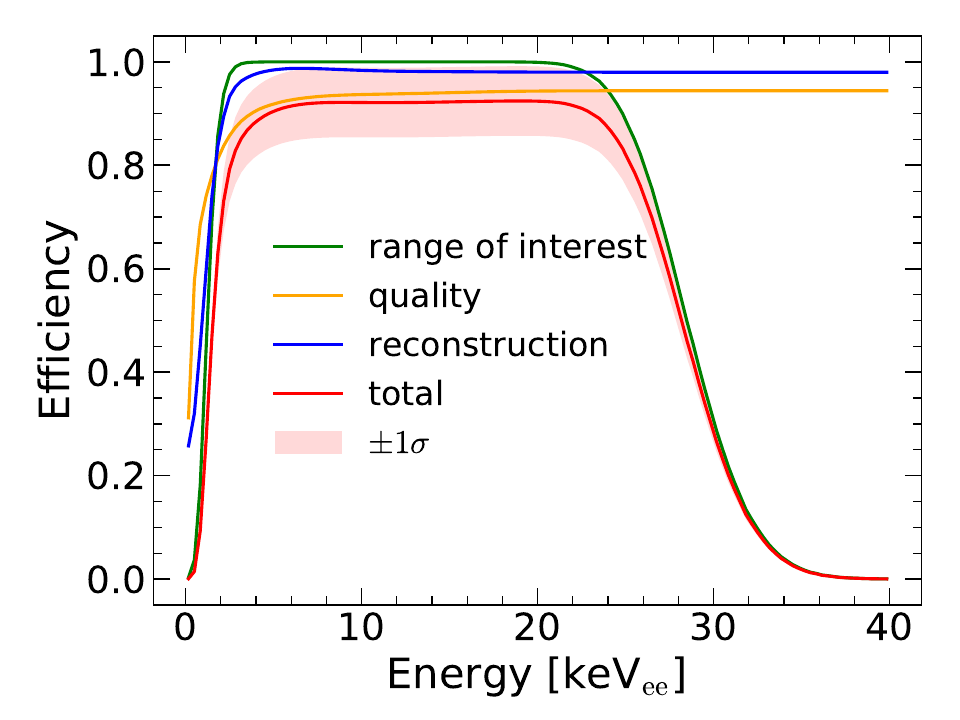}
    \caption{
    The detection efficiency (green line), reconstruction efficiency (blue line), data quality efficiency (orange line), and the total efficiency (red, with the shaded band representing the uncertainty ) as a function of NR energy (upper panels) and ER energy (lower panels).
    The left and right panels show the efficiencies for Run0 and Run1, respectively.
    % \textcolor{red}{(@TY Quality, reconstruction efficiencies as a function $S1$, $S2$, and energy.)}
    % \textcolor{red}{(To reviewers: we need to update pairing efficiency, and the error band as well.)}
    }
    \label{fig:efficiency}
\end{figure*}

% efficiency part
Signal reconstruction can also contribute to a direct loss in efficiency, particularly during the processes of pulse classification and the pairing of $S1$ and $S2$ signals. The assessment of these efficiencies is performed using the waveform simulation framework~\cite{pandax4t_wf}.
Efficiency values, characterized as a function of energy, are presented in Fig.~\ref{fig:efficiency}.

\section{Signal correction}
\label{sec:signal_correction}

%\subsection{PMT gain correction}

% \textcolor{red}{(QL: This subsection too technical, and not used in signal model. I suggest to remove.)}
% \textcolor{blue}{TY: This gain calibration step can be reduced to one sentence and move to the signal reconstruction part.}

% LED light calibration is performed once a week to routinely update the gain response of PMTs.
% Data taken around light calibration runs are reprocessed in order to obtain more accurate detected charges.
% However, later we find that the auto-script may give bad fit result of hit area from time to time and such LED gain calibration is neither timely nor accurate.
% Thus we change to a self-calibration strategy based on \insitu single photon peak measurement (so-called ``rolling gain'') to run-by-run correct gain of all PMT channels.
% Besides, the auto-script may give bad fit result of hit area from time to time, thus we adopt the so-called ``rolling gain'' approach to further correct those specific runs.
% The ``rolling gain'' correction is quite necessary during radioactive source calibration data taking when PMTs would suffer larger light intensity, leading to the 1-hit area (corresponding to single photon) peak deviates away from $1$~PE (Fig.~\ref{fig:single_photon_fit}).

% \begin{figure}
%     \centering
%     \includegraphics{plots/rolling_gain/single_photon_peak_fit.png}
%     \caption{
%     \textcolor{black}{TY: Single photon peak fit of PMT channel 12201 in run 5108.}
%     }
%     \label{fig:single_photon_fit}
% \end{figure}

\subsection{Spatial uniformity correction}
\label{subsec:signal_spatial_correction}
\begin{figure*}[htp]
    \centering
    \includegraphics[width=0.95\textwidth]{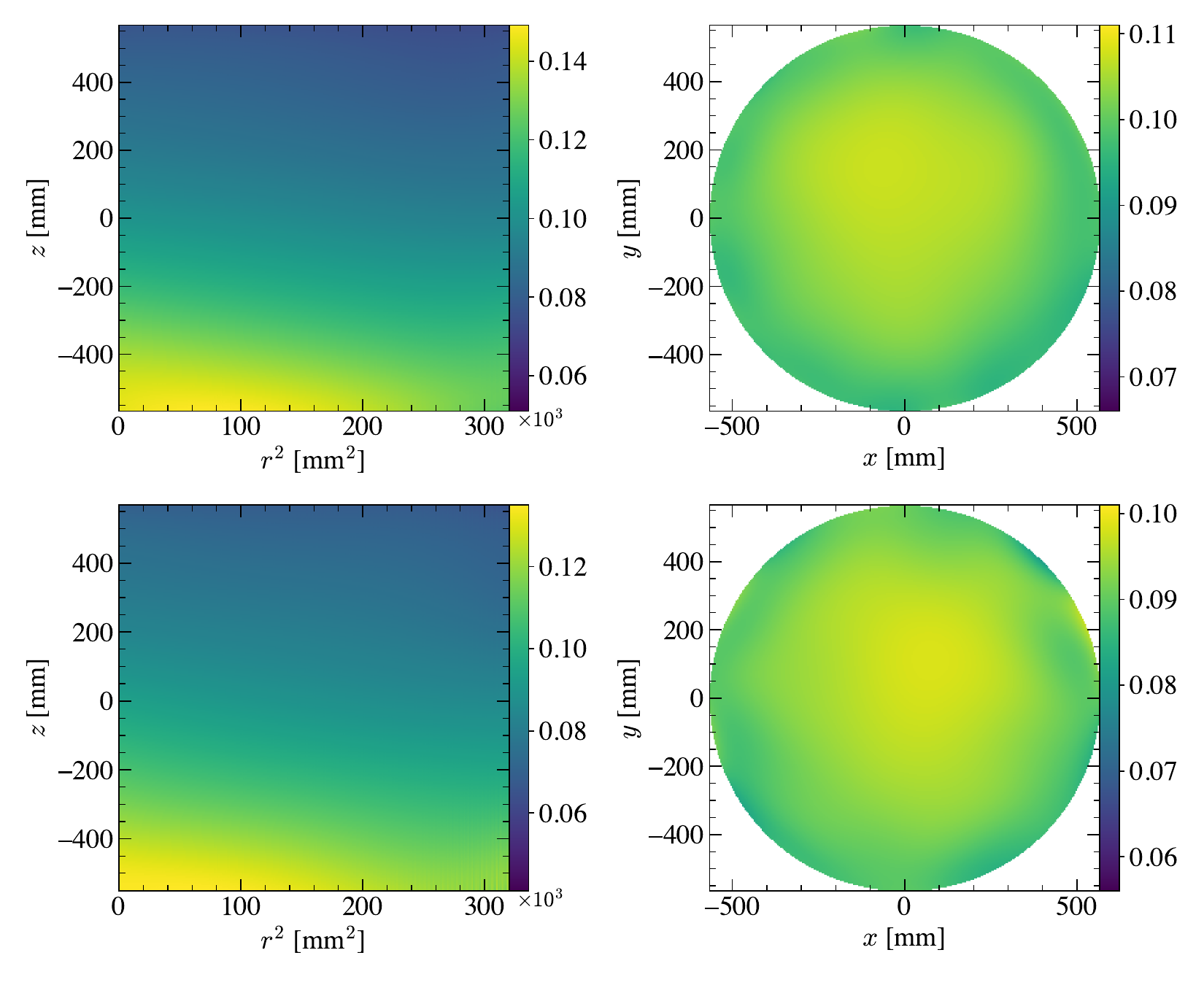}
    \caption{
    The 3-D PDE $\varepsilon_{\textrm{PDE}}$ map projected on ($r^2$, $z$) and ($x$, $y$) planes are shown in the left and right panels, respectively.
    The top and bottom panels show the maps for Run0 and Run1, respectively.
    The color axes give the values of the position-dependent PDE $\varepsilon_{\textrm{PDE}}(x,y,z)$ (Eq.~\ref{eq:pde}).
    % \textcolor{red}{@TY Need to add Run1 maps, and convert to PDE.}
    }
    \label{fig:s1_correction_map}
\end{figure*}

\begin{figure*}[htp]
    \centering
    \includegraphics[width=0.95\textwidth]{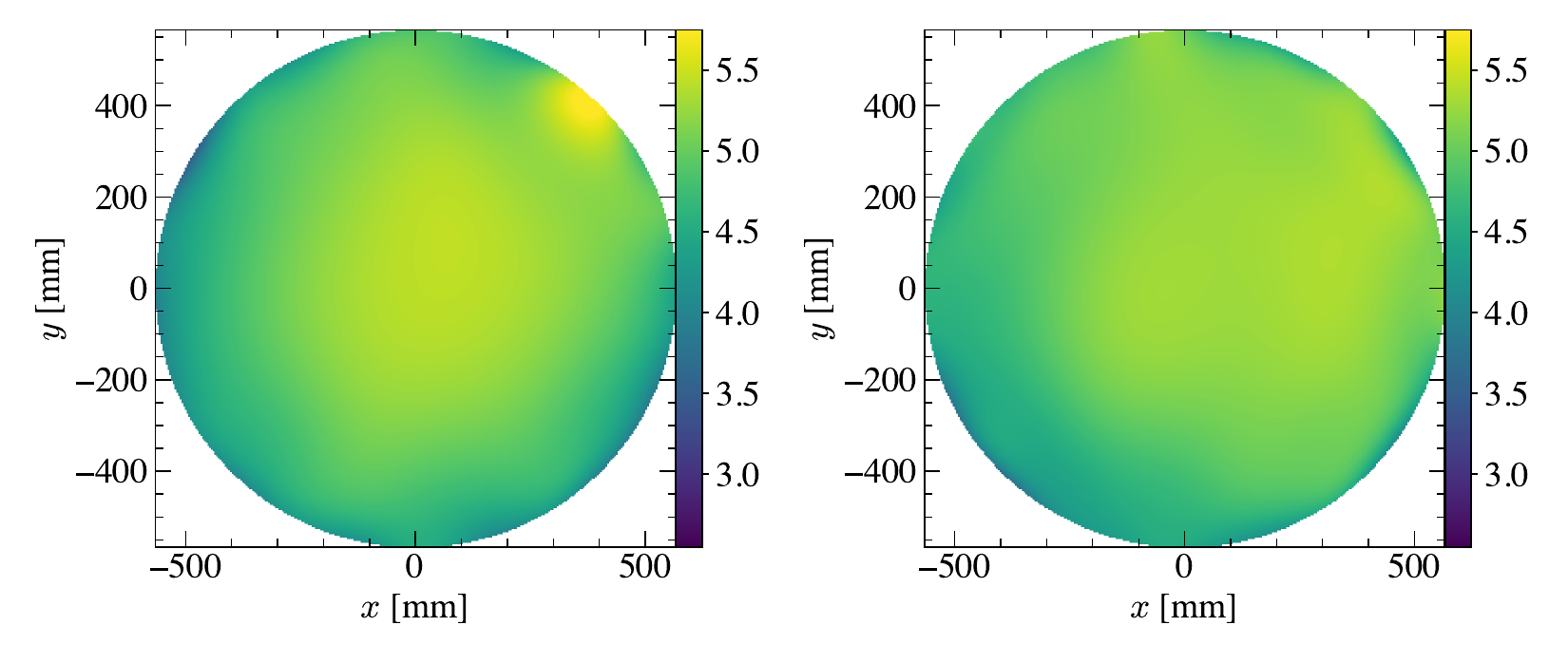}  
    \caption{
    The $S2$ gain maps as a function of horizontal position ($x$, $y$) for Run0 and Run1. 
    The bright spot that appears in the right top of the left plot is due to one noisy PMT channel during Run0.
    % \textcolor{red}{@TY need to scale to SEG mapping. }
    }
    \label{fig:s2_correction_map}
\end{figure*}

% \subsection{$S1$ correction}
The spatial non-uniformity of the $S1$ and $S2$ signals in the PandaX-4T detector is primarily attributed to several factors. 
These include the unevenness of the electric field, the levelness of the liquid-gas surface, and the optical solid angle. 
Additionally, operational conditions of the PMTs and impurity concentration in the LXe can also contribute to these non-uniformities.
These spatial non-uniformities have the potential to degrade the energy resolution of the detector, thereby impacting its overall detection sensitivity.
To correct for the spatial non-uniformities, a method utilizing an injected radioactive source, specifically \nclm{Kr-83}, is employed.
It is assumed that the 41.5 keV X-ray-induced ER events from \nclm{Kr-83} are uniformly distributed throughout the sensitive volume of the PandaX-4T TPC.
The correction maps, expressed as the photon detection efficiency $\varepsilon_{\textrm{PDE}}$ and charge amplification factor $\kappa_\mathrm{b}$ for $S1$ and $S2_\mathrm{b}$, respectively, are obtained through a fitting procedure using a three-variable 9th-degree polynomial function ($\sum_{ijk}c_{ijk}x^i y^j z^k$, where $i,j,k$ are integers from 0 to 9).
Fig.~\ref{fig:s1_correction_map} shows the $\varepsilon_{\textrm{PDE}}$ on ($r^2$, $z$) and ($x$, $y$) for Run0 and Run1.
Fig.~\ref{fig:s2_correction_map} shows $\kappa_\mathrm{b}$ on ($x$, $y$) for Run0 and Run1.
The reconstructed positions in \nclm{Kr-83} data are approximated as the true positions since the position reconstruction resolution is small at the $S2$ size for \,\nclm{Kr-83} ($>$1000 detected electrons).

As mentioned in Sec.~\ref{sec:signal_collection}, the number of electrons gets reduced due to the attachment to electro-negative impurities in LXe during the drift process.
The 5.6~MeV \ncl{Rn-222} $\alpha$ events are also used for obtaining the electron lifetime $\tau_e$.
All physical $S2$s adopt this z-dependent charge correction by a factor of $e^{-z/\tau_e/v_{\textrm{drift}}}$.

% \subsection{$S2$ correction}

\subsection{Temporal variation correction}

The magnitude of the detected $S1$ and $S2$ signals in the PandaX-4T experiment is known to be affected by variations in the detector conditions, such as the liquid level. 
These variations occur over time, particularly during Run1 when the overflow tube experienced a failure. 
To mitigate the impact of this instability, a temporal correction is applied to the $S1$ and $S2$ signals.
To derive the run-by-run time-correction factors for the $S1$ and $S2$ signals, 5.6 MeV $\alpha$ decay events from \ncl{Rn-222} are utilized.
These events are used to determine the correction factors, which are then applied to signals at all energy levels.
% To avoid PMT saturation effects and influences from other radioactive materials, only ~\ncl{Rn-222} $\alpha$ events with drift times ranging from 200\,$\mu$s to 550\,$\mu$s are considered.
Only ~\ncl{Rn-222} $\alpha$ events with drift times ranging from 200\,$\mu$s to 550\,$\mu$s are considered.
The lower and upper limits of the drift time range are set to avoid influences from other radioactive impurities and PMT saturation, respectively.
The reference points for the correction factors are determined based on the average values of the last 10 DM runs for both Run0 and Run1.
Set 1-3 of Run0 are further corrected set-by-set based on the 163.9~keV $\gamma$ peak from \nclm{Xe-131} in a similar approach.
% The temporal variation in the $S1$ and bottom $S2$ signals, with respect to these referce points, can reach up to $0.3\%$ and $0.4\%$, respectively.
The variations before such temporal correction are $0.6\%$($0.7\%$) in $S1$ and $1.6\%$($4.6\%$) in $S2_\mathrm{b}$ on average for Run0(Run1).

\subsection{Position correction}

\begin{figure*}[htp]
    \centering
    \includegraphics[width=0.98\textwidth]{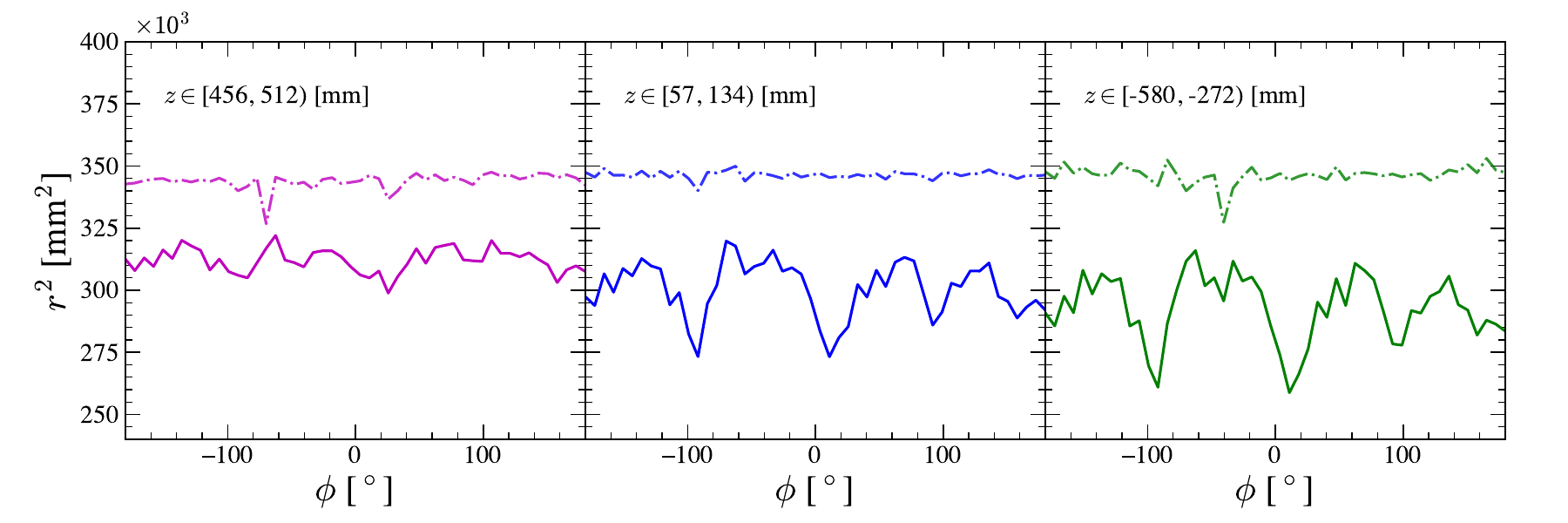}
    \caption{
    The median of radius square  of ~\ncl{Po-210} $\alpha$ events before (solid line) and after (dash-dotted line) the azimuth-angle-dependent scaling at different $z$.
    }
    \label{fig:po210_pos}
\end{figure*}

Due to several detector effects including the distortion of drift electric field and the segmented coverage of the top PMTs, events, especially for those close to the PTFE wall, tend to be reconstructed towards the interior in terms of horizontal position.
For a direct comparison in the MC simulation, an azimuth-angle-dependent as well as z-dependent horizontal radial position affine scaling is necessary.
The scaling factors are derived based on the uniformity of ~\nclm{Kr-83} ER events. 
Fig.~\ref{fig:po210_pos} shows the significant effect of this radial scaling using ~\ncl{Po-210} $\alpha$ events, which are mostly from the PTFE surface and selected through dedicated criteria involving event's $S1$ charge and TBA.

\subsection{Model of signal correction in signal response model}

In the PandaX-4T TPC, both the $S1$ and $S2$ signals exhibit spatial dependence, as discussed in previous subsections.
% % These dependencies are calibrated using a dedicated calibration involving the injection of a \nclm{Kr-83} source~\cite{}.
% The spatial dependence of the $S1$ signal arises from the spatially varying $\varepsilon_{\textrm{PDE}}$ as a function of truth position ($x$, $y$, $z$). 
% The $\varepsilon_{\textrm{PDE}}$ values for the $r$-$z$ and $x$-$y$ planes are shown in Figure \ref{fig:s1_correction_map}. 
% Within the selected FV of PandaX-4T, the maximum difference in $\varepsilon_{\textrm{PDE}}$ is approximately \textcolor{red}{xx\%}.
% The $S2$ signal's dependence on the $z$ coordinate is attributed to the attachment of ionized electrons to impurities during the drifting process, as described in Section~\ref{sec:signal_collection}. 
% Additionally, the $S2$ signal's dependence on the $x$-$y$ plane is likely influenced by spatial variations in the $S2$ light collection efficiency, the warping of the anode electrode, and the unevenness of the TPC. 
% The maximum difference in the amplification factor $\kappa$ within the FV is approximately xx\%, and the $\kappa$ values on the $x$-$y$ plane are depicted in Figure \ref{fig:s2_correction_map}.
To account for the spatial dependence of signals, the $S1$ and $S2$ charges can be corrected through:
\begin{equation}
   \centering
   \begin{aligned}
       Q^c_{S1} & = Q_{S1} \langle \varepsilon_{\textrm{PDE}} \rangle / \varepsilon_{\textrm{PDE}} (x_{\textrm{rec}}, y_{\textrm{rec}}, z_{\textrm{rec}}), \\
       Q^c_{S2} & = Q_{S2}  e^{z/\tau_e/\nu_{\textrm{drift}}} \langle \kappa \rangle / \kappa (x_{\textrm{rec}},y_{\textrm{rec}}).
   \end{aligned}
\label{eq:correction}
\end{equation}
% where $\langle \kappa \rangle$ and $\langle \varepsilon_{\textrm{PDE}} \rangle $ are the average values of $\kappa$ and $\varepsilon_{\textrm{PDE}}$, respectively, within the FV.
% $\nu_{\textrm{drift}}$ represents the electron drift velocity.
It should be noted that the position coordinates in Eq.~\ref{eq:correction} are the reconstructed coordinates.
These reconstructed coordinates are susceptible to fluctuations due to the inherent resolution limitations of the position reconstruction algorithms.
The extent of these fluctuations is influenced by the size of the $S2$ signal, with smaller $S2$ signals resulting in more pronounced fluctuations. 
The $S2$-dependent position reconstruction resolution, denoted as $\sigma_{\textrm{pos}}$, is illustrated in Fig.~\ref{fig:position_resolution}.
Assuming identical position resolutions in the $x$ and $y$ directions, the reconstructed transverse positions $x_{\textrm{rec}}$ and $y_{\textrm{rec}}$ can be expressed as
\begin{equation}
    x_{\textrm{rec}}  =  G(x, \sigma_{\textrm{pos}}), 
    y_{\textrm{rec}}  =  G(y, \sigma_{\textrm{pos}}). 
\label{eq:position_reconstruction_sample}
\end{equation}

\section{Signal selection}
\label{sec:signal_selection}

In order to maintain a high level of data purity and eliminate spurious events, a series of data selections are applied.
These selections involve quality assessments of the $S1$ and $S2$ signals, correlation between $S1$ and $S2$, waveform ``dirtiness'', and other considerations. 
More details are given in the following subsections.

\subsection{Data quality}

\begin{figure}[htp]
    \centering
    \includegraphics[width=0.95\columnwidth]{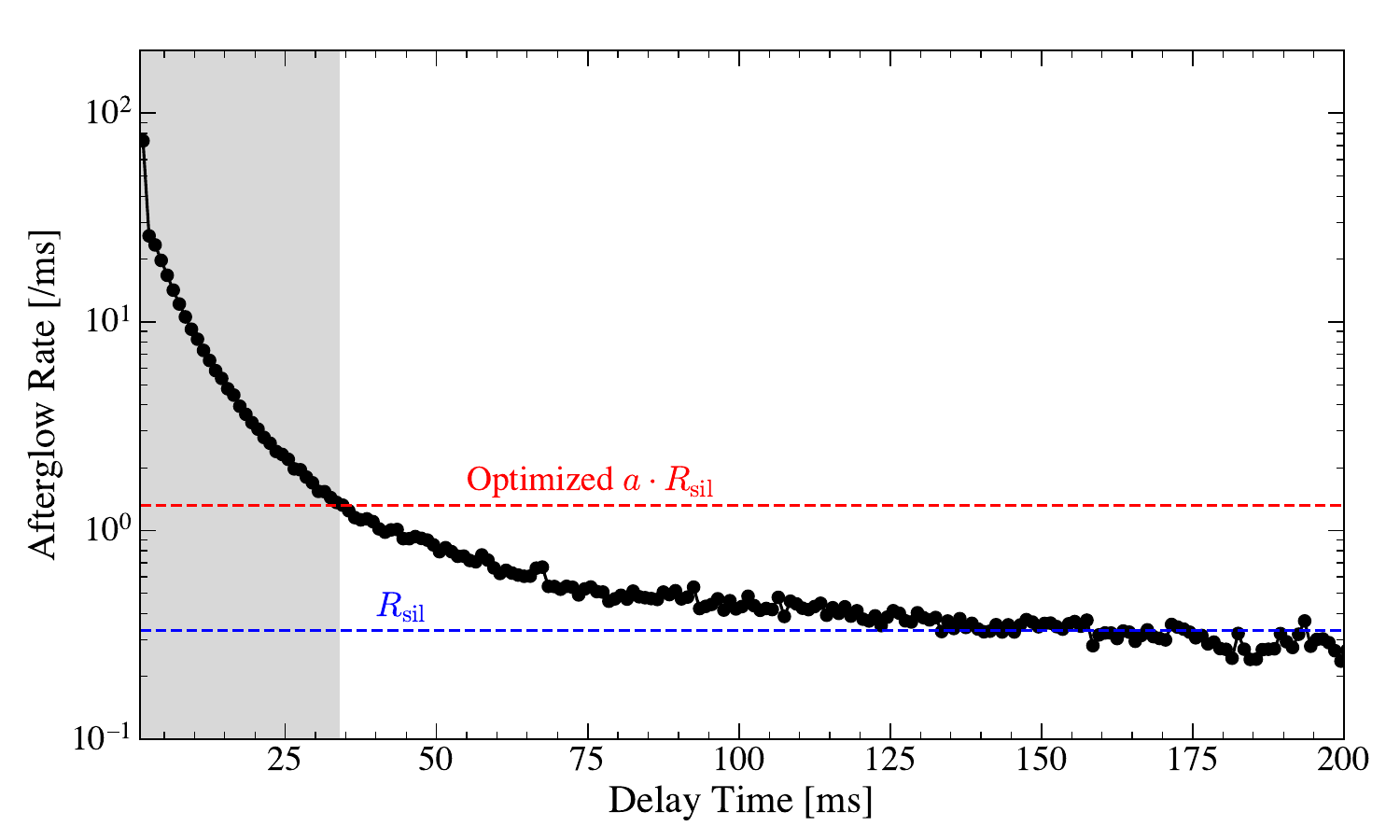}
    \caption{
    % \textcolor{red}{
    % (Is the unit on Y-axis correct? This is differential rate. Maybe it should be kHz/ms?)
    % }
    An example of the $S2$ afterglow rate as a function of the delay time following one large $S2$ is shown by the black dots.
    The blue dashed line represents the calculated ``silent rate'' level. 
    The red dashed line corresponds to the optimized cut-off rate $a\cdot R_{\textrm{sil}}$. The gray shaded region gives the time window for cut-off, which is determined based on the cut-off rate.
    }
    \label{fig:deadtime_cut}
\end{figure}

To ensure data quality in the PandaX-4T experiment, certain measures are implemented to remove exposure times that exhibit high rates of either $S1$ or $S2$ signals.
The first step involves identifying and removing data files, each containing approximately $15$ seconds of data, that exhibit significantly higher rates of $S1$ signals within a continuous period of time where the $S1$ rate exceeds the normal rate (10 to 15 Hz) by 2 standard deviations (2 to 3 Hz).
$1\%$ and $7\%$ of live time are removed in Run0 and Run1, respectively.
A specific treatment is applied to address the issue of afterglow, which refers to the presence of delayed electrons following a large signal. This treatment is performed on an event-by-event basis. 
After each pulse with a charge exceeding 10000\,PE, a certain length of the recorded data time window is vetoed and excluded from analysis.
To determine the appropriate length of the vetoed window, the concept of ``silent rate'' is introduced. 
The silent rate ($R_{\textrm{sil}}$) refers to the average rate of signals in non-vetoed windows, which are significantly delayed  ($\gtrsim$ 200~ms.) following large signals, within a given data file.
The window length is determined such that the charge density (total $S2$ charge per unit time) falls below a threshold of $a \cdot R_{\textrm{sil}}$, as illustrated in Fig.~\ref{fig:deadtime_cut}.
The scaling factor $a$ is optimized by maximizing a figure-of-merit (FoM) defined as:
\begin{equation}
\label{{eq:fom-deadtime}}
    \textrm{FoM} = \frac{\epsilon_{\textrm{aft}} (a) }{\sqrt{ R_{\textrm{aft}} (a) }},
\end{equation}
where $\epsilon_{\textrm{aft}}$ and $R_{\textrm{aft}}$ are the fraction of exposure time left and average $S2$ rate, respectively, after applying the aforementioned veto on data time window.
The exact value of $a$ depends on the $S2$ ROI for the analysis which influences the values of $R_{\textrm{sil}}$ and $R_{\textrm{aft}}$.

\subsection{Individual signal quality}

\begin{table*}[htp]
    \centering
    \begin{tabular}{c|p{8cm}|c}
    \hline\hline
    Symbol          & Description  &  Remarks for major noise\\
    \hline\hline
    $A_\mathrm{hit,bottom}^{\max}(S1)$  & The largest hit area at the bottom of $S1$ & Remove sparking\\
    $A_\mathrm{hit,top}^{\max}(S1)$     & The largest hit area at the top of $S1$ & Remove sparking\\
    $\langle Q_{S1} \rangle_\mathrm{PMT}$    & The average charge per PMT channel of $S1$ & Remove sparking\\
    $N_{\mathrm{peak}}(S1)$             & The number of peaks identified in the $S1$ waveform & Remove $S1$-like single electron\\
    $w_\mathrm{CDF}^{90-10}(S1)$        & The CDF width of $S1$ waveform & Remove $S1$-like single electron\\
    TBA$_{S1}$                          & The TBA of $S1$ & Remove accidental coincidence\\
    $w_\mathrm{CDF}^{90-10}(S2)$        & The CDF width of $S2$ waveform & Remove accidental coincidence\\
    $h_{S2}$                            & The height of $S2$ waveform & Remove sparking\\
    $\sigma_{A_\mathrm{hit}}$           & The standard deviation of the hit areas of $S2$ & Remove sparking \\
    TBA$_{S2}$                          & The TBA of $S2$ & Remove gas event\\
    $\sigma^*_\mathrm{TM}$              & \RaggedRight The charge-weighted standard deviation of the reconstructed position by TM algorithm & Poorly reconstructed position\\
    $\Delta r^2_\mathrm{TM-PAF}$        & The square of the distance between two reconstructed positions by TM and PAF algorithms & Poorly reconstructed position\\
    % $Q_{S2} / w_\mathrm{CDF}^{10-90}(S2) / h_{S2}$ & The normalized $S2$ charge by the product of its width and height \\
    \hline\hline 
    \end{tabular}
    \caption{
    The parameters that are used in the signal quality selections, together with their descriptions and the types of noise they are intended to remove.
    }
    \label{tab:cut_parameters}
\end{table*}

\begin{figure*}[htp]
    \centering
    \includegraphics[width=0.95\textwidth]{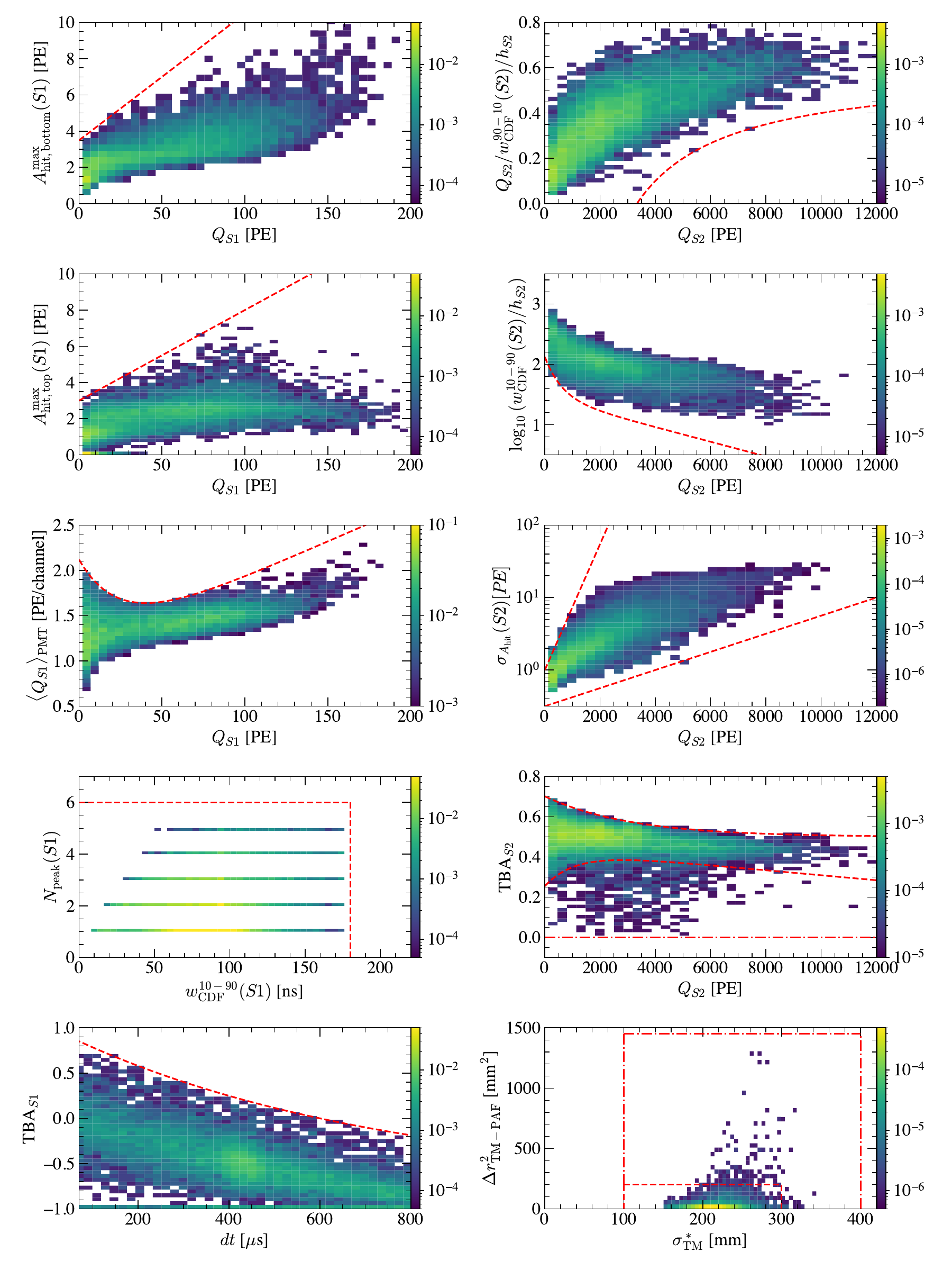}
    \caption{
    The parameter spaces of all the $S1$- and $S2$-related quality selections.
    The data shown are the sum of all the \,\ncl{Rn-220}, \ncl{Am-241}Be and DD calibration data.
    The red dashed lines give the selection boundaries in each specific parameter space.
    The red dash-dotted lines indicate the relaxed selection boundaries for the ``off-PMT'' region.
    Table~\ref{tab:cut_parameters} describes the meaning of the variables used here.
    }
    \label{fig:s1s2_qc_dist}
\end{figure*}

\begin{figure*}[htp]
    \centering
    \includegraphics[width=0.95\textwidth]{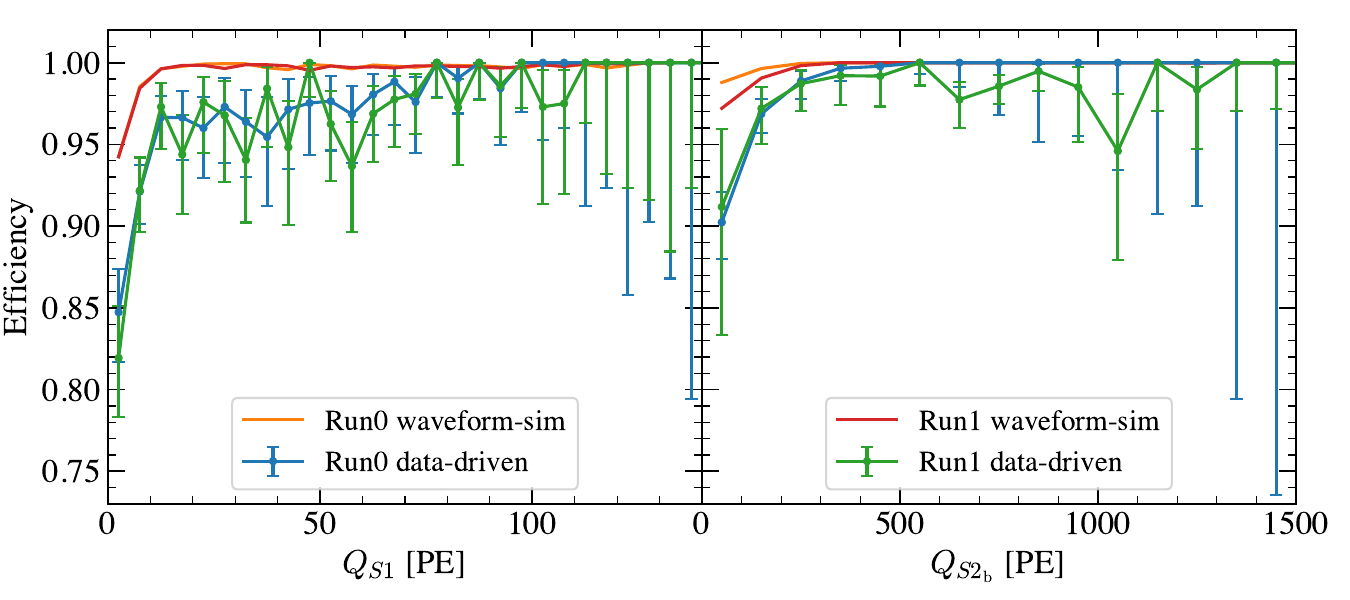}
    \caption{
     Efficiency curves as functions of $Q_{S1}$ and $Q_{S2_\mathrm{b}}$ for $S1$-related selection cuts and $S2$-related selection cuts, respectively. The difference between waveform simulation and data-driven (see text for detailed description) results is considered to be a systematic uncertainty. 
    }
    \label{fig:s1s2_eff}
\end{figure*}

The reconstruction quality of individual physical signals is ensured by applying range selections on the parameters that are related to the $S1$ and $S2$ pulse shapes, as well as the signal distributions among PMTs.
All the relevant parameters and their descriptions are listed in Table~\ref{tab:cut_parameters}.
% For $S1$s, based on the quality criteria mentioned in Sec.~\ref{subsec:pairing}, more stringent selection thresholds were set during the event selection.
% For $S2$s, several charge-dependent cuts on the TBA, waveform shape such as width-to-height ratio ($\log_{10}(w_\mathrm{CDF}^{10-90}(S2) / h_{S2})$), and the charge partition on PMT arrays are applied.
% Since the horizontal position of a physical event is reconstructed based on the charge distribution on the top PMT array of the $S2$, a few position parameters used in the TM and PAF algorithm provide extra constraints for a good quality $S2$.
It's worth noting that for events occurring near the PMTs that were turned off due to malfunctions (referred to as ``off-PMT'' region), the selection criteria based on TBA and reconstructed position parameters are appropriately relaxed to ensure that the acceptance in this region remains consistent with the rest of the TPC. 
The detail for signal quality selections on each parameter is presented in Fig.~\ref{fig:s1s2_qc_dist}, showing the low-energy events distribution from the sum of all the \,\ncl{Rn-220}, \ncl{Am-241}Be and DD calibration data in each specific parameter space, along with dashed colored lines represent the cut boundaries.
The data points that deviate from the charge-dependent TBA$_{S2}$ and fall below the lower boundary of the cut are associated with events that occur in the ``off-PMT'' region.

The efficiency of these individual signal quality selections are derived both from a data-driven approach using all of \,\ncl{Rn-220}, \ncl{Am-241}Be and DD calibration data and an approach based on the simulated samples from the waveform simulation~\cite{pandax4t_wf}, shown in Fig.~\ref{fig:s1s2_eff}.
In the data-driven approach, the events in these low energy calibration data that pass the single scatter cuts (see Sec.~\ref{subsec:single_scatter}) are selected to evaluate the efficiencies of the signal quality selections. 
Particularly, we require these events to have $S2$ charge and $w_{\textrm{CDF}}^{90-10}$ values within $15\%$-$85\%$ ($25\%$-$75\%$) quantiles of the expected $S2$ charge and width distributions for NR (ER) calibration data to ensure the purity of the data sample.
Considering the correlation between the selection criteria, efficiency of the selections that concern $S1$-related ($S2$-related) parameters are modeled as a function of $Q_{S1}$ ($Q_{S2}$).
The results of Run0 and Run1 are consistent with each other. 
The data-driven derived efficiency is taken as the nominal results, whilst the difference between waveform-simulation and data-driven results is considered as the systematic uncertainty.

\subsection{$S1$-$S2$ correlation}
\label{subsec:s1s2_correlation}

\begin{figure*}[htp]
    \centering
    \includegraphics[width=0.85\textwidth]{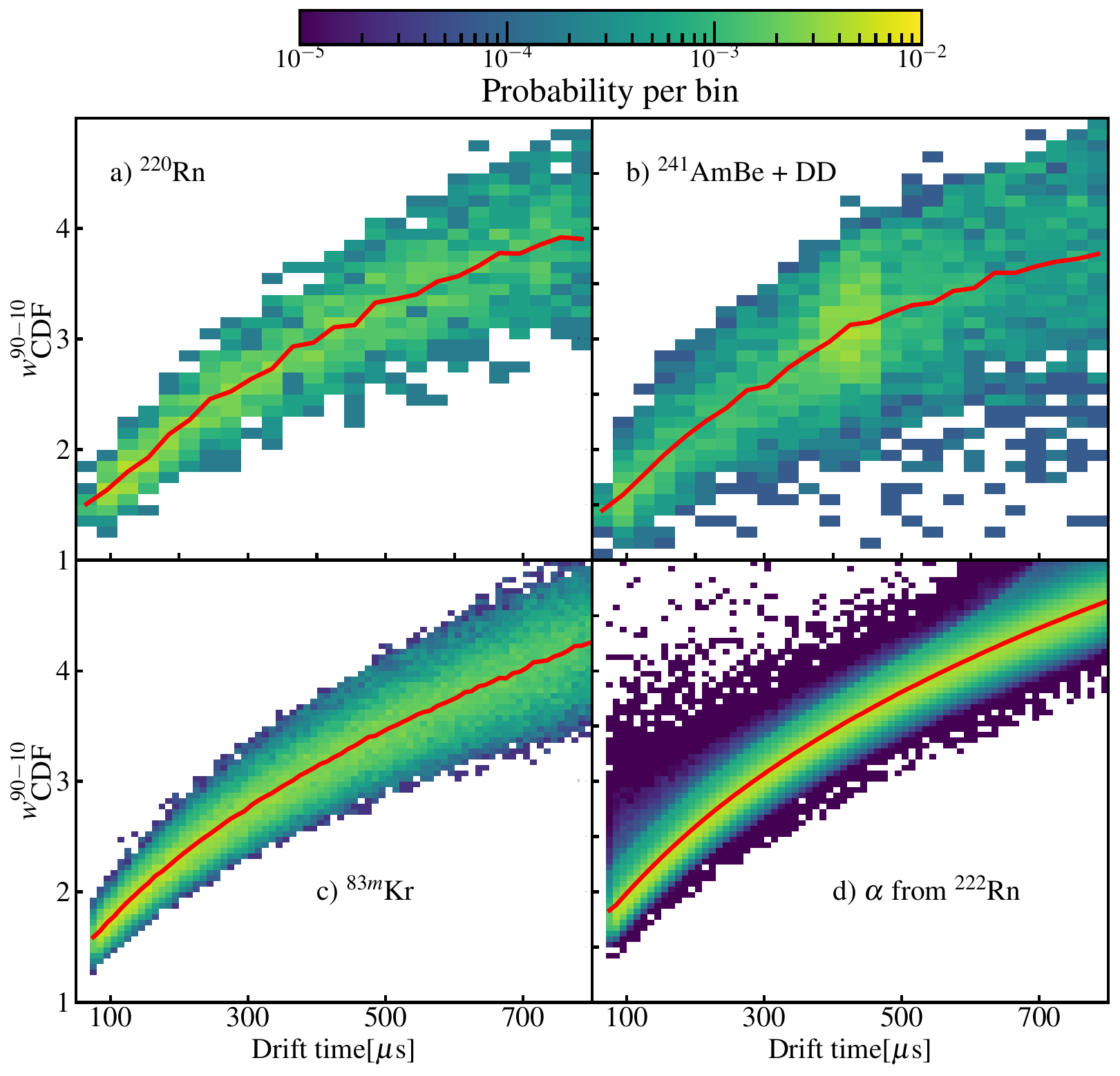}
    \caption{
    The normalized distributions of the $S2$'s $w^{90-10}_{\textrm{CDF}}$ over the drift time for four types of data: a) the \,\ncl{Rn-220} calibration, b) the neutron calibration using \ncl{Am-241}Be and DD, c) the \nclm{Kr-83} calibration, and d) the $\alpha$ events originated from \ncl{Rn-222} impurities.
    The red solid lines represent the means of the distributions.
    }
    \label{fig:diffusion}
\end{figure*}

\begin{figure*}
    \centering
    \includegraphics[width=0.9\columnwidth]{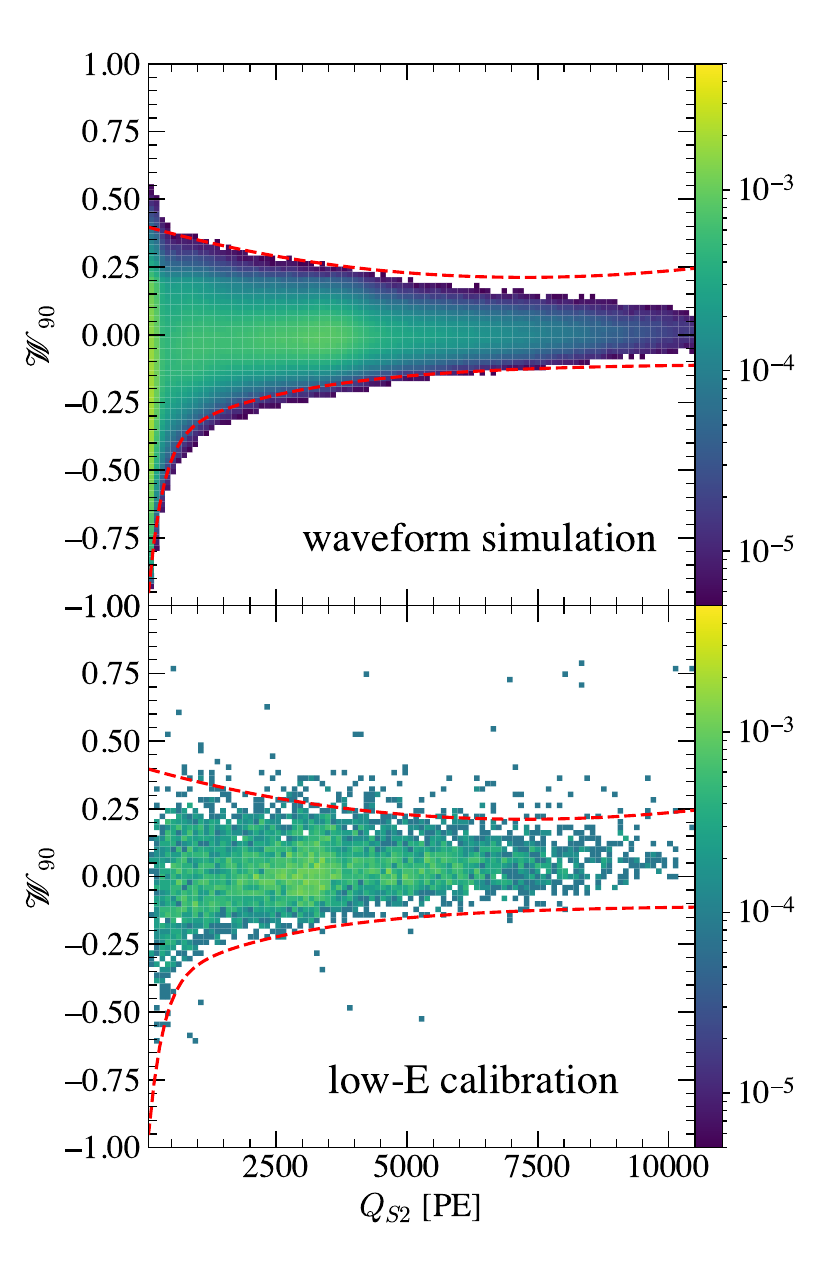}
    \includegraphics[width=0.9\columnwidth]{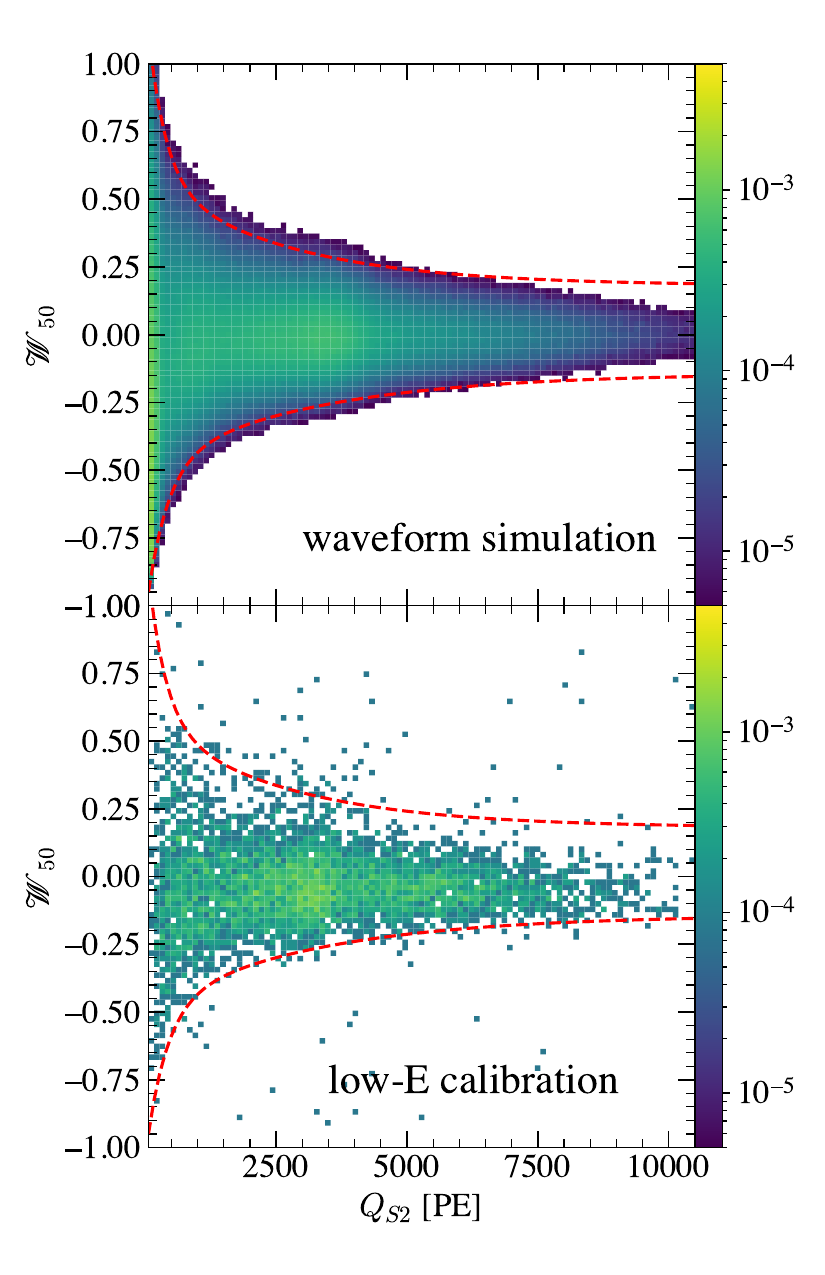}
    \caption{
    The distributions of the normalized flattened CDF widths.
    The top and bottom panels give the distributions from the waveform simulation data and  Run1 calibration data (~\ncl{Rn-220}, \ncl{Am-241}Be and DD), respectively.
    The left and right panels give the distributions of $\mathscr{W}_{90}$ and $\mathscr{W}_{50}$, respectively.
    The red dashed lines indicate the selected parameter spaces.
    }
    \label{fig:diffusion-cut}
\end{figure*}

In liquid xenon detectors, ionized electron clusters experience a diffusion effect during their drift process.
As a result, clusters with longer drift times will have larger widths, and specifically, the size in the vertical direction is reflected in the waveform's width.
In Fig.~\ref{fig:diffusion}, we show the distributions of the CDF width $w_{\textrm{CDF}}^{90-10}$ versus the drift time from the \nclm{Kr-83}, \ncl{Rn-220}, and neutron calibrations, as well as the $\alpha$ events from \ncl{Rn-222} in the background data.
Therefore a direct correlation between the primary width of the waveform and the drift time is established, known as the diffusion relation.
Moreover, the broadening of the width distribution for smaller $S2$s suffers more pronounced binomial fluctuations in the number of electrons contributing to the small signals.
A charge-dependent selection criterion is  applied on the normalized flattened parameter spaces of the 10\%-to-90\% ($w_{\textrm{CDF}}^{90-10}$) and 10\%-to-50\% ($w_{\textrm{CDF}}^{50-10}$) CDF widths.
The flattening process reduces the dependence on the drift time.
The distributions of the normalized flattened CDF widths $\mathscr{W}_{90} \equiv (w_{\textrm{CDF}}^{90-10} - M(w_{\textrm{CDF}}^{90-10}) ) / M(w_{\textrm{CDF}}^{90-10})$ and $\mathscr{W}_{50} \equiv (w_{\textrm{CDF}}^{50-10} - M(w_{\textrm{CDF}}^{50-10}) ) / M(w_{\textrm{CDF}}^{50-10})$, where $M$ function gets the median of the variable from simulation, are displayed in Fig.~\ref{fig:diffusion-cut}.
The parameter region between the 0.5\% and 99.5\% contours are selected.

The scintillation light can be generated anywhere in the sensitive volume and emit isotropically.
A recoil taking place near the cathode/gate will collect more light on the bottom/top PMT array due to the solid angle.
Hence such a geometric relation of the scintillation light is illustrated by a selection cut in terms of $S1$ TBA as a function of the electron cloud drift time.

\subsection{Single scatter}
\label{subsec:single_scatter}

Neutrons generated through spontaneous fission and the ($\alpha$, n) reaction of radioactive materials within the detector can undergo MS interactions within the sensitive volume.
A crucial parameter for categorizing a single-scattering (SS) event is the count of $S2$ signals that meet specific quality selection criteria.
Only $S2$s with a full width greater than 0.8~$\mu$s that fall within certain TBA parameter ranges for $S2$ will contribute to the count.
To distinguish from the afterglow effect, any $S2$ signal that exhibits a charge value above a predetermined threshold is considered in the analysis
\begin{equation} 
%\leftalignedlabel{test}
    \left\{ 
    \begin{aligned}
    & Q_{S2, i} > 75\,\textrm{PE} \\
    & Q_{S2, i} > 0.06 Q_{S2, \textrm{max}}
    \end{aligned}
    \right.
    \textrm{, for all } i, 
    \label{eq:single_s2}
\end{equation}
where $Q_{S2, i}$ and $Q_{S2, \textrm{max}}$ are uncorrected charges of the $i$-th largest and the largest $S2$, respectively, in the event. 

Besides, in order to further suppress the background due to the material neutrons, MS events with one interaction in the veto region are rejected.
The veto region is between the inner vessel and TPC's PTFE side panel reflector.
These MS events are identified as non-zero photo charges within the $S1$ window in the veto region, and excluded from the analysis.
Approximately 20\% of the neutron events are removed by this veto based on the neutron calibration data.

\subsection{Waveform ``dirtiness''}

The actual recorded waveform in real data within an event time window contains not only the $S1$ signal induced by scintillation light and the $S2$ signal caused by electroluminescence from ionized electrons but also various noise signals.
These noise signals originate from PMT dark counts, micro-discharges, delayed single-electron extractions, and so on, which can degrade the quality of physical signal reconstruction.
Thus, each event must adhere to a basic duty cycle requirement within the event time window.
In order to reduce the interference of noise signals on the reconstruction of the waveform and charge amplitude of physical $S1$ and $S2$ signals, the proportion of physical signals with respect to the entire event total charge is required to reach a certain threshold.
The noise level depends on the data taking status, leading to a different threshold of 72\% and 59\% for ER and NR calibration data, respectively.
ER calibration data have an extra requirement that the charge proportion of nonphysical signals before the major $S2$ must be smaller than 6\%, with respect to the entire event total charge.  
Furthermore, for WIMP search analysis, events with more than one $S1$ signals that can be paired with an $S2$ are removed to eliminate any ambiguity in pairing.
The optimized pulse classification strategy for low energy analysis is not proper for $S1$ much greater than the low energy region of interest ($>500$ PE), and it happens to be classified to a specific signal type called ``Unknown'', which leads to its $S2$ wrongly paired with an small isolated $S1$.
In order to specifically remove such events, the charge of all ``Unknown'' signal before the major $S2$ are required to be less than $4.2$ times the charge of the mis-identified major $S1$.
% For a similar purpose, a higher energy event with $Q_{S1} > 500$~PE becomes wrongly paired lower energy candidate when its $S1$ is mistagged and another isolated S1 takes place before its $S2$, and thus such events are specifically removed.

\subsection{Model of signal selection in signal response model}

The efficiencies of the majority of these selections can be modeled as functions of the $S1$ or $S2$ signals. 
However, certain selections may also depend on additional variables, such as the drift time. 
For example, we require a correlation between the $S2$ width and drift time to satisfy the diffusion principle.
For each simulated event in the signal response model, a weight $w$ is assigned corresponding to its expected efficiency:
\begin{equation}
    w = \epsilon_q \cdot \epsilon_r  \cdot \epsilon_{\textrm{rec}}\cdot \epsilon_\mathrm{ss},
    \label{eq:efficiency}
\end{equation}
where $\epsilon_q$ and $\epsilon_r$ are the products of the quality data selections and the selections that require the $S1$ and $S2$ be in the ROI, respectively.
$\epsilon_{\textrm{rec}}$ is the efficiency of signal reconstruction which is illustrated in Sec.~\ref{sec:signal_reconstruction}.
The single scatter selection efficiency is denoted as $\epsilon_\mathrm{ss}$. 
This selection requires that the largest $S2$ charge ($Q_{S2}^{1st}$) in an event is greater than a certain threshold value. 
The single scatter selection is a distinct data selection criterion designed to retain a majority of the genuine single scatter events while potentially misclassifying a portion of the true multiple scatter events as single scatters.
Due to its unique nature, a simple efficiency measure is insufficient to fully characterize its impact on data purity. 
Thus, the SS selection is directly incorporated into the fast MC simulation of the signal response model.
This enables the extraction of the $S1$ and $S2$ signal charges associated with each deposition cluster, providing a more accurate representation of the detector response to the GEANT4-simulated (G4-simulated)~\cite{Agostinelli2003} event.
In the case of $S1$ signals, their charges are combined since the simulated particles typically have sufficient speed, resulting in energy depositions occurring within a time frame shorter than the scintillation light propagation time and the dimer decay constant.
Unless the $S2$ signals of a particular simulated event satisfy the SS selection criteria given in Eq.~\ref{eq:single_s2}, the efficiency $\epsilon_\mathrm{ss}$ is set to 1 instead of 0. 
Additional details regarding the G4-based simulation and the clustering of energy depositions will be provided in Sec.~\ref{sec:fit_to_data}. 

\section{Fit to data}
\label{sec:fit_to_data}

The signal response model of PandaX-4T is tuned by matching the ($Q^c_{S1}$, $Q^c_{S2_\mathrm{b}}$) distributions between the calibration data and the fast-MC simulation.
To expedite the fitting process, acceleration is employed for the simulation, leveraging the computational capabilities of GPUs.
% Furthermore, the nuisance parameters present in the signal model, which are determined through independent studies, are constrained based on the corresponding uncertainties associated with those studies. 
In this section, we provide a comprehensive description of the tuning process.

\subsection{Calibration data}

The tuning process of the PandaX-4T signal response model incorporates calibration data from three sources: injected ~\ncl{Rn-220} source, external ~\ncl{Am-241}Be neutron source, and a DD neutron generator.
The total numbers of events used in fit are 1921 (2838), 1823 (935), and 1049 (1770) for \ncl{Rn-220}, \ncl{Am-241}Be, and D-D calibration data, respectively, in Run0 (Run1).
For the \,\ncl{Rn-220} calibration, the energy spectrum is assumed to be ``flat'' in the low-energy region due to the dominant decay process of $\beta$ decay from \,\ncl{Pb-212}, which has a relatively high Q value of 584\,keV.
The spatial distribution of \,\ncl{Pb-212} is also assumed to be uniform within the TPC, considering that several hours of data after injection has been remove to allow for sufficient diffusion.
The ~\ncl{Am-241}Be source is positioned outside the stainless steel container of PandaX-4T, at a radial distance of about 80\,cm from the center of the PandaX-4T TPC.
Three separate ~\ncl{Am-241}Be runs are conducted with varying vertical positions of the source relative to the TPC center.
This arrangement ensures neutron events are captured in the top, middle, and bottom regions of the PandaX-4T TPC. 
The emitted neutron energy spectrum from ~\ncl{Am-241}Be is continuous, with the neutron energy ranging from several keV to a few MeV.
In the case of the DD neutron generator, neutrons are transported from outside the water tank to the TPC through a stainless steel pipe surrounded by water during detector operation. 
% The end of the pipe stops a few centimeters away from the outer vessel wall of the cryostat.
% The thickness of this water layer has uncertainty due to the potential expansion of the water tank when fully loaded. 
% The thickness is effectively determined to be approximately 6.25\,cm through a primary matching of energy spectra between the G4-based simulation and the DD calibration data. 
DD calibration is performed with the generator tube oriented perpendicular to the stainless steel pipe, resulting in monoenergetic neutron energy of approximately 2.45\,MeV.
% The neutron/$\beta$ energy spectra, the ($x$, $y$) distributions, and the $z$ distributions of the aforementioned calibration data can be found in Fig.~\ref{fig:calibration_data}.

% \begin{figure}
%     \centering
%     \includegraphics{}
%     \caption{
%     \textcolor{red}{@LYY Four figures: source energy spectra,  ($x$, $y$) distributions, and $z$ distributions of Rn220, AmBe, and DD.}
%     }
%     \label{fig:calibration_data}
% \end{figure}

\subsection{GEANT4 simulation}

\begin{figure}[htp]
    \centering
    \includegraphics[width=0.95\columnwidth]{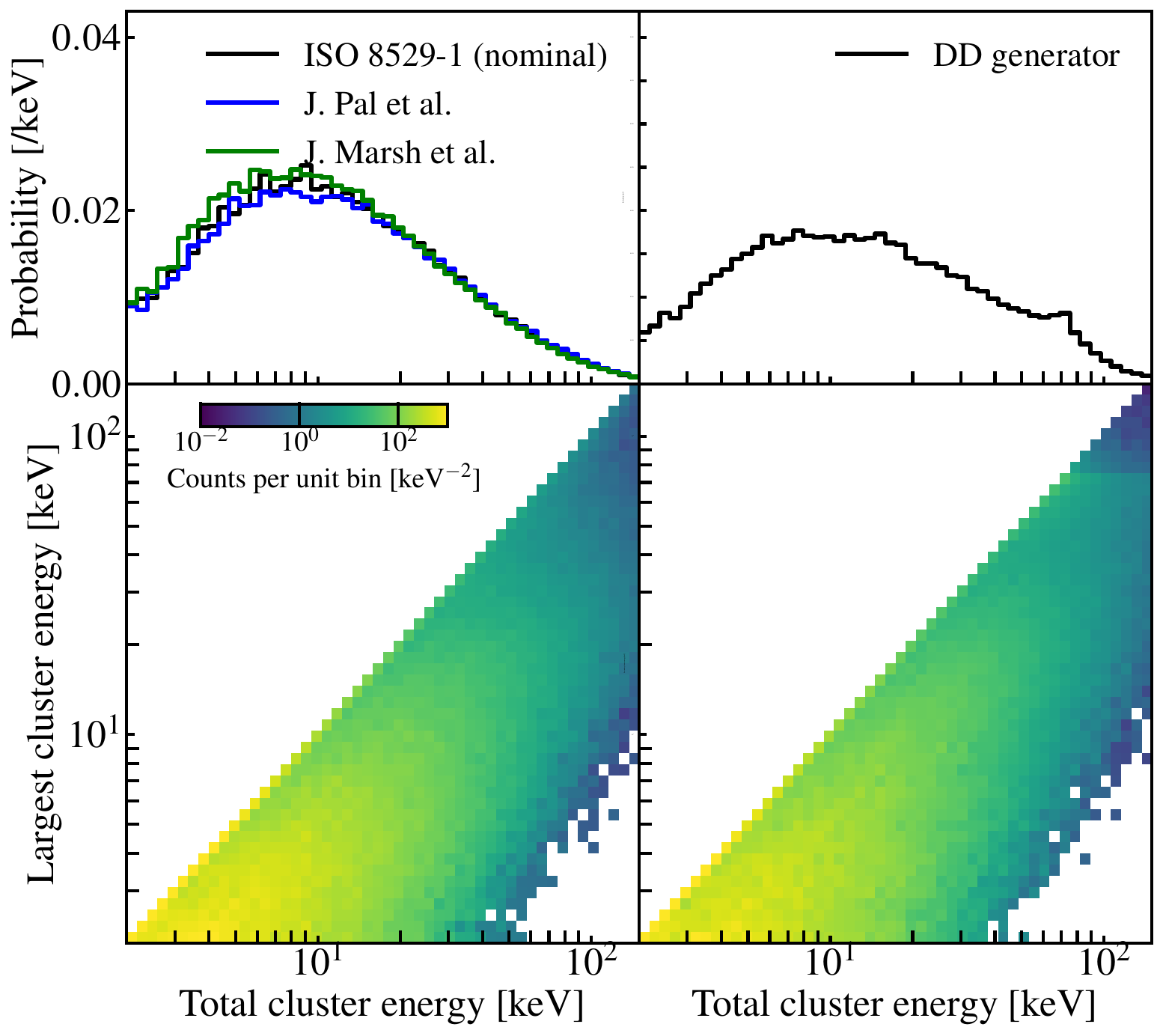}
    \caption{
    The simulation outputs (energy deposition inside the FV) by GEANT4 toolkit are shown.
    The left and right panels display the simulation results for $^{241}$AmBe source and DD generator, respectively.
    The top panels give the total deposit energy spectra, and the bottom panels display the distribution of the largest cluster energy over the total energy.
    The black lines are the nominal spectra used, while the blue and green solid lines are the alternative AmBe spectra from Ref.~\cite{weber2013gentle}.
    }
    \label{fig:ambe_dd_g4_spectra}
\end{figure}

The signal response model for NRs in PandaX-4T is tuned using neutron calibration data.
Due to the large size of the PandaX-4T TPC, neutrons have a considerable probability of undergoing multiple scattering within the detector. 
In the low-energy region, the selection efficiency for SS and the purity of rejecting MS are not optimal.
Thus, the contribution of MS in the selected ``SS'' events is not negligible.
To accurately model the contamination of MS, a dedicated simulation is conducted using the PandaX BambooMC framework~\cite{chen2021bamboomc} based on the GEANT4 toolkits~\cite{Agostinelli2003}.
Comparing the simulation result with NR calibration data, the relative difference is less than $8.5\%$.
This simulation, referred to as the G4-based simulation in the manuscript, takes into account neutron propagation within the TPC.
Furthermore, the capability of SS/MS discrimination is dictated by the $z$ resolution in the TPC, which is determined by the complex signal reconstruction process, including clustering and $S2$ reclustering (as described in Sec.~\ref{sec:signal_reconstruction}). 
The $z$ resolution could depend on various factors such as the size and width of the $S2$ signal, as well as the position in the $(x,y)$ plane.
To incorporate these effects, a specific procedure is followed after the G4-based simulation.
The simulated data are first subjected to a primary clustering algorithm that combines energy depositions with $z$ positions closer than 0.5 mm.
These primary energy clusters then are fed into waveform simulation~\cite{pandax4t_wf}, and subsequently undergo data reconstruction that is the same as used for real data. 
%The output energy clusters depend on the input clusters of the waveform simulation. 
%Thus, we feed the output clusters back to the waveform simulation, and repeat the process.
%This iterative process continues until a converged clustering is achieved. 
The resulting energy clusters for ~\ncl{Am-241}Be and DD calibration data are utilized in the signal response model for model parameter fitting.
Fig.~\ref{fig:ambe_dd_g4_spectra} displays the total deposited energy spectra in the TPC for the ~\ncl{Am-241}Be and DD calibration data. The neutron energies from the ~\ncl{Am-241}Be source are associated with uncertainties, and additional deposit energy spectra are shown for  alternative models with different initial neutron energy spectra, similar to~\cite{weber2013gentle}. Negligible differences in energy spectrum are observed.
In the lower panel of Fig.~\ref{fig:ambe_dd_g4_spectra}, the 2-D distribution of the second largest energy versus the largest energy in the simulated data is also given.
In addition, the end of the pipe used in DD neutron calibration stops a few centimeters away from the outer vessel wall of the cryostat.
The thickness of this water layer has uncertainty due to the potential expansion of the water tank when fully loaded. 
The thickness is effectively determined to be approximately 6.25\,cm through a primary matching of energy spectra between the G4-based simulation and the DD calibration data.

\subsection{Parametrization in signal response model}
\label{subsec:parameterization}

\begin{figure*}
    \centering
    \includegraphics[width=0.95\textwidth]{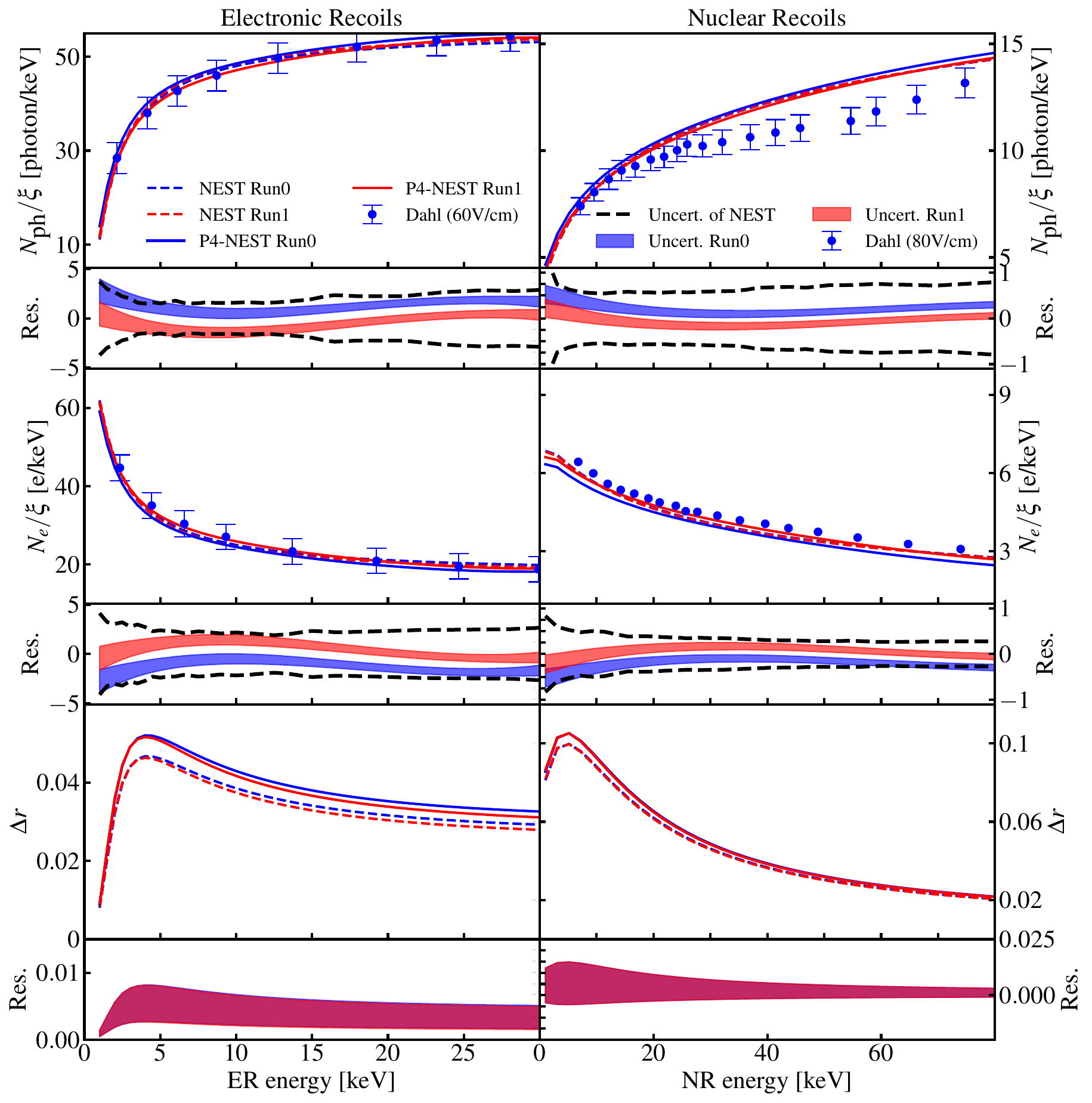}
    \caption{
    The light yield, charge yield, and recombination fluctuation $\Delta r$ as a function of the deposit energy are shown in the top, middle, and bottom panels, respectively.
    The left and right panels correspond to the results for ER and NR, respectively.
    The blue and red lines give the corresponding results for Run0 and Run1 electric field configurations.
    The dashed and solid lines represent the nominal predictions from NESTv2~\cite{NESTv2} and the tuned results in this analysis.
    Small panels beneath each major panels give the uncertainty bands of the residual difference between the tuned results and nominal NEST prediction.
    The black dashed lines indicate the $\pm1\sigma$ uncertainties of the nominal NEST predictions. 
    We also show the light and charge yield measurements from Eric Dahl~\cite{szydagis2023review,CarlEric} with the electric fields of 60 and 80\,V/cm.
    }
    \label{fig:ly_cy}
\end{figure*}

% fixed parameters
The energy and field dependencies of key parameters of intrinsic light and charge generation in LXe for the signal response model are taken from the NESTv2 effective model, as described in Ref.~\cite{NESTv2}. 
These parameters include the excited-atom-to-ion ratio ($\alpha$), the Lindhard factor for nuclear recoils ($L$), the initial mean recombination fraction ($\langle r \rangle_0$), and the initial recombination fluctuation ($\Delta r_0$). 
The exact expressions used can be found in Appendix~\ref{sec:appendix_A}.

% tuned parameters
The signal response model for ERs and NRs incorporates additional degrees of freedom for tuning the light and charge yields.
Specifically, the mean recombination fraction $\langle r \rangle$ is adjusted by adding a 3rd-order Legendre polynomial $P$ multiplied by an exponential function, and the recombination fluctuation $\Delta r$ is scaled by a factor $\lambda$, respectively.
The exponential function is applied to suppress the tuning in high-energy region, which is well understood with global measurements.
Their dependence on the deposit energy $\xi$ can be expressed as: 
\begin{equation}
    \begin{aligned}
    \langle r \rangle (\xi) & =  \langle r \rangle_0 (\xi) + P_3 (\xi /\xi_\mathrm{norm}; p_0, p_1, p_2, p_3)\cdot e^{-\xi/\xi_\mathrm{norm}}, \\
    \Delta r (\xi) & =  \Delta r_0 (\xi) \cdot \lambda,\\
    \xi_\mathrm{norm}^\mathrm{NR} & = 30\,\mathrm{keV},\\
    \xi_\mathrm{norm}^\mathrm{ER} & = 150\,\mathrm{keV},\\
    \end{aligned}
    \label{eq:ER_NR_parametrization}
\end{equation}
where $p_0$, $p_1$, $p_2$, and $p_3$ are the coefficients of the 3rd-order Legendre polynomial function.
The $\xi_{\textrm{norm}}$ is the exponential constant, which are fixed for ERs and NRs, respectively.
The orders of the Legendre polynomial functions are determined so that adding more degree of freedom brings no significant improvement in data/model comparison.
Independent sets of free parameters $p_0$, $p_1$, $p_2$, $p_3$, and $\lambda$ are assigned to ER and NR.
To compare the performance of the NESTv2 nominal model with the PandaX-tuned model (referred to as P4-NEST), we present the mean photon yields $N_{\textrm{ph}}/\xi$ and charge yields $N_e/\xi$, along with the corresponding $\Delta r$, as functions of energy for both ER and NR. 
These results are depicted in Fig.~\ref{fig:ly_cy}.

\subsection{Penalty constraints}
\label{subsec:energy_penalty}

The nuisance parameters employed to characterize the detector effect within the signal response model are acquired through independent studies and are associated with corresponding uncertainties.  
A comprehensive inventory of the free, constrained, and fixed parameters employed in the signal response model can be found in Table~\ref{tab:run0run1_fit_pars}. 
It is worth noting that a specific constraint is applied to the parameters $g_1$ and $g_{2{\textrm{b}}}$. 
The mean reconstructed energies for the three monoenergetic peaks (\,\nclm{Kr-83}, \nclm{Xe-129}, and \nclm{Xe-131}) are required to closely align with their corresponding true energies:
\begin{equation}
    \Lambda = \sum_i \left( E_{\textrm{rec}, i} - E_{\textrm{tr}, i} \right)^2 / (2 \sigma^2_{E_{\textrm{rec},i}}),
\label{eq:constraint_on_rec_energy}
\end{equation}
where $i$ represents the index of the monoenergetic peak.
$E_{\textrm{rec}, i}$ and $E_{\textrm{tr}, i}$ are the reconstructed (following Eq.~\ref{eq:energy_recon}) and true energies, respectively.
$\sigma_{E_{\textrm{rec},i}}$ is the statistical uncertainty of the reconstructed energy following the error propagation $\sigma_{E_{\textrm{rec},i}} = W \sqrt{(\sigma_{cS1, i}/g_1)^2 + (\sigma_{cS2_\mathrm{b}, i}/g_{2\mathrm{b}})^2}$, where $\sigma_{cS1, i}$ and $\sigma_{cS2_\mathrm{b}, i}$ are the statistical uncertainties of the mean corrected signals $Q^c_{S1}$ and $Q^c_{S2_\mathrm{b}}$ for the $i$-th monoenergetic peak.

\subsection{Contamination in the calibration data}
\label{subsec:contamination_neutron_calibration}

Neutrons can scatter inelastically or be captured by atoms in material, resulting in the emission of high-energy gamma rays. 
As a consequence, neutron calibration data are contaminated by ER events.
Additionally, intrinsic gamma rays are emitted alongside neutrons in the case of the \ncl{Am-241}Be source.
To account for these effects, we introduce one degree of freedom for the ratio of ER contamination to NR in each neutron calibration data. 
The energy spectrum of such ERs is assumed to be flat in low-energy region, since they are basically caused by small-angle Compton scatters of MeV gamma rays.
In Run0, we also have remnant tritiated methane during the \ncl{Rn-220} calibration.
A tritium $\beta$ component with a contrained rate is added to the \ncl{Rn-220} simulation in the parameter fitting, with total event of 14$^{+13}_{-0}$ counts.
The uncertainty is estimated as the difference between the tritium rates estimated using the background data taken during and before the \ncl{Rn-220} calibration.

The accidental coincidence (AC) rate can increase in calibration data, particularly during neutron calibrations, due to high event rates.
We estimate the AC rate and its spectral shape using the isolated $S1$ and $S2$ rates~\cite{meng2021dark}. 
% Figure~\ref{fig:ac} illustrates the estimated AC distributions for the \ncl{Am-241}Be and DD calibrations in Run0 and Run1.

% \begin{figure}[htp]
%     \centering
%     \includegraphics[scale=0.4]{plots/AC/AC_distribution.pdf}
%     \caption{
%     \textcolor{red}{
%     (@LYY AC Log10(cS2b/cS1) vs cS1 distributions of AC for AmBe and DD in Run0 and Run1, totally four plots.)
%     }
%     }
%     \label{fig:ac}
% \end{figure}

% \subsection{Fit to calibration data in Run0}

\subsection{Combined fit to calibration data in Run0+Run1}
\label{subsec:run0run1_fit}

% \clearpage
\begin{sidewaystable*}
    \centering
    \small
    \begin{tabular}{c|c|c|c|c|l}
    \hline\hline
    Parameters & Description & Constrain & Nominal & Best-fit & Note \& reference \\
    \hline\hline
    %%%%%%%%%%%%%%%%%
    $p_{\textrm{dpe}}$              &
    Double-PE probability           &
    fixed                           &
    0.22                            &
    -                               &
    (Eq.~\ref{eq:dpe})              \\
    %%%%%%%%%%%%%%%%%
    $\tau_e$                        &
    Electron lifetime               &
    fixed                    &
    -                               &
    -                               &
    Time dependent (Eq.~\ref{eq:tau_e}, Fig.~\ref{fig:tau_e})                          \\
    %%%%%%%%%%%%%%%%%
    $\varepsilon_{\textrm{ext}}$    &
    Electron extraction efficiency  &
    fixed                    &
    -                               &
    -                               &
    Correlated with g$_2$ and $\kappa$ (Eq.~\ref{eq:extraction_efficiency})    \\
    %%%%%%%%%%%%%%%%%
    $\kappa$                        & 
    Electron amplification factor   &
    fixed                    &
    -                               & 
    -                               &
    Time dependent (Eq.~\ref{eq:electron_amplification} \& Fig.~\ref{fig:seg_evolution})                           \\
    %%%%%%%%%%%%%%%%%%
    $g_1$                           &
    $S1$ gain                       & 
    free                            &
    -                               &
    $0.0997^{+0.0002}_{-0.0005}$    &
    \multirow{4}{*}{Extra constraint on reconstructed energy (Eq.~\ref{eq:energy_recon} \& Sec.~\ref{subsec:energy_penalty})}                          \\
    %%%%%%%%%%%%%%%%%%%
    $g_{2\mathrm{b}}$               &
    $S2$ gain                       &
    free                            &
    -                               &
    $4.12^{+0.06}_{-0.04}$          &
                                    \\
    %%%%%%%%%%%%%%%%%%
    $f_{g_1}$                       &
    Scale factor of $S1$ gain in Run1 &
    fixed                           &
    0.90985                         &
    -                               &
                                    \\
    %%%%%%%%%%%%%%%%%%
    $f_{g_{2\mathrm{b}}}$           &
    Scale factor of $S2$ gain in Run1 &
    fixed                           &
    1.22067                         &
    -                               &
                                    \\
    %%%%%%%%%%%%%%%%%%%
    $\varepsilon_{\textrm{hit}}$    &
    Loss probability of 1 hit due to clustering &
    fixed                     &
    -                               &
    -                               &
    $S1$ dependent (Eq.~\ref{eq:hit_clustering_loss} \& Fig.~\ref{fig:s1_s2_bias})                                               \\
    %%%%%%%%%%%%%%%%%%
    $\delta_{S1}^{\textrm{self}}$   &
    self-trigger bias on $S1$       &
    fixed                    &
    -                               &
    -                               &
    \multirow{2}{*}{$S1$ dependent (Eq.~\ref{eq:self_trigger_bias} \& Fig.~\ref{fig:s1_s2_bias})} \\
    %%%%%%%%%%%%%%%%%%
    $\Delta \delta_{S1}^{\textrm{self}}$        &
    Standard deviation of self-trigger bias     &
    fixed$^{\ast}$                  &
    -                               &
    -                               \\
    %%%%%%%%%%%%%%%%%%
    $\delta_{S2}$                   &
    Mean $S2$ reconstruction bias   &
    fixed                    &
    -                               &
    -                               &
    \multirow{2}{*}{$S2$ dependent (Eq.~\ref{eq:s2_bias} \& Fig.~\ref{fig:s1_s2_bias})} \\
    %%%%%%%%%%%%%%%%%%
    $\Delta \delta_{S2}$            &
    Standard deviation of reconstruction bias &
    fixed$^{\ast}$                  &
    -                               &
    -                               \\
    %%%%%%%%%%%%%%%%%%
    $\sigma_{\textrm{pos}}$         &
    Position reconstruction resolution &
    fixed                    &
    -                               &
    -                               &
    $S2$ dependent (Eq.~\ref{eq:position_reconstruction_sample} \& Fig.~\ref{fig:position_resolution}) \\
    %%%%%%%%%%%%%%%%%%
    $\epsilon_q$                    &
    Quality cut efficiency          &
    fixed                    &
    -                               &
    -                               &
    Depend on various variables (Eq.~\ref{eq:efficiency} \& Fig.~\ref{fig:efficiency}) \\
    %%%%%%%%%%%%%%%%%%%
    $\epsilon_r$                    &
    ROI efficiency                  &
    fixed                    &
    -                               & 
    -                               &
    (Eq.~\ref{eq:efficiency} \& Fig.~\ref{fig:efficiency}) \\
    %%%%%%%%%%%%%%%%%%%
    $\epsilon_{\textrm{rec}}$       &
    Signal reconstruction efficiency &
    fixed                    &
    -                               &
    -                               &
    Depend on various variables (Eq.~\ref{eq:efficiency} \& Fig.~\ref{fig:efficiency}) \\
    %%%%%%%%%%%%%%%%%%%%
    $\epsilon_\mathrm{ss}$          &
    Single scatter cut efficiency   &
    fixed                    &
    -                               &
    -                               &
    Special implementation in fast MC (Sec.~\ref{sec:signal_selection}) \\
    %%%%%%%%%%%%%%%%%%%%
    $p_0^{\textrm{ER}}$             &
    \multirow{4}{*}{3rd-order Legendre coefficients for ER} &
    free          &
    -             &
    $1.1\pm0.4$                     &
    \multirow{10}{*}{(Eq.~\ref{eq:ER_NR_parametrization} \& Sec.~\ref{subsec:parameterization})} \\
    %%%%%%%%%%%%%%%%%%%%
    $p_1^{\textrm{ER}}$             &
                                    &
    free          &
    -             &
    $-3.1\pm1.2$                    &
                                    \\
    %%%%%%%%%%%%%%%%%%%%
    $p_2^{\textrm{ER}}$             &
                                    &
    free          &
    -             &
    $2.2^{+0.6}_{-0.8}$             &
                                    \\
    %%%%%%%%%%%%%%%%%%%%
    $p_3^{\textrm{ER}}$             &
                                    &
    free          &
    -             &
    $-1.7^{+0.7}_{-0.6}$            &
                                    \\
    %%%%%%%%%%%%%%%%%%%%
    $p_0^{\textrm{NR}}$             &
    \multirow{4}{*}{3rd-order Legendre coefficients for NR} &
    free          &
    -             &
    $0.7\pm0.3$                     &
                                   \\
    %%%%%%%%%%%%%%%%%%%%
    $p_1^{\textrm{NR}}$             &
                                    &
    free          &
    -             &
    $-1.6^{+0.9}_{-0.7}$            &
                                    \\
    %%%%%%%%%%%%%%%%%%%%
    $p_2^{\textrm{NR}}$             &
                                    &
    free          &
    -             &
    $1.3^{+0.5}_{-0.7}$             &
                                    \\
    %%%%%%%%%%%%%%%%%%%%
    $p_3^{\textrm{NR}}$             &
                                    &
    free          &
    -             &
    $-0.6\pm0.4$                    &
                                    \\
    %%%%%%%%%%%%%%%%%%%%
    $A^{\textrm{ER}}$               &
    Recombination fluctuation scaling for ER &
    free          &
    -             &
    $1.11^{+0.06}_{-0.04}$          &
                                    \\
    %%%%%%%%%%%%%%%%%%%%
    $A^{\textrm{NR}}$                &
    Recombination fluctuation scaling for NR &
    free          &
    -             &
    $1.06^{+0.09}_{-0.10}$          &
                                    \\
    %%%%%%%%%%%%%%%%%%%%
    $R^{\textrm{AmBe,Run0}}_{\textrm{ER}}$   &
    Ratio of ER contamination to NR in Run0 AmBe       &
    free          &
    -             &
    $0.031\pm0.011$                 &
    \multirow{4}{*}{(Sec.~\ref{subsec:contamination_neutron_calibration})}                   \\
    %%%%%%%%%%%%%%%%%%%%
    $R^{\textrm{AmBe,Run1}}_{\textrm{ER}}$   &
    Ratio of ER contamination to NR in Run1 AmBe       &
    free          &
    -             &
    $0.028^{+0.016}_{-0.010}$       &
    \multirow{4}{*}{(Sec.~\ref{subsec:contamination_neutron_calibration})}                   \\
    %%%%%%%%%%%%%%%%%%%%
    $R^{\textrm{DD,Run0}}_{\textrm{ER}}$   &
    Ratio of ER contamination to NR in Run0 DD         &
    free          &
    -             &
    $0.029^{+0.016}_{-0.014}$       &
                       \\
    %%%%%%%%%%%%%%%%%%%%
    $R^{\textrm{DD,Run1}}_{\textrm{ER}}$   &
    Ratio of ER contamination to NR in Run1 DD         &
    free          &
    -             &
    $0.029^{+0.008}_{-0.012}$       &
                       \\
    %%%%%%%%%%%%%%%%%%%%
    $R^{\textrm{Rn,Run0}}_{\textrm{T}}$   &
    Ratio of Tritium's contamination in Run0 Rn         &
    constrained                     &
    0.007$\pm$0.006                 &
    $0.010^{+0.006}_{-0.004}$       &
                       \\
    %%%%%%%%%%%%%%%%%%%%
    $R^{\textrm{AmBe,Run0}}_{\textrm{AC}}$   &
    Ratio of AC contamination to NR in Run0 AmBe       &
    fixed                           &
    0.0038                          &
    -                               &
                        \\
    %%%%%%%%%%%%%%%%%%%%
    $R^{\textrm{AmBe,Run1}}_{\textrm{AC}}$   &
    Ratio of AC contamination to NR in Run1 AmBe       &
    fixed                           &
    0.0076                          &
    -                               &
                        \\
    %%%%%%%%%%%%%%%%%%%%
    $R^{\textrm{DD,Run0}}_{\textrm{AC}}$   &
    Ratio of AC contamination to NR in Run0 DD         &
    fixed                           &
    0.0021                          &
    -                               &
                        \\
    %%%%%%%%%%%%%%%%%%%%
    $R^{\textrm{DD,Run1}}_{\textrm{AC}}$   &
    Ratio of AC contamination to NR in Run1 DD         &
    fixed                           &
    0.0044                          &
    -                               &
                        \\
    %%%%%%%%%%%%%%%%%%%%%                                
    $d_{\textrm{ER}}$               &
    Recombination fraction shift for ER in Run1 &
    \multirow{2}{*}{free}           &
    \multirow{2}{*}{-}              &
    $-0.032^{+0.004}_{-0.004}$      &
    \multirow{2}{*}{(Sec.~\ref{subsec:run0run1_fit})}                               \\
    %%%%%%%%%%%%%%%%%%%%%                                
    $d_{\textrm{NR}}$               &
    Recombination fraction shift for NR in Run1 &
                                    &
                                    &
    $-0.043^{+0.006}_{-0.07}$       &
                                   \\
    
    \hline\hline
    \end{tabular}
    \caption{
    Parameters of the combined fit with Run0 and Run1.
    }
    \label{tab:run0run1_fit_pars}
\end{sidewaystable*}

\begin{figure*}[htp]
    \centering
    \includegraphics[width=0.95\textwidth]{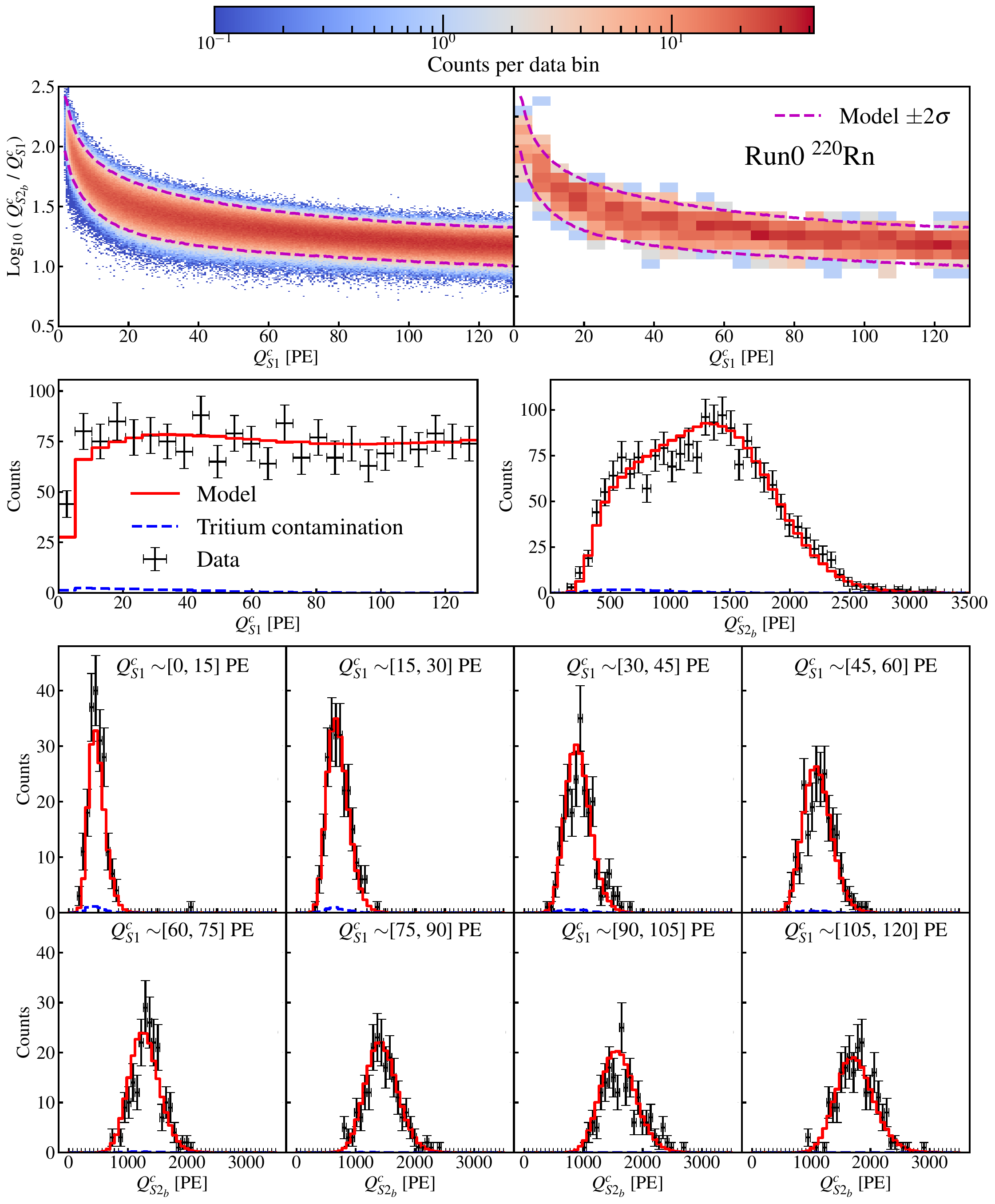}
    \caption{
    \nohyphens{
    The comparison between $^{220}$Rn calibration data in Run0 and best-fit results of P4-NEST model from the simultaneous fit using Run0 and Run1 data.
    % The figure layout and the line denotations are the same as Fig.~\ref{fig:run0_best_fit_comparison_rn}.} 
    The top panels show the distribution of Log$_{10}$($Q_{S2_\mathrm{b}}^c/Q_{S1}^c$) over $Q_{S1}^c$ for the P4-NEST model (left) and the data (right).
    The magenta dashed lines encircles the $\pm$2$\sigma$ region of the NR from the P4-NEST model.
    The middle panels give the $Q_{S1}^c$ and $Q_{S2_\mathrm{b}}^c$ spectra from the data (black error bars) and the P4-NEST model (red solid lines), respectively.
    Blue dashed lines represent the contribution of the residual tritium impurities in the calibration run.
    The lower eight panels give the $Q_{S2_\mathrm{b}}^c$ spectra in different $Q_{S1}^c$ ranges.
    }
    }
    \label{fig:run0run1combined_best_fit_comparison_run0_rn}
\end{figure*}

\begin{figure*}[htp]
    \centering
    \includegraphics[width=0.95\textwidth]{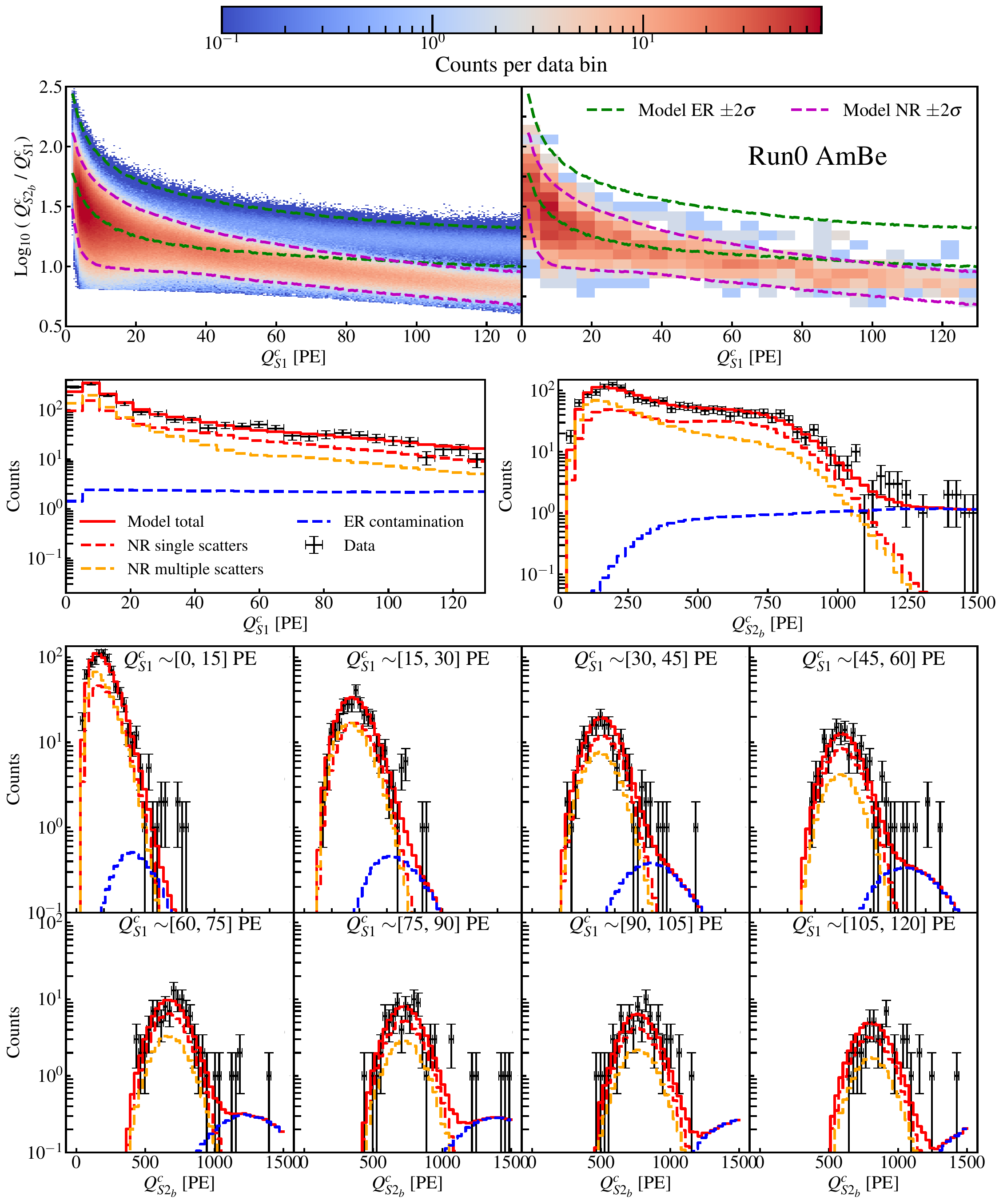}
    \caption{
    The comparison between $^{241}$AmBe neutron calibration data in Run0 and best-fit results of P4-NEST model from the simultaneous fit using Run0 and Run1 data.
    The figure layout and the line denotations are the same as Fig.~\ref{fig:run0run1combined_best_fit_comparison_run0_rn}.
    The green dashed lines are the $\pm$2$\sigma$ boundaries of the ER from P4-NEST.
    The red, orange, and blue dashed lines in the middle and bottom panels are the model spectra of SS NR, MS NR, and ER contamination in the neutron calibration, respectively.
    }
    \label{fig:run0run1combined_best_fit_comparison_run0_ambe}
\end{figure*}

\begin{figure*}[htp]
    \centering
    \includegraphics[width=0.95\textwidth]{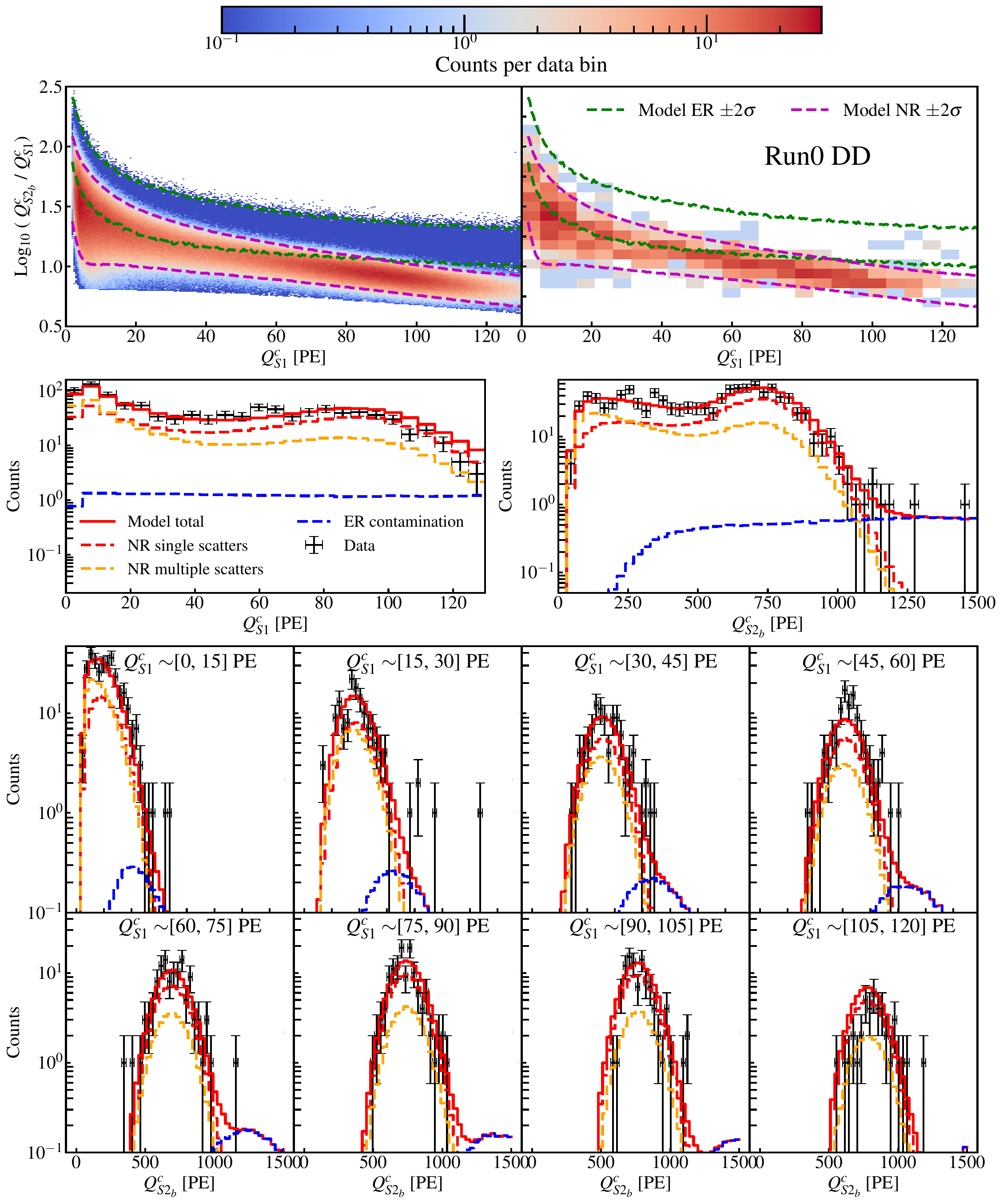}
    \caption{
    The comparison between DD neutron calibration data in Run0 and best-fit results of P4-NEST model from the simultaneous fit using Run0 and Run1 data.
    The figure layout and the line denotations are the same as Fig.~\ref{fig:run0run1combined_best_fit_comparison_run0_ambe}.
    }
    \label{fig:run0run1combined_best_fit_comparison_run0_dd}
\end{figure*}

\begin{figure*}[htp]
    \centering
    \includegraphics[width=0.95\textwidth]{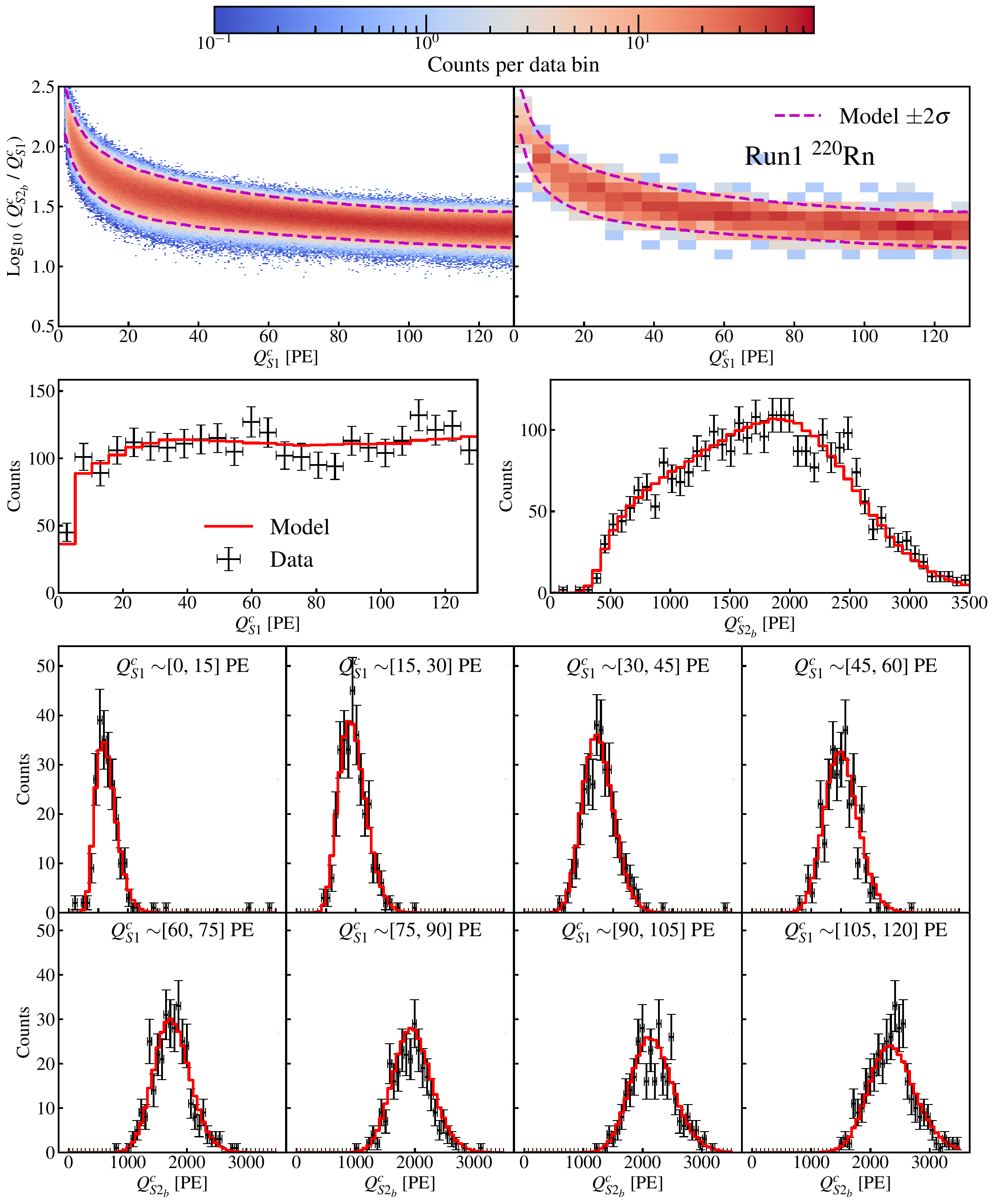}
    \caption{
    The comparison between $^{220}$Rn calibration data in Run1 and best-fit results of P4-NEST model from the simultaneous fit using Run0 and Run1 data.
    The figure layout and the line denotations are the same as Fig.~\ref{fig:run0run1combined_best_fit_comparison_run0_rn}.
    }
    \label{fig:run0run1combined_best_fit_comparison_run1_rn}
\end{figure*}

\begin{figure*}[htp]
    \centering
    \includegraphics[width=0.95\textwidth]{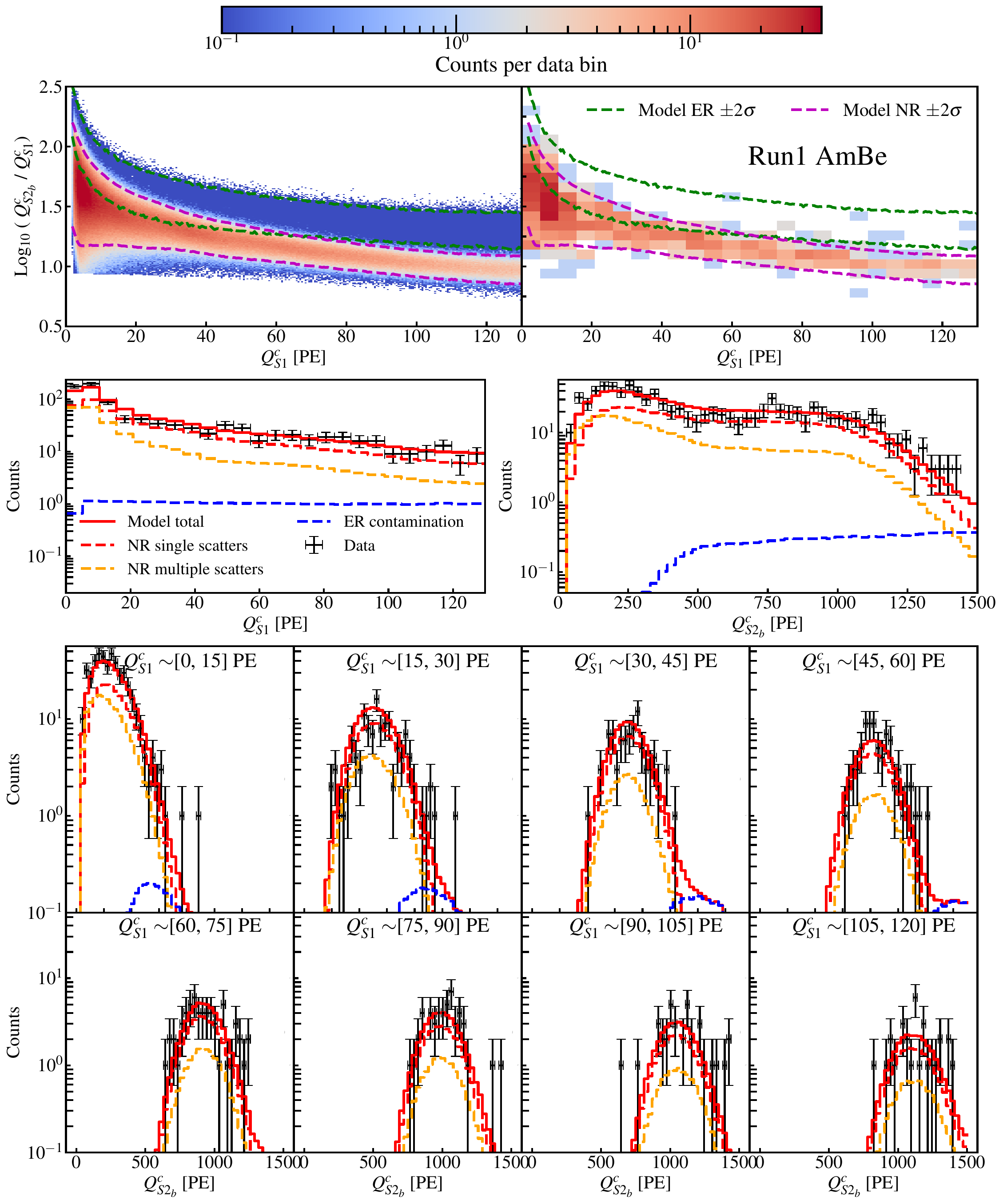}
    \caption{
    The comparison between $^{241}$AmBe neutron calibration data in Run1 and best-fit results of P4-NEST model from the simultaneous fit using Run0 and Run1 data.
    The figure layout and the line denotations are the same as Fig.~\ref{fig:run0run1combined_best_fit_comparison_run0_ambe}.
    } 
    \label{fig:run0run1combined_best_fit_comparison_run1_ambe}
\end{figure*}

\begin{figure*}[htp]
    \centering
    \includegraphics[width=0.95\textwidth]{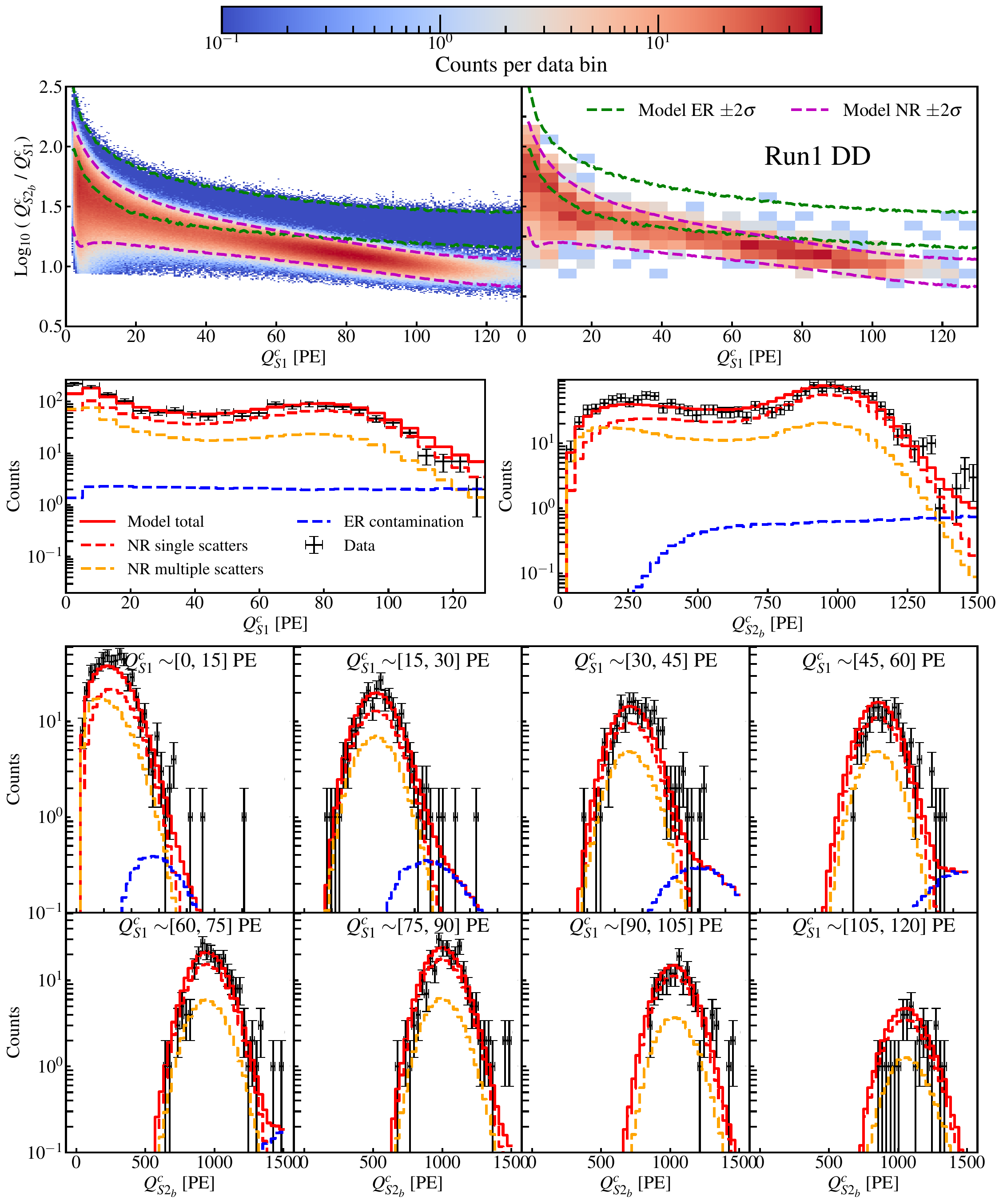}
    \caption{
    The comparison between DD neutron calibration data in Run1 and best-fit results of P4-NEST model from the simultaneous fit using Run0 and Run1 data.
    The figure layout and the line denotations are the same as Fig.~\ref{fig:run0run1combined_best_fit_comparison_run0_ambe}.
    }
    \label{fig:run0run1combined_best_fit_comparison_run1_dd}
\end{figure*}

% how the fit is done
The fit is performed by maximizing the likelihood function:
\begin{equation}
\centering
    \mathcal{L} = \prod_{\alpha,\,\beta} \frac{\lambda_{\alpha \beta}^{N_{\alpha \beta}}}{N_{\alpha \beta} !} e^{-\lambda_{\alpha \beta}} \cdot e^{\Lambda} ,
\label{eq:likelihood}
\end{equation}
where $\alpha$ and $\beta$ represent the indices of the used calibration dataset and the data binning, respectively. 
In our analysis, the 2-D distribution of the $Q_{S2_\mathrm{b}}$ over $Q_{S1}^c$ is fit for tuning the signal model, with 30 bins for the $S1$ from 0 to 135\,PE and 30 bins for the $Q_{S2_\mathrm{b}}$ from 0 to 3000\,PE by default.
$\lambda_{\alpha \beta}$ and $N_{\alpha \beta}$ are the expectation and observed number, respectively, for $\beta$-th bin of the distribution from $\alpha$-th calibration data.
% The fit includes a Gaussian penalty term $G (\boldsymbol{\theta})$, which accounts for a set of constrained nuisance parameters $\boldsymbol{\theta}$.
$\Lambda$ is the constraint on the reconstructed energy which is given by Eq.~\ref{eq:constraint_on_rec_energy}.
To optimize the likelihood, we employ the emcee toolkit~\cite{Foreman-Mackey2013}, which utilizes Goodman \& Weare's Affine Invariant Markov chain Monte Carlo Ensemble sampler~\cite{Goodman2010}. 
During the optimization process, the expected values $\lambda_{\alpha \beta}$ are calculated using dedicated signal response simulation data with sufficient statistics. Specifically, for each iteration, we use a substantial number of events (default: $10^{7}$ for ER calibration and $2\times 10^{6}$ for NR calibration) to ensure accurate estimation.
Due to this iteration-by-iteration simulation causing statistical fluctuation of the simulated samples, the likelihood is further corrected:
\begin{equation}
    \mathcal{L^\prime} = \mathcal{L} / \sqrt{1 + \sigma_{\mathcal{L}}^2},
\end{equation}
where $\sigma_{\mathcal{L}}$ is the relative statistical fluctuation of the likelihood which is typically 1-2 for the ER and NR simulations with default MC statistics. 
They are updated every 500 iterations during the fitting.

% Describe the change in Run1
A simultaneous fit of all the calibration data (~\ncl{Rn-220}, \ncl{Am-241}Be, and DD calibration) in Run0 and Run1 was conducted\footnote{The fit results using only Run0 data can be found in Appendix~\ref{sec:appendix_B}}.
The optimization process by emcee is determined to be converged when the Gelman-Rubin test statistics~\cite{Gelman1992} drops below 1.1.
The parameterization of $\langle r \rangle$ in Run1 is identical to those in Run0, except for that we consider there a constant difference between the $\langle r \rangle$ values of Run1 and Run0. 
This difference is introduced to accommodate the slightly different electric field conditions experienced during Run0 and Run1 in the sensitive regions of the experiment.
Regarding the recombination fluctuation $\Delta r$, it is assumed to have a weak dependence on the electric field. 
Therefore, the field dependency modeled in NESTv2~\cite{NESTv2} is directly adopted for $\Delta r$.
All fixed parameters associated with the detector effects, as discussed in previous sections, have been updated for Run0+Run1 combined fit and are listed in Table~\ref{tab:run0run1_fit_pars}.
As described in Section~\ref{sec:signal_collection}, the values of $g_1$ and $g_{2\mathrm{b}}$ in Run1 were found to exhibit slight deviations from those in Run0. 
This difference could be attributed to various factors, such as the disabling of some PMTs, differences in liquid level, and variations in the electric field configuration during Run1. 
The differences in the mean values of $Q_{S1}^c$ and $Q_{S2_\mathrm{b}}^c$ for the $\alpha$ events between Run0 and Run1 were utilized to constrain the ratio of $g_1$ and $g_{2\mathrm{b}}$ between the two runs.
% \begin{equation}
%     \left\{
%     \begin{aligned}
%     & g_{1, \textrm{Run1}} & =\quad & d_{g_1} + g_{1, \textrm{Run0}}  \\
%     & g_{2\mathrm{b}, \textrm{Run1}} & =\quad & d_{g_{2\mathrm{b}}} + g_{2\mathrm{b}, \textrm{Run0}}.
%     \end{aligned}
% \end{equation}
% \textcolor{red}{LYY:}
\begin{equation}
    \left\{
    \begin{aligned}
    & g_{1, \textrm{Run1}} & =\quad & f_{g_1} \cdot g_{1, \textrm{Run0}}  \\
    & g_{2\mathrm{b}, \textrm{Run1}} & =\quad & f_{g_{2\mathrm{b}}} \cdot g_{2\mathrm{b}, \textrm{Run0}}
    \end{aligned}
    \right.
\end{equation}
The comparisons of the $Q_{S1}^c$ distribution, the $Q_{S2_\mathrm{b}}^c$ distribution, and the $Q_{S2_\mathrm{b}}^c$ distributions at different $Q_{S1}^c$ between the data and the best-fit model (all calibration data  with Run0 and Run1 combined) are shown from Fig.~\ref{fig:run0run1combined_best_fit_comparison_run0_rn} to ~\ref{fig:run0run1combined_best_fit_comparison_run1_dd}.

% \begin{figure}
%     \centering
%     \includegraphics[width=0.95\textwidth]{plots/DataModelComparison/Run0Run1Combined_Run0_Rn.pdf}
%     \includegraphics[width=0.95\textwidth]{plots/DataModelComparison/Run0Run1Combined_Run1_Rn.pdf}
%     \includegraphics[width=0.95\textwidth]{plots/DataModelComparison/Run0Run1Combined_Run0_AmBe.pdf}
%     \includegraphics[width=0.95\textwidth]{plots/DataModelComparison/Run0Run1Combined_Run1_AmBe.pdf}
%     \includegraphics[width=0.95\textwidth]{plots/DataModelComparison/Run0Run1Combined_Run0_DD245.pdf}
%     \includegraphics[width=0.95\textwidth]{plots/DataModelComparison/Run0Run1Combined_Run1_DD245.pdf}
%     \caption{
%     \textcolor{red}{(Comparison of the Log10(S2/S1) vs S1, and S1 spectra,sliced S2 spectra, for \ncl{Rn-220}, \ncl{Am-241}Be and DD. Run0 and Run1 combined.)}
%     }
%     \label{fig:run0run1_best_fit_comparison_1}
% \end{figure}

\section{Summary}
\label{sec:summary}

The article provides a comprehensive description of the signal response model employed in the PandaX-4T experiment. The model encompasses various processes, including the generation of intrinsic photons and electrons within the LXe, the detection and collection of photon and electron signals, as well as the subsequent signal reconstruction, correction, and selection processes.
Considerable effort has been made to ensure a realistic representation of these processes within the signal response model. 
However, it is important to note that the production of photon and electron signals in LXe, particularly in the low-energy range, is subject to significant uncertainty, due to limited measurements in this range.
To address this, the parameters of the signal response model have been refined through a simultaneous fit utilizing all available calibration data from both Run0 and Run1.
Comparing the observed data with the signal response model, a good agreement has been achieved.
Furthermore, future data taking in the PandaX-4T experiment is expected to yield additional information with an upgraded detector. 
This anticipated increase in data volume and improved detector capabilities will enable more stringent constraints on the signal response model, thereby reducing the associated uncertainty.

\section*{Acknowledgements}

%This project is supported in part by grants from National Natural Science
%Foundation of China (Nos. 12090060, 12090061, 12005131, 11905128, 11925502, 12222505, 11835005, 12205181), and by the Office of Science and
%Technology, Shanghai Municipal Government (grant No. 22JC1410100). We thank supports from Double First Class Plan of
%the Shanghai Jiao Tong University.
%We also thank the sponsorship from 
%%the Chinese Academy of Sciences Center for Excellence in Particle
%%Physics (CCEPP), 
%the Hongwen Foundation in Hong Kong, Tencent
%Foundation in China, and Yangyang Development Fund. Finally, we thank the CJPL administration and
%the Yalong River Hydropower Development Company Ltd. for
%indispensable logistical support and other help.

This project is supported in part by grants from National Science
Foundation of China (Nos. 12090060, 12090061, 12205181), a grant from the Ministry of Science and Technology of China (No. 2023YFA1606200),
and by Office of Science and
Technology, Shanghai Municipal Government (grant No. 22JC1410100). 
We thank for the support from Double First Class Plan of
the Shanghai Jiao Tong University. 
We also thank the sponsorship from the Chinese Academy of Sciences Center for Excellence in Particle Physics (CCEPP), Hongwen Foundation in Hong Kong, Tencent
Foundation in China, and Yangyang Development Fund. Finally, we thank the CJPL administration and
the Yalong River Hydropower Development Company Ltd. for
indispensable logistical support and other help.

% % We thank Kaixuan Ni and Yuehuan Wei for helpful discussions. 
% Q.L. is supported by One Thousand Talent Program for Young Scholars.

%\input{main.bbl}
%\bibliographystyle{apssamp}
%\bibliography{reference.bib}

%\input{appendix}
\appendix

\section{Parameterization of recombination}
\label{sec:appendix_A}

The key parameters, which describe the recombination process in LXe, include the excited-atom-to-ion ratio ($\alpha$), the Lindhard factor for nuclear recoils ($L$), the initial mean recombination fraction ($\langle r \rangle_0$), and the initial recombination fluctuation ($\Delta r_0$).
The expressions for these parameters to ER and NR, respectively, are shown in Eq.~\ref{eq:fixed_par_er} and~\ref{eq:fixed_par_nr}, with $\rho_{\textrm{Xe}}$ and $E$ being the LXe density and field strength.
$\Phi$ represents the error function.

\begin{figure*}
\centering 
\begin{equation}
\centering
% \fontsize{8}{10}
\begin{aligned}
& \left\{
    \begin{aligned}
    \alpha& =  \left(0.067366 + 0.093963 \rho_{\textrm{Xe}} \right) \cdot \Phi(0.05\xi) \\
    \langle r \rangle_0^{\textrm{ER}}& = 1 - \frac{\langle N_\textrm{e} \rangle_{\textrm{ER}}}{\langle N_i \rangle_{\textrm{ER}}} \\
    \Delta r_0^{\textrm{ER}}& =  A e^{-(\zeta_{\textrm{ER}}-0.5)^2/0.084} \left( 1+\Phi(-0.6899(\zeta_{\textrm{ER}} - 0.5) )\right)
    \end{aligned} \right. \\
    \textrm{for ER, where} &
    \left\{
    \begin{aligned}
    \eta & = 1 + \frac{0.4607}{(1+(E/621.74)^{-2.2717})^{53.503}} \\
    Y_0 & = \frac{1000}{W} + 6.5\left( 1 - \frac{1}{1+(E/47.408)^{1.9851}} \right) \\
    Y_1 & = 32.99 \eta \left( 1 - \frac{1}{1+(E/(0.02672 e^{\rho_{\textrm{Xe}}/0.3393}))^{0.6705}} \right)\\
    \tau & = \left( 1652.264+\frac{1.145935e10-1652.3}{1+(E/0.02673)^{1.564691}} \right) \xi^{-2} \\
    \langle N_\textrm{e} \rangle_{\textrm{ER}} & = \xi \left( Y_1+
    \frac{Y_0 - Y_1}{(1+1.304\xi^{2.1393})^{0.35535}} +
    \frac{28}{1+\tau} \right) \\
    \langle N_i \rangle_{\textrm{ER}} & = 1000 \xi / (W \alpha) \\
    A & = 0.1383 - 0.09583 / \left( 1+(E/1210.)^{1.25}\right) \\
    \zeta_{\textrm{ER}} & = \langle N_\textrm{e} \rangle_{\textrm{ER}} W / (1000\xi)
    \end{aligned} \right.
\end{aligned}
\label{eq:fixed_par_er},
\end{equation}

\begin{equation}
\centering
% \fontsize{8}{10}
\begin{aligned}
& \left\{
    \begin{aligned}
    \alpha & = \frac{(\langle N_\textrm{e} \rangle_{\textrm{NR}}+ \langle N_{\textrm{ph}} \rangle_{\textrm{NR}})\varsigma }{\langle N_i \rangle_{\textrm{NR}}} - 1\\
    L & = (\langle N_{\textrm{ph}}\rangle_{\textrm{NR}}+\langle N_\textrm{e}\rangle_{\textrm{NR}}) W/\xi\\
    \langle r \rangle_0^{\textrm{NR}} & = 1 - \frac{\langle N_\textrm{e} \rangle_{\textrm{NR}}}{\langle N_i \rangle_{\textrm{NR}}}\\
    \Delta r_0^{\textrm{NR}} & = 0.1 e^{-(\zeta_{\textrm{NR}}-0.5)^2/0.0722}
    \end{aligned} \right. \\
    \textrm{for NR, with} &
\left\{
    \begin{aligned}
    \varsigma & = 0.0480E^{-0.0533}(\rho_{\textrm{Xe}}/2.90)^{0.30} \\
    \langle N_\textrm{e} \rangle_{\textrm{NR}} & = \xi \left( 1 - \frac{1}{1+(\xi/0.3)^2} \right) / (\varsigma \sqrt{\xi + 12.6}) \\
    \langle N_{\textrm{ph}} \rangle_{\textrm{NR}} & = \left( 11.0\xi^{1.1} - \langle N_\textrm{e} \rangle_{\textrm{NR}}\right) \left( 1 - \frac{1}{1+(\xi/0.3)^2}\right) \\
    \langle N_i \rangle_{\textrm{NR}} & = 4 \left( e^{\langle N_\textrm{e} \rangle_{\textrm{NR}} \varsigma / 4} - 1 \right) / \varsigma \\
    \zeta_{\textrm{NR}} & = \langle N_\textrm{e} \rangle_{\textrm{NR}} W / (1000\xi)
    \end{aligned} \right.,
\end{aligned}
\label{eq:fixed_par_nr}
\end{equation}

\end{figure*}

\section{Fit results with Run0 data only}
\label{sec:appendix_B}

\begin{figure*}[htp]
    \centering
    \includegraphics[width=0.95\textwidth]{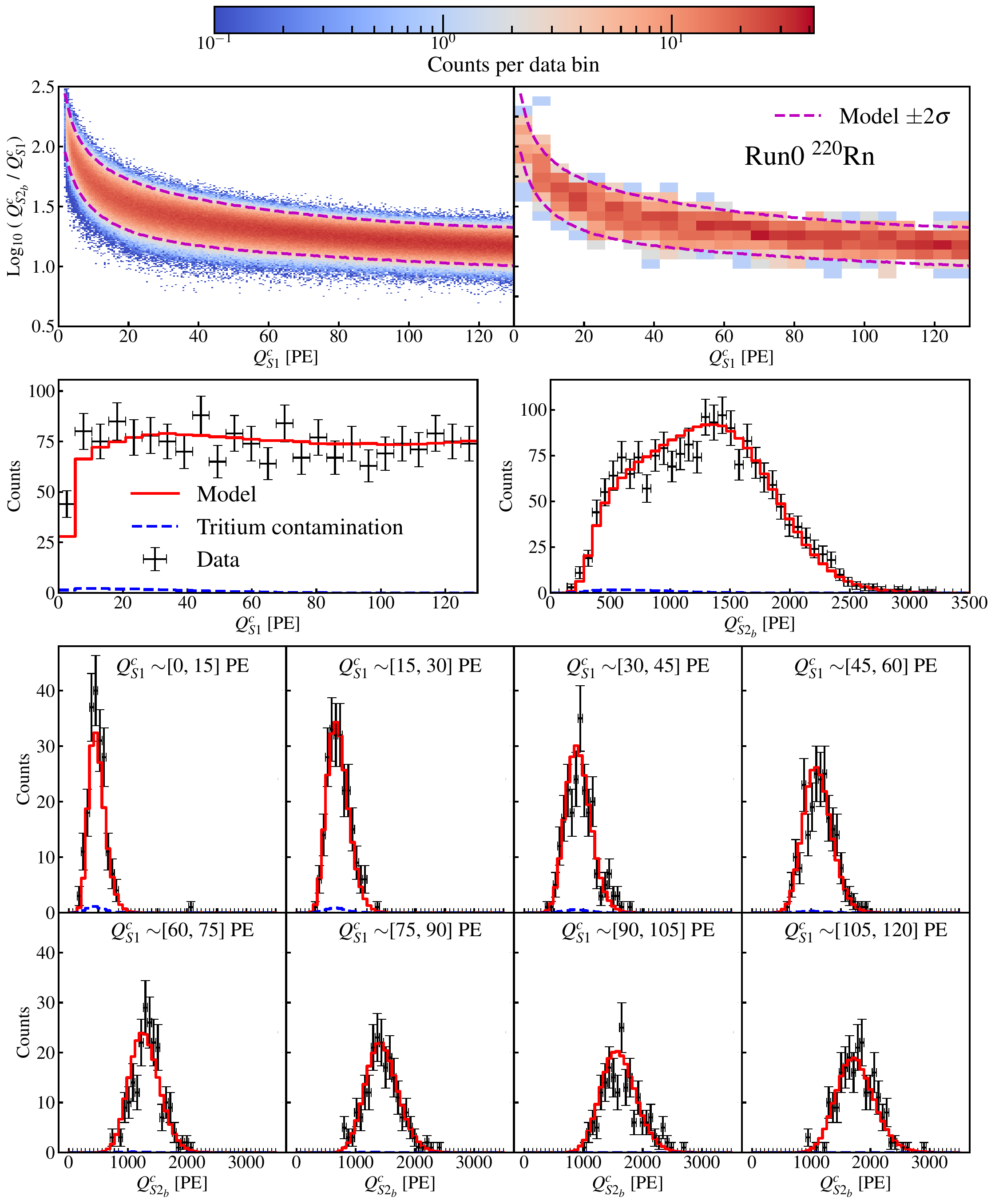}
    \caption{
    The comparison between $^{220}$Rn calibration data in Run0 and best-fit results of P4-NEST model.
    The figure layout and the line denotations are the same as Fig.~\ref{fig:run0run1combined_best_fit_comparison_run0_rn}.
    }
    \label{fig:run0_best_fit_comparison_rn}
\end{figure*}

\begin{figure*}[htp]
    \centering
    \includegraphics[width=0.95\textwidth]{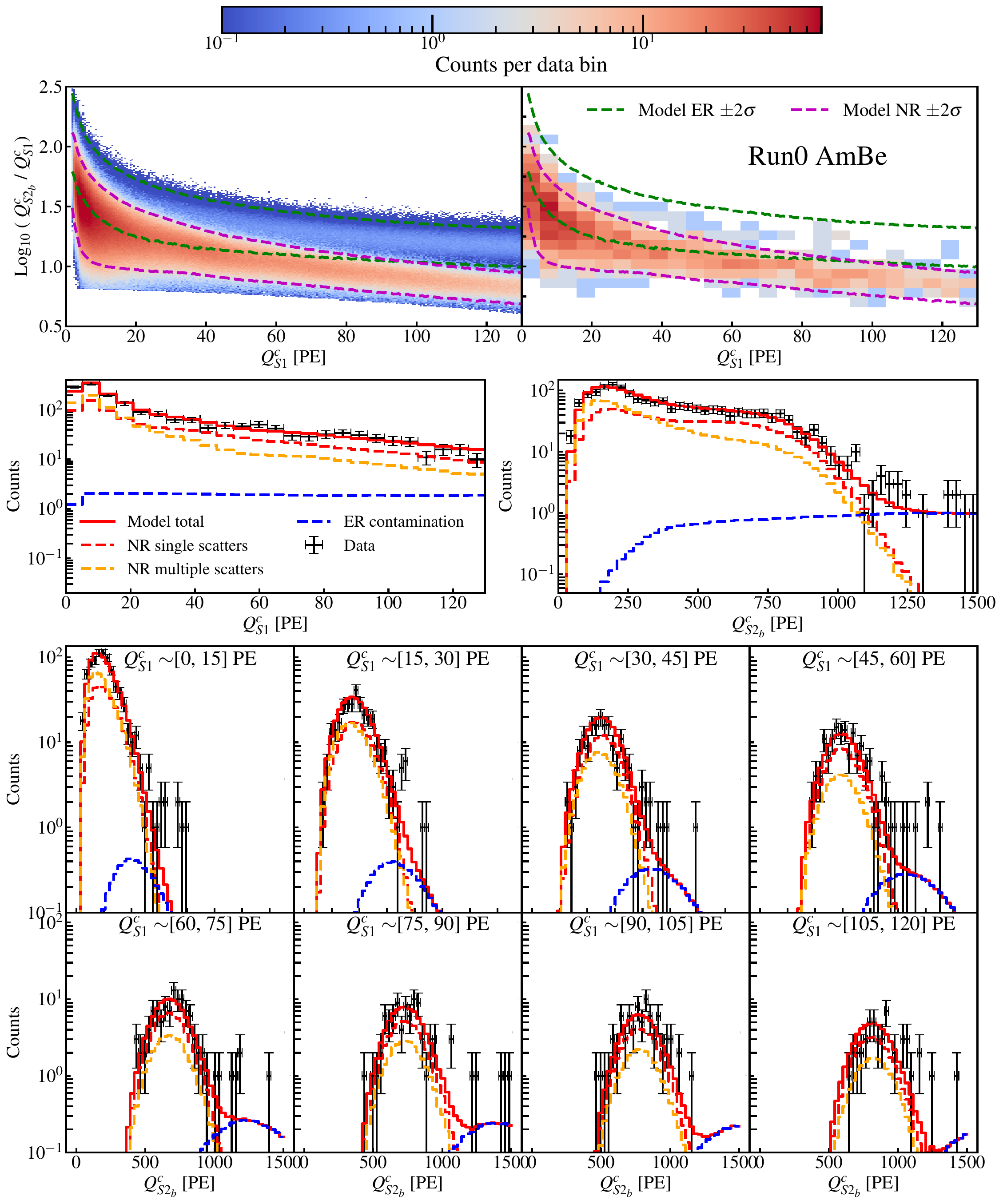}
    \caption{
    The comparison between $^{241}$AmBe calibration data in Run0 and best-fit results of P4-NEST model.
    The figure layout and the line denotations are the same as Fig.~\ref{fig:run0run1combined_best_fit_comparison_run0_ambe}.
    }
    \label{fig:run0_best_fit_comparison_ambe}
\end{figure*}

\begin{figure*}[htp]
    \centering
    \includegraphics[width=0.95\textwidth]{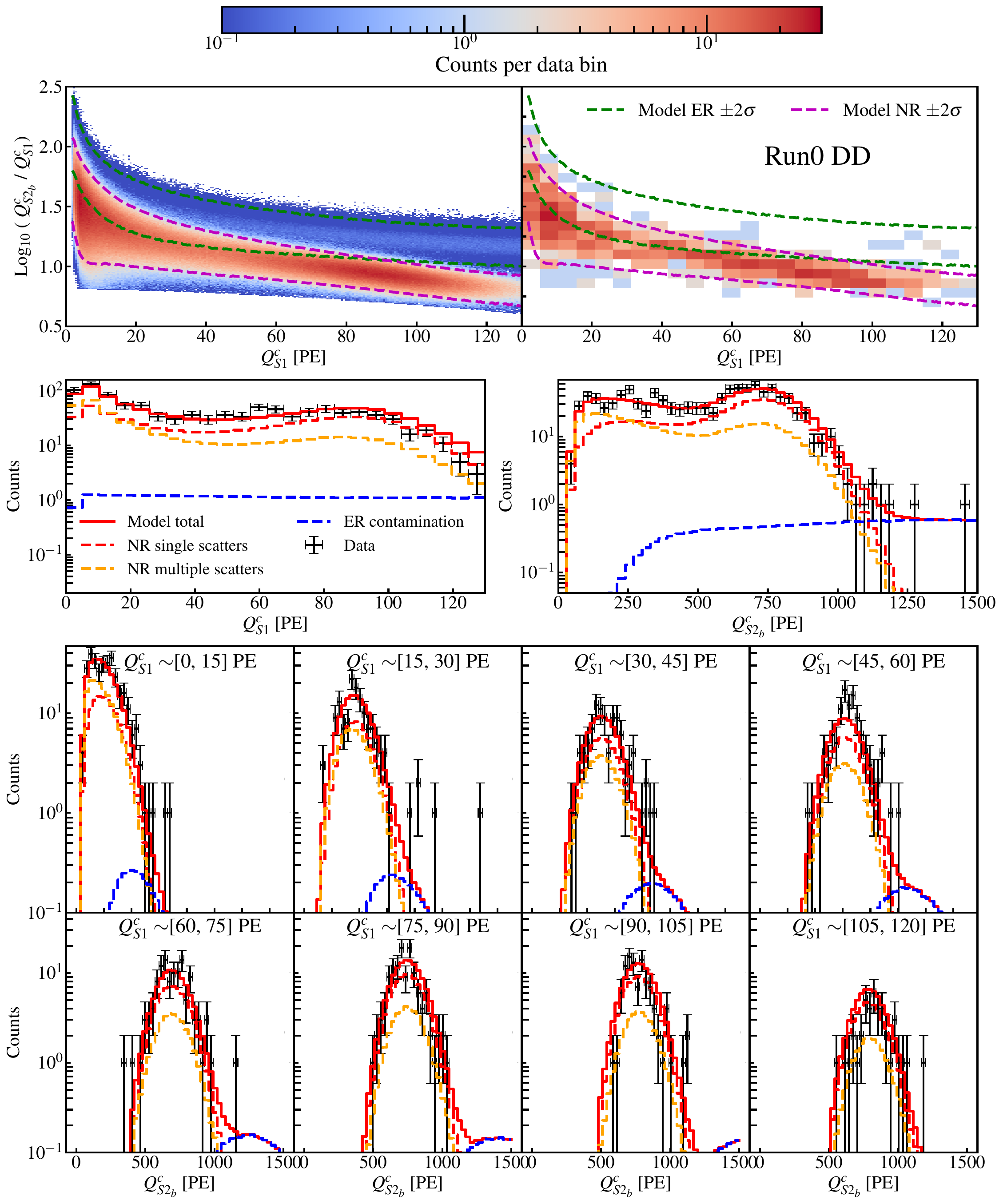}
    \caption{
    The comparison between DD neutron calibration data in Run0 and best-fit results of P4-NEST model.
    The figure layout and the line denotations are the same as Fig.~\ref{fig:run0run1combined_best_fit_comparison_run0_ambe}.
    }
    \label{fig:run0_best_fit_comparison_dd}
\end{figure*}

A combined fit to all the calibration data (~\ncl{Rn-220}, \ncl{Am-241}Be, and D-D calibration) in Run0 is performed.
Table~\ref{tab:run0_fit_pars} summarizes the nominal values and best-fit values of both the free and constrained parameters obtained from the fits.
The best-fit parameters are in good consistency with the nominals.
The comparisons of the $Q_{S1}^c$ distribution, the $Q_{S2_\mathrm{b}}^c$ distribution, and the $Q_{S2_\mathrm{b}}^c$ distributions at different $Q_{S1}^c$ between the data and the best-fit model (all calibration data combined) are shown in Fig.~\ref{fig:run0_best_fit_comparison_rn}, Fig.~\ref{fig:run0_best_fit_comparison_ambe} and Fig.~\ref{fig:run0_best_fit_comparison_dd}.

\clearpage
\begin{sidewaystable*}
    \centering
    \small
    \begin{tabular}{c|c|c|c|c|l}
    \hline\hline
    Parameters & Description & Constrain & Nominal & Best-fit & Note \& reference \\
    \hline\hline
    %%%%%%%%%%%%%%%%%
    $p_{\textrm{dpe}}$              &
    Double-PE probability           &
    fixed                           &
    0.22                            &
    -                               &
    (Eq.~\ref{eq:dpe})              \\
    %%%%%%%%%%%%%%%%%
    $\tau_e$                        &
    Electron lifetime               &
    fixed$^\ast$                    &
    -                               &
    -                               &
    Time dependent (Eq.~\ref{eq:tau_e}, Fig.~\ref{fig:tau_e})                          \\
    %%%%%%%%%%%%%%%%%
    $\varepsilon_{\textrm{ext}}$    &
    Electron extraction efficiency  &
    fixed$^\ast$                    &
    -                               &
    -                               &
    Correlated with g$_2$ and $\kappa$ (Eq.~\ref{eq:extraction_efficiency})   \\
    %%%%%%%%%%%%%%%%%
    $\kappa$                        & 
    Electron amplification factor   &
    fixed$^\ast$                    &
    -                               & 
    -                               &
    Time dependent (Eq.~\ref{eq:electron_amplification} \& Fig.~\ref{fig:seg_evolution})                           \\
    %%%%%%%%%%%%%%%%%%
    $g_1$                           &
    $S1$ gain                       & 
    free                            &
    -                               &
    $0.0998^{+0.0011}_{-0.0009}$    &
    \multirow{2}{*}{Extra constraint on reconstructed energy (Eq.~\ref{eq:energy_recon} \& Sec.~\ref{subsec:energy_penalty})}                          \\
    %%%%%%%%%%%%%%%%%%%
    -                               &
    $S2$ gain                       &
    free                            &
    -                               &
    $3.95^{+0.12}_{-0.14}$         
                                    \\
    %%%%%%%%%%%%%%%%%%%
    $\varepsilon_{\textrm{hit}}$    &
    Loss probability of 1 hit due to clustering &
    fixed$^\ast$                     &
    -                               &
    -                               &
    $S1$ dependent (Eq.~\ref{eq:hit_clustering_loss} \& Fig.~\ref{fig:s1_s2_bias})                                               \\
    %%%%%%%%%%%%%%%%%%
    $\delta_{S1}^{\textrm{self}}$   &
    self-trigger bias on $S1$       &
    fixed$^\ast$                    &
    -                               &
    -                               &
    \multirow{2}{*}{$S1$ dependent (Eq.~\ref{eq:self_trigger_bias} \& Fig.~\ref{fig:s1_s2_bias})} \\
    %%%%%%%%%%%%%%%%%%
    $\Delta \delta_{S1}^{\textrm{self}}$        &
    Standard deviation of self-trigger bias     &
    fixed$^{\ast}$                  &
    -                               &
    -                               \\
    %%%%%%%%%%%%%%%%%%
    $\delta_{S2}$                   &
    Mean $S2$ reconstruction bias   &
    fixed$^\ast$                    &
    -                               &
    -                               &
    \multirow{2}{*}{$S2$ dependent (Eq.~\ref{eq:s2_bias} \& Fig.~\ref{fig:s1_s2_bias})} \\
    %%%%%%%%%%%%%%%%%%
    $\Delta \delta_{S2}$            &
    Standard deviation of reconstruction bias &
    fixed$^{\ast}$                  &
    -                               &
    -                               \\
    %%%%%%%%%%%%%%%%%%
    $\sigma_{\textrm{pos}}$         &
    Position reconstruction resolution &
    fixed$^\ast$                    &
    -                               &
    -                               &
    $S2$ dependent (Eq.~\ref{eq:position_reconstruction_sample} \& Fig.~\ref{fig:position_resolution}) \\
    %%%%%%%%%%%%%%%%%%
    $\epsilon_q$                    &
    Quality cut efficiency          &
    fixed$^\ast$                    &
    -                               &
    -                               &
    Depend on various variables (Eq.~\ref{eq:efficiency} \& Fig.~\ref{fig:efficiency}) \\
    %%%%%%%%%%%%%%%%%%%
    $\epsilon_r$                    &
    ROI efficiency                  &
    fixed$^\ast$                    &
    -                               &
    -                               &
    (Eq.~\ref{eq:efficiency} \& Fig.~\ref{fig:efficiency}) \\
    %%%%%%%%%%%%%%%%%%%
    $\epsilon_{\textrm{rec}}$       &
    Signal reconstruction efficiency &
    fixed$^\ast$                    &
    -                               &
    -                               &
    Depend on various variables (Eq.~\ref{eq:efficiency} \& Fig.~\ref{fig:efficiency}) \\
    %%%%%%%%%%%%%%%%%%%%
    $\epsilon_\mathrm{ss}$          &
    Single scatter cut efficiency   &
    fixed$^\ast$                    &
    -                               &
    -                               &
    Special implementation in fast MC (Sec.~\ref{sec:signal_selection}) \\
    %%%%%%%%%%%%%%%%%%%%
    $p_0^{\textrm{ER}}$             &
    \multirow{4}{*}{3rd-order Legendre coefficients for ER} &
    \multirow{12}{*}{free}          &
    \multirow{12}{*}{-}             &
    $1.0^{+0.7}_{-0.5}$             &
    \multirow{10}{*}{(Eq.~\ref{eq:ER_NR_parametrization} \& Sec.~\ref{subsec:parameterization})} \\
    %%%%%%%%%%%%%%%%%%%%
    $p_1^{\textrm{ER}}$             &
                                    &
                                    &
                                    &
    $-2.8^{+1.5}_{-2.3}$            &
                                    \\
    %%%%%%%%%%%%%%%%%%%%
    $p_2^{\textrm{ER}}$             &
                                    &
                                    &
                                    &
    $2.0^{+1.5}_{-0.9}$             &
                                    \\
    %%%%%%%%%%%%%%%%%%%%
    $p_3^{\textrm{ER}}$             &
                                    &
                                    &
                                    &
    $-1.5^{+0.8}_{-1.2}$            &
                                    \\
    %%%%%%%%%%%%%%%%%%%%
    $p_0^{\textrm{NR}}$             &
    \multirow{4}{*}{3rd-order Legendre coefficients for NR} &
                                    &
                                    &
    $0.4^{+0.4}_{-0.5}$             &
                                   \\
    %%%%%%%%%%%%%%%%%%%%
    $p_1^{\textrm{NR}}$             &
                                    &
                                    &
                                    &
    $-0.7\pm-1.1$                   &
                                    \\
    %%%%%%%%%%%%%%%%%%%%
    $p_2^{\textrm{NR}}$             &
                                    &
                                    &
                                    &
    $0.6\pm0.8$                     &
                                    \\
    %%%%%%%%%%%%%%%%%%%%
    $p_3^{\textrm{NR}}$             &
                                    &
                                    &
                                    &
    $-0.2^{+0.5}_{-0.6}$            &
                                    \\
    %%%%%%%%%%%%%%%%%%%%
    $A^{\textrm{ER}}$                &
    Recombination fluctuation scaling for ER &
                                    &
                                    &
    $1.16\pm0.07$                 &
                                    \\
    %%%%%%%%%%%%%%%%%%%%
    $A^{\textrm{NR}}$                &
    Recombination fluctuation scaling for NR &
                                    &
                                    &
    $1.09^{+0.12}_{-0.09}$          &
                                    \\
    %%%%%%%%%%%%%%%%%%%%
    $R^{\textrm{AmBe}}_{\textrm{ER}}$   &
    Ratio of ER contamination to NR in AmBe       &
                                    &
                                    &
    $0.027^{+0.008}_{-0.009}$       &
    \multirow{4}{*}{(Sec.~\ref{subsec:contamination_neutron_calibration})}                   \\
    %%%%%%%%%%%%%%%%%%%%
    $R^{\textrm{DD}}_{\textrm{ER}}$   &
    Ratio of ER contamination to NR in DD         &
                                    &
                                    &
    $0.027\pm0.013$                 &
                       \\
    %%%%%%%%%%%%%%%%%%%%
    $R^{\textrm{Rn,Run0}}_{\textrm{T}}$   &
    Ratio of Tritium's contamination in Run0 Rn         &
    constrained                                &
    0.010$\pm$0.006                 &
    $0.015^{+0.005}_{-0.004}$       &
                       \\
    %%%%%%%%%%%%%%%%%%%%
    $R^{\textrm{AmBe}}_{\textrm{AC}}$   &
    Ratio of AC contamination to NR in AmBe       &
    fixed                           &
    0.0038                          &
    -                               &
                        \\
    %%%%%%%%%%%%%%%%%%%%
    $R^{\textrm{DD}}_{\textrm{AC}}$   &
    Ratio of AC contamination to NR in DD         &
    fixed                           &
    0.0021                          &
    -                               &
                        \\

    \hline\hline
    \end{tabular}
    \caption{
    Parameters of the Run0-only fit.
    }
    \label{tab:run0_fit_pars}
\end{sidewaystable*}

\end{document}